\documentclass[accepted]{article}

\usepackage{aistats2024}

\usepackage{amsmath,amssymb,amsfonts}
\allowdisplaybreaks[4]
\usepackage{algorithm,algpseudocode}
\usepackage{bm,bbm}
\usepackage{booktabs}
\usepackage{graphicx}
\usepackage{textcomp}
\usepackage{xcolor}
\usepackage{subfigure}

\usepackage[normalem]{ulem}
\newtheorem{lemma}{Lemma}
\newtheorem{theorem}{Theorem}

\newtheorem{Cor}{Corollary}
\usepackage{cancel}
\usepackage{array}

\usepackage{amsmath,amsfonts,bm}

\def\eqref#1{equation~\ref{#1}}

\def\1{\bm{1}}

\DeclareMathAlphabet{\mathsfit}{\encodingdefault}{\sfdefault}{m}{sl}
\SetMathAlphabet{\mathsfit}{bold}{\encodingdefault}{\sfdefault}{bx}{n}

\usepackage{makecell}
\usepackage{graphicx}
\usepackage{multirow}
\usepackage[export]{adjustbox}
\usepackage{hyperref}
\usepackage{url}
\usepackage{wrapfig,lipsum,booktabs}
\usepackage{longtable}
\usepackage{colortbl}
\usepackage{enumitem}
\usepackage[round]{natbib}
\renewcommand{\bibname}{References}
\renewcommand{\bibsection}{\subsubsection*{\bibname}}
\allowdisplaybreaks
\newcommand\blfootnote[1]{%
  \begin{NoHyper}
  \renewcommand\thefootnote{}\footnote{#1}%
  \addtocounter{footnote}{-1}%
  \end{NoHyper}
}

\begin{document}

\twocolumn[

\aistatstitle{Solving General Noisy Inverse Problem via Posterior Sampling: \\A Policy Gradient Viewpoint}

\aistatsauthor{Haoyue Tang\textsuperscript{1,*}\And Tian Xie\textsuperscript{1}\And Aosong Feng\textsuperscript{2}\And Hanyu Wang\textsuperscript{3,$\dag$}\And Chenyang Zhang\textsuperscript{1}\And Yang Bai\textsuperscript{1} }

\aistatsaddress{ \textsuperscript{1}Meta AI \And \textsuperscript{2}Yale University  \And \textsuperscript{3}University of Maryland, College Park} 

]

\begin{figure*}[h]
\centering
\begin{tabular}{c@{\hspace{0.1cm}}c@{\hspace{0.1cm}}c@{\hspace{0.1cm}}c@{\hspace{0.1cm}}}
\includegraphics[width=.1\textwidth]{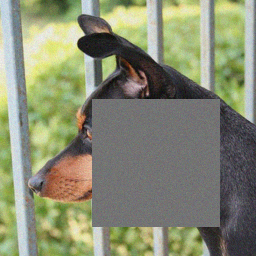}\includegraphics[width=.1\textwidth]{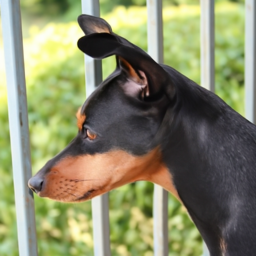}&\includegraphics[width=.1\textwidth]{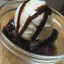}\includegraphics[width=.1\textwidth]{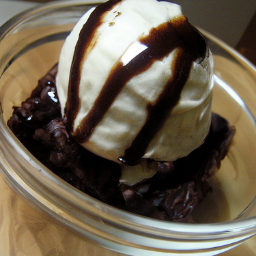}&\includegraphics[width=.1\textwidth]{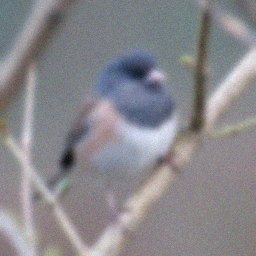}\includegraphics[width=.1\textwidth]{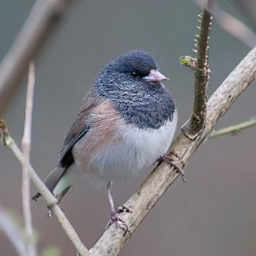}&\includegraphics[width=.1\textwidth]{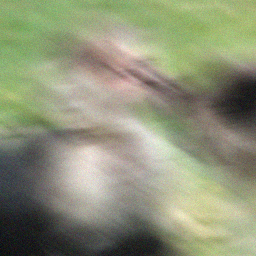}\includegraphics[width=.1\textwidth]{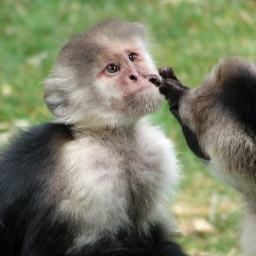}\\
(a) Inpainting & (b) Super-Resolution & (c) Gaussian deblurring & (d) Motion deblurring\\
&\includegraphics[width=.1\textwidth]{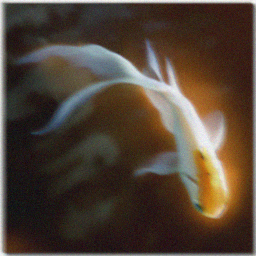}\includegraphics[width=.1\textwidth]{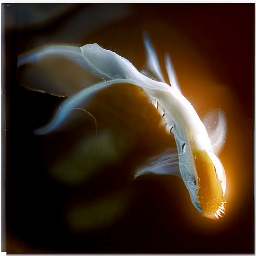}& \multicolumn{2}{c}{\includegraphics[width=.1\textwidth]{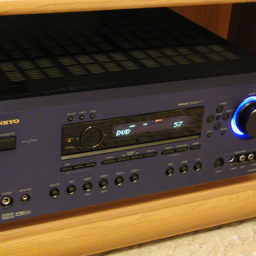}\includegraphics[width=.2\textwidth]{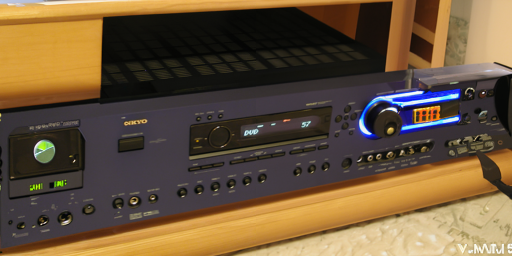}}\\
& (e) Non-linear deblurring & \multicolumn{2}{c}{(f) Uncropping}
\end{tabular}
\caption{{\small Examples on solving noisy image inverse problems on ImageNet validation set using our proposed method without task specific model finetuning or training.}}
\label{fig:intro}
\end{figure*}

\begin{abstract}
\hspace{-12pt}\blfootnote{\hspace{-17pt}\textsuperscript{$\dag$}Work done during an internship at Meta AI. \\ \textsuperscript{*}Email: tanghaoyue13@tsinghua.org.cn}
Solving image inverse problems (e.g., super-resolution and inpainting) requires generating a high fidelity image that matches the given input (the low-resolution image or the masked image). By using the input image as guidance, we can leverage a pretrained diffusion generative model to solve a wide range of image inverse tasks without task specific model fine-tuning. To precisely estimate the guidance score function of the input image, we propose Diffusion Policy Gradient (DPG), a tractable computation method by viewing the intermediate noisy images as policies and the target image as the states selected by the policy. Experiments show that our method is robust to both Gaussian and Poisson noise degradation on multiple linear and non-linear inverse tasks, resulting into a higher image restoration quality on FFHQ, ImageNet and LSUN datasets. 
\end{abstract}

\section{Introduction and Problem Formulation}
Denoising Diffusion Probabilistic Models (DDPM) \cite{ho2020denoising,sohl2015deep} provide tractable solutions to model an unknown high quality image distribution. Their modeling and generation capabilities have been exploited in a wide range of image inverse problems \cite{dhariwal2021diffusion,https://doi.org/10.48550/arxiv.2204.11824,rombach2021highresolution,kawar2022denoising}, where the objective is to recover a high-quality image corresponding to a given low-resolution or blurred image. However, training a diffusion model from scratch for each inverse task is time-consuming. Alternatively, one can use the input image as a guidance,  and recover the high-quality image using guided diffusion using a pretrained diffusion generative model \cite{ho2021classifierfree,dhariwal2021diffusion}. Nonetheless, when the input image is distorted by random noise, this guidance signal becomes inaccurate. Therefore, solving such noisy inverse problems is challenging. 

We now describe the noisy image inverse problem in more details. Consider ${\bf x}_0$ as a high-quality image with distribution $p_0({\bf x}_0)$. Let ${\bf y}$ be a noisy input image obtained by applying an operator $\mathcal{A}$ to image ${\bf x}_0$, i.e.,
\begin{equation}
{\bf y} = \mathcal{A}({\bf x}_0) + {\bf n},\label{eq:geny}
\end{equation}
where ${\bf n}$ represents distorted random noise. The operator $\mathcal{A}$ is dependent on the specific image inverse tasks. For example, in super-resolution tasks, $\mathcal{A}$ is the downsampling operator; in inpainting problems, the operator $\mathcal{A}$ extracts the unmasked pixels of an image. However, the operator ${\mathcal{A}}$ is often low-rank or invertible, making direct computation of the inverse ${\bf y}$ impossible. Alternatively, one can leverage the Bayes' rule $p_0({\bf x}_0|{\bf y})\propto p_0({\bf x}_0)p_0({\bf y}|{\bf x}_0)$ to sample from both the prior $p_0({\bf x}_0)$ and the likelihood $p_0({\bf y}|{\bf x}_0)$, where the prior $p_0({\bf x}_0)$ is implicitly modeled by a pre-trained diffusion generative model.

There are currently two lines of work in utilizing pre-trained diffusion generative models to solve image inverse problems. The first line of work utilizes the low rank structure of the operator $\mathcal{A}$, and directly plugs the known information ${\bf y}$ into the corresponding space of ${\bf x}_0$. SDEdit \cite{meng2022sdedit} solves image inpainting and stroke based generation tasks by plugging a noisy ${\bf y}$ into a selected starting point of the diffusion generation process. Blended Diffusion \cite{Avrahami_2022_CVPR,avrahami2023} and DiffEdit \cite{couairon2023diffedit} enhance image inpainting and editing performance by substituting the unmasked pixels of the generated image with the noisy pixels on ${\bf y}$ in every diffusion generation step. To solve a wider range of tasks such as super-resolution and deblurring, researchers further decompose $\mathcal{A}$ using the singular value decomposition (SVD) to obtain its column and null space \cite{song2021solving,wang2023ddnm,kawar2022denoising}. The null space contents is refined with the help of both the pre-trained diffusion generative model and the known column space contents from ${\bf y}$. Specifically, in each diffusion generation step $i$, \cite{kawar2022denoising} fills the column space contents in ${\bf x}_i$ with a noisy input image ${\bf y}$ and then predict ${\bf x}_{i-1}$; \cite{wang2023ddnm} refines the column space of ${\bf x}_i$ using both the current prediction ${\bf x}_i$ and the input ${\bf y}$, and then use the refined ${\bf x}_i$ to denoise and obtain ${\bf x}_{i-1}$. However, those plug-in approaches can only work for linear inverse problems, and each task requires an SVD decomposition of the operator $\mathcal{A}$. To solve a wider range of non-linear inverse problem when the SVD decomposition becomes impossible, another line of research directly use the conditional probability $p_i({\bf y}|{\bf x}_i)$ to guide the generation process \cite{chung2022improving,chung2023diffusion,meng2022diffusion,song_2023_icml,song2023solving,rout2023solving,song2023pseudoinverseguided,yujie2023dear}. Notice that the guidance score function $\nabla_{{\bf x}_i}\log\mathbb{E}_{p_{0|i}({\bf x}_0|{\bf x}_i)}[p_0({\bf y}|{\bf x}_0)]$ is the gradient of the expected cost taken over distribution $p_{0|i}({\bf x}_0|{\bf x}_i)$, and computing $p_{0|i}({\bf x}_0|{\bf x}_i)$ requires run the diffusion generation process from step $i$. To relieve the computation burden, the diffusion posterior sampling (DPS) method \cite{chung2023diffusion} approximates $p_{0|i}({\bf x}_0|{\bf x}_i)$ with a Gaussian distribution  $q_{0|i}({\bf x}_0|{\bf x}_i)=\mathcal{N}({\bm \mu}_i({\bf x}_i), r_i^2{\bf I})$, where the mean ${\bm \mu}_i({\bf x}_i)$ is the minimum mean squared error (MMSE) estimation of the clean image ${\bf x}_0$ given the current noisy image ${\bf x}_i$. The score function, denoted as $\nabla_{{\bf x}_i}\log\mathbb{E}{p_{0|i}({\bf x}_0|{\bf x}_i)}[p_0({\bf y}|{\bf x}_0)]$, is approximated by computing the gradient of $\log p_0({\bf y}|{\bf x}_0)$ at sample ${\bf x}_0={\bm \mu}_i({\bf x}_i)$ with the highest density mass in distribution $q_{0|i}({\bf x}_0|{\bf x}_i)$, i.e., $\nabla_{{\bf x}_i}\log \mathbb{E}_{p_{0|i}({\bf x}_0|{\bf x}_i)}[p_0({\bf y}|{\bf x}_0)]$. In this work, we further improve the estimation of DPS via the policy gradient method by using multiple samples from $q_{0|i}({\bf x}_0|{\bf x}_i)$. 

{\bf Our Contributions}. We propose a new method to estimate the guidance score for solving image inverse problems in each diffusion step, which results in a better image restoration quality and fidelity
 compared with the ground truth image. Our contributions are summarized as follows:
\begin{itemize}[itemsep=2pt,topsep=0pt,parsep=0pt]
\item We redefine each noisy image as a policy, where the predicted clean image serves as a state selected by the policy. DPG is a new approach for estimating the guidance score function given the input image ${\bf y}$.
\item DPG eliminates the need for computing a closed-form pseudo-inverse or performing SVD decomposition. Leveraging a pre-trained diffusion generative model, our method can address a broad range of image inverse problems without requiring task-specific model fine-tuning. Selected image inverse results generated by DPG are presented in Fig.~\ref{fig:intro}.
\item The score function estimated by DPG is theoretically more accurate than DPS, particularly in the earlier stages of the generation process. In experiments, DPG excels in restoring high-frequency details in images. Quantitative evaluations conducted on FFHQ, ImageNet, and LSUN image restoration tasks demonstrate that our proposed method achieves improvements in both image restoration quality and consistency compared to the ground truth.
\end{itemize}

\section{Methodology}
\subsection{Preliminaries: Diffusion Models}\label{sec:diffusionbackground}
Let ${\bf x}_0$ be a high-fidelity image that follows an unknown distribution $p_0({\bf x}_0)$. Distribution $p_0({\bf x}_0)$ can be turned into a Gaussian by gradually adding Gaussian noise by iteratively for $N$ steps. The noise injection in each step $i$ is as follows:
\begin{equation}
    {\bf x}_i=\sqrt{1-\beta_i}{\bf x}_{i-1}+\sqrt{\beta_i}{\bf z}_i, {\bf z}_i\sim\mathcal{N}(0, {\bf I}). 
\end{equation}
When the iterations $N\rightarrow\infty$, the evolution of ${\bf x}_{0:T}$ is then a result of a continuous time stochastic differential equation. Let $T$ be a constant, denote $t_i:=\frac{i}{N}T$ and $\beta(t_i):=\frac{\beta_{i+1}}{\Delta t}$, the value of each noisy image ${\bf x}_i$ is the result of ${\bf x}(t_i)$ obtained by the following stochastic differential equation \cite[Eq.~(11)]{song2021scorebased}:
\begin{equation}
    \text{d}{\bf x}=-\frac{1}{2}\beta(t){\bf x}\text{d}t+\sqrt{\beta(t)}\text{d}{\bf w},t\in[0, T],\label{eq:forwarddiffusion}
\end{equation}
where ${\bf w}$ is a Wiener process. Denote $\overline{\alpha}(t):=\exp\left(-\int_0^t\beta(s)\text{d}s\right)$ and let $\sigma(t):=\sqrt{1-\overline{\alpha}(t)}$, the conditional probability $p_{t|0}({\bf x}(t)|{\bf x}(0))$ is a Gaussian distribution, i.e., 
\begin{equation}p_{t|0}({\bf x}(t)|{\bf x}(0))=\mathcal{N}({\bf x}(t)|\sqrt{\overline{\alpha}(t)}{\bf x}(0), (1-\overline{\alpha}(t)){\bf I}). \label{eq:forwardcond}
\end{equation}At the final step $T$, $\overline{\alpha}(T)\rightarrow 0$ and therefore, ${\bf x}_T\sim \mathcal{N}(0, {\bf I})$ is a Gaussian distribution. 

Generating high resolution image ${\bf x}_0$ is equivalent to sample from distribution $p_0({\bf x}_0)$. An intuitive solution is to run the reverse process of \eqref{eq:forwarddiffusion} by starting from the final step ${\bf{x}}(T)\sim\mathcal{N}(0, {\bf I})$. According to \cite[Eq.~(29)]{song2021scorebased}, the reverse SDE of \eqref{eq:forwarddiffusion} is as follows:
\begin{equation}
    \text{d}{\bf x}=\left[-\frac{\beta(t)}{2}{\bf x}-\beta(t)\nabla_{{\bf x}}\log p_t({\bf x})\right]\text{d}t+\sqrt{\beta(t)}\text{d}{\bf w}, \label{eq:reverseSDE}
\end{equation}
where $p_t({\bf x}(t):=\int p_{t|0}({\bf x}(t)|{\bf x}_0)p({\bf x}_0)\text{d}{\bf x}_0$ is the marginal distribution of noisy image ${\bf x}_t$. 

A high quality image ${\bf x}_0\sim p_0({\bf x}_0)$ can be generated by numerically solving the reverse time SDE \eqref{eq:reverseSDE}. The earliest DDPM solver interpolates $[0, T]$ with $N$ discrete timestamps. Let $t_i:=\frac{i}{N}T$, ${\bf x}_i:={\bf x}(t_i)$ be the $i$-th interpolated time point and value, DDPM generate ${\bf x}_0$ via the following equation:
\begin{align}
    {\bf x}_{i-1}=\frac{1}{\sqrt{\alpha}_i}{\bf x}_t+\frac{1-\alpha_i}{\sqrt{1-\overline{\alpha}_i}}(\sigma_i\nabla_{\bf x}\log p_i({\bf x}_i))+\sqrt{\beta_i}\overline{{\bf z}}_i, \label{eq:DDPM}
\end{align}
where $\overline{{\bf z}}_i\sim\mathcal{N}(0, {\bf I})$ is a random Gaussian noise sampled in step $i\in[N]$, $\beta_i:=\beta(t_i)\Delta t$ and $\overline{\alpha}_i:=\overline{\alpha}(t_i)$, $\sigma_i:=\sqrt{1-\overline{\alpha}_i}$. Coefficient $\alpha_{i-1}:=1-\beta_{i-1}=\exp(-\beta_{i-1})=\exp\left(-\int_{t_{i-1}}^{t_i}\beta(s)\text{d}s\right)=\frac{\overline{\alpha}_i}{\overline{\alpha}_{i-1}}$. 

The Denoising Diffusion Implicit Model (DDIM) accelerates DDPM by generating ${\bf x}_0$ using:
\begin{align}
    {\bf x}_{i-1}=&\frac{\sqrt{\overline{\alpha}_{i-1}}}{\sqrt{\overline{\alpha}_i}}{\bf x}_i+\sqrt{\overline{\alpha}_{i-1}}\left(\sqrt{\frac{1-\overline{\alpha}_{i-1}}{\overline{\alpha}_{i-1}}}-\sqrt{\frac{1-\overline{\alpha}_i}{\overline{\alpha}_i}}\right)\nonumber\\
    &\times(\sigma_i\log p_i({\bf x}_i)). \label{eq:DDIM}
\end{align}

\subsection{Solving Inverse Problems as Posterior Sampling} 
Our goal is to find the high quality image ${\bf x}_0$ that matches the input ${\bf y}$, i.e., find the root of \eqref{eq:geny}. With the priori $p_0({\bf x}_0)$, generating ${\bf x}_0$ that matches ${\bf y}$ can be viewed as sampling from the following conditional probability:
\begin{equation}
    p_0({{\bf x}_0|{\bf y}})=\frac{p_0({\bf x}_0)p_0({\bf y}|{\bf x}_0)}{p({\bf y})}\propto p_0({\bf x}_0)p_0({\bf y}|{\bf x}_0). 
\end{equation}
The conditional probability $p_0({\bf y}|{\bf x}_0)$ is a function of the reconstruction loss denoted by $\ell_{{\bf y}}({\bf x}_0)$, i.e., 
\begin{equation}
    p_0({\bf y}|{\bf x}_0)\propto\exp\left(-\frac{1}{Z}\ell_{{\bf y}}({\bf x}_0)\right), 
\end{equation}
where $Z$ is a constant independent of ${\bf x}_0$ and ${\bf y}$. For image ${\bf y}$ that is distorted by Gaussian random noise, i.e.,  ${\bf n}\sim\mathcal{N}(0, \sigma^2_{\bf y}{\bf I})$, the reconstruction loss $\ell_{{\bf y}}({\bf x}_0)=\Vert {\bf y}-\mathcal{A}({\bf x}_0)\Vert_2^2$. 

We then discuss sampling from $p_0({\bf x}_0|{\bf y})$ by running the reverse of the forward diffusion~\eqref{eq:forwarddiffusion}. Let $p_t({\bf x}(t)|{\bf y}):=\int p_{t|0}({\bf x}(t)|{\bf x}_0)p({\bf x}_0|{\bf y})\text{d}{\bf x}_0$ be the marginal distribution of ${\bf x}_t$ given input image ${\bf y}$. By replacing $\nabla_{\bf x}\log p_t({\bf x})$ in \eqref{eq:reverseSDE} with $\nabla_{\bf x}\log p_t({\bf x}_t|{\bf y})$, we can sample ${\bf x}_0\sim p_0({\bf x}_0|{\bf y})$ through the following SDE:
\begin{equation}
    \text{d}{\bf x}=\left[-\frac{\beta(t)}{2}{\bf x}-\beta(t)\nabla_{\bf{x}}\log p_t({\bf x}(t)|{\bf y})\right]\text{d}t+\sqrt{\beta(t)}\text{d}{\bf w}. \label{eq:condreverseSDE}
\end{equation}
To solve \eqref{eq:condreverseSDE} numerically through the DDPM \eqref{eq:DDPM} or DDIM \eqref{eq:DDIM} method, we need to compute the score function of the ${\bf s}_i({\bf x}_i, {\bf y}):=\nabla_{{\bf x}(t_i)}\log p_{t_i}({\bf x}(t_i)|{\bf y})=\nabla_{{\bf x}_i}\log p_i({\bf x}_i|{\bf y})$, which can be decomposed by:
\begin{align}
&{\bf s}_i({\bf x}_i, {\bf y}):=\sigma_i\nabla_{{\bf x}_i}\log p_{i}({\bf x}_i|{\bf y})=\sigma_i\nabla_{{\bf x}_i}\log p_i({\bf x}_i, {\bf y})\nonumber\\
=&\underbrace{\sigma_i\nabla_{{\bf x}_i}\log p_i({\bf x}_i)}_{{\bm \epsilon}_{\bm \theta}({\bf x}_i, i)}+\sigma_i\nabla_{{\bf x}_i}\log p_i({\bf y}|{\bf x}_i).\label{eq:score}
\end{align}
The first term $\nabla_{{\bf x}_i}\log p_i({\bf x}_i)$ is the score function of the marginal distribution $p_i({\bf x}_i)$, which is learned through the generative model parameterized by ${\bm \theta}$, i.e., $\sigma_i\nabla_{{\bf x}_i}\log p_i({\bf x}_i)={\bm \epsilon}_{\bm \theta}({\bf x}_i, i)$. Computing the second term $\nabla_{{\bf x}_i}\log p_i({\bf y}|{\bf x}_i)$ in \eqref{eq:score} is hard because $p_i({\bf y}|{\bf x}_i)=\int_{{\bf x}_0}p_{0|i}({\bf x}_0|{\bf x}_i) p_0({\bf y}|{\bf x}_0)\text{d}{\bf x}_0$ requires integral over posterior $p_{0|i}({\bf x}_0|{\bf x}_i)$. 

\subsection{Computing $\nabla_{\bf x}\log p_i({\bf x}_i|{\bf y})$ as Policy Gradient}
The computation of the gradient $\nabla_{{\bf x}_i}\log p_i({\bf y}|{\bf x}_i)$ can be decomposed as follows:
\begin{align}
    &\nabla_{{\bf x}_i}\log p_i({\bf y}|{\bf x}_i)\nonumber\\
    =&\nabla_{{\bf x}_i}\left(\log\int p_{0|i}({\bf x}_0|{\bf x}_i)p_0({\bf y}|{\bf x}_0)\text{d}{\bf x}_0\right)\nonumber\\
    =&\frac{1}{\int p_{0|i}({\bf x}_0|{\bf x}_i)p_0({\bf y}|{\bf x}_0)\text{d}{\bf x}_0}\nonumber\\
    &\times\left(\nabla_{{\bf x}_i}\int p_{0|i}({\bf x}_0|{\bf x}_i)p_0({\bf y}|{\bf x}_0)\text{d}{\bf x}_0\right)\nonumber\\
    \propto&\nabla_{{\bf x}_i}\int \underbrace{p_{0|i}({\bf x}_0|{\bf x}_i)}_{\text{State Density Function}}\underbrace{p_0({\bf y}|{\bf x}_0)}_{\text{Cost}}\text{d}{\bf x}_0=:\tilde{{\bf s}}_i({\bf x}_i, {\bf y}). \label{eq:pgdef}
\end{align}
Notice that ${\tilde{\bf s}}_i({\bf x}_i, {\bf y})$ contains the directional information about the guidance score $\nabla_{{\bf x}_i}\log p_i({\bf y}|{\bf x}_i)$. To compute $\tilde{{\bf s}}_i({\bf x}_i, {\bf y})$, notice that the generated image ${\bf x}_0$ is determined by the intermediate noisy image ${\bf x}_i$, and the conditional probability $p_0({\bf y}|{\bf x}_0)$ can be viewed as a cost of ${\bf x}_0$. Therefore, the computation of the score function $\tilde{\bf s}_i({\bf x}_i, {\bf y})$ in~\eqref{eq:pgdef} is closely related to policy gradient in reinforcement learning, where $p_{0|t}({\bf x}_0|{\bf x}_t)$ is the state density function by choosing policy ${\bf x}_i$, and $p_0({\bf y}|{\bf x}_0)$ is the cost. The following theorem enables us to compute the score function~\eqref{eq:pgdef} from the policy gradient perspective:
\begin{theorem}[Leibniz Rule]\label{thm:leibniz}
    Suppose $p_0({\bf x}_0)$ is the probability measure of $N_{\text{train}}, N_{\text{train}}<\infty$ high quality training images. Then for all $i\in[N]$, we can compute the score function ${\tilde{\bf s}}_i({\bf x}_i, {\bf y})$ from \eqref{eq:pgdef} as follows:
\begin{align}
    \hspace{-15pt}\tilde{\bf{s}}_i({\bf x}_i, {\bf y})\!=\!\mathbb{E}_{p_{0|i}({\bf x}_0|{\bf x}_i)}\left[p_0({\bf y}|{\bf x}_0)\nabla_{{\bf x}_i}\log p_{0|i}({\bf x}_0|{\bf x}_i)\right]\!.\! \label{eq:vanillapg}
\end{align}
\end{theorem}
Notice that the diffusion model is usually trained with a finite number of images, Theorem~\ref{thm:leibniz} works for a wide range of pretrained diffusion models. Proof of Theorem~\ref{thm:leibniz} is provided in Appendix~5.1. A benefit for using \eqref{eq:vanillapg} is that we do not need to compute the derivative of $p_0({\bf y}|{\bf x}_0)$ (i.e., the reconstruction loss $\ell_{{\bf y}}({\bf x}_0)$) and therefore do not require $p_0({\bf y}|{\bf x}_0)$ to be differentiable. 

\subsection{Implementation Details}
\subsubsection{Tractable Monte Carlo sampling}Computing the score function in~\eqref{eq:vanillapg} requires sampling clean images ${\bf x}_0$ from a complex distribution $p_{0|i}({\bf x}_0|{\bf x}_i)$ and then compute its density. To facilitate the sampling and computation, 
similar to \cite{chung2023diffusion,song_2023_icml}, we select a Gaussian distribution $q_{0|i}({\bf x}_0|{\bf x}_i)=\mathcal{N}({\bm \mu}_i({\bf x}_i), r_i^2{\bf I})$ to approximate $p_{0|i}({\bf x}_0|{\bf x}_i)$. The mean ${\bm \mu}_i({\bf x}_i)$ and variance $r_i$ for $q_{0|i}$ should be selected so that $q_{0|i}$ is close to $p_{0|i}$, i.e., the KL divergence between $p_{0|i}$ and $q_{0|i}$ is small. The following Lemma provides us with a good mean and variance selection principle:
\begin{lemma}\label{lemma:klgauss}
    For any distribution $p_{0|i}({\bf x}_0|{\bf x}_i)$ whose density function is absolutely continuous on $\mathbb{R}^{d_{\bf x}}$, the optimum Gaussian distribution $q_{0|i}({\bf x}_0|{\bf x}_i)=\mathcal{N}({\bm \mu}_t({\bf x}_i), r_i^2{\bf I})$ that minimizes the KL divergence $D(p_{0|i}({\bf x}_0|{\bf x}_i)||q_{0|i}({\bf x}_0|{\bf x}_i))$ is:
    \begin{equation}
        {\bm \mu}_i({\bf x}_i)=\mathbb{E}_{p_{0|i}}[{\bf x}_0|{\bf x}_i], r_i^2=\frac{1}{d_{\bf x}}\mathbb{E}_{p_{0|i}}\left[\Vert{\bf x}_0-{\bm \mu}_i({\bf x}_i)\Vert_2^2\right]. \label{eq:tweedie}
    \end{equation}
\end{lemma}
We then select the mean ${\bm \mu}_i({\bf x}_i)$ and variance $r_i$ using Lemma~\ref{lemma:klgauss}. \\
(1). ${\bm \mu}_i({\bf x}_i)$: According to \eqref{eq:tweedie}, the optimum mean ${\bm \mu}_i({\bf x}_i)$ is the MMSE estimation of ${\bf x}_0$ given ${\bf x}_i$, which requires an integral over $p_{0|i}({\bf x}_0|{\bf x}_i)=\frac{p_0({\bf x}_0)p_{i|0}({\bf x}_i|{\bf x}_0)}{p_i({\bf x}_i)}$. The Tweedie's estimator \cite{efron2011tweedie,kim2021noisescore} is capable of estimating ${\bm \mu}_i({\bf x}_i)$ using the conditional $p_{i|0}({\bf x}_i|{\bf x}_0)=\mathcal{N}({\bf x}_i; \sqrt{\overline{\alpha}_i}{\bf x}_0, \sqrt{1-\overline{\alpha}_i}{\bf I})$ and the score function $\nabla_{{\bf x}_i}\log p_i({\bf x}_i)=\frac{{\bm \epsilon}_{\bm \theta}({\bf x}_i, i)}{\sqrt{1-\overline{\alpha}_i}}$, i.e.,   
\begin{align}
    &{\bm \mu}_i({\bf x}_i)\nonumber\\
    :=&\sqrt{\frac{1-\overline{\alpha}_i}{\overline{\alpha}_i}}\left(\frac{1}{\sqrt{1-\overline{\alpha}_i}}{\bf x}_i\!-\!\sqrt{1\!-\!\overline{\alpha}_i}\nabla_{{\bf x}_i}\log p_i({\bf x}_i)\right)\nonumber\\
    \overset{(a)}{=}&\frac{1}{\sqrt{\overline{\alpha}_i}}\left({\bf x}_i-\sqrt{1-\overline{\alpha}_i}{\bm \epsilon}_{{\bm \theta}}({\bf{x}}_i, i)\right),
\end{align}
where equality $(a)$ is obtained because ${\bm \epsilon}_{\bm \theta}({\bf x}_i, i)=\sqrt{1-\overline{\alpha}_i}\nabla_{{\bf x}_i}\log p_i({\bf x}_i)$. \\
    (2). $r_i$: According to \eqref{eq:tweedie}, $r_i$ should be the average pixel-wise prediction error of the clean image ${\bf x}_0$. However, computing $r_i$ using the Tweedie's formula requires computing the Hessian of $\log p_i({\bf x}_i)$, which is computational challenging in diffusion models. As an alternative, notice that the reconstruction error between the input ${\bf y}$ and the current prediction $\mathcal{A}({\bm \mu}_i({\bf x}_i))$, i.e., $\Vert{\bf y}-\mathcal{A}({\bm \mu}_i({\bf x}_i))\Vert_2$ can also reflect the prediction error of the clean image. A small reconstruction error indicates our prediction ${\bm \mu}_i({\bf x}_i)$ is accurate and the error $\Vert{\bf x}_0-{\bm \mu}_i({\bf x}_i)\Vert_2^2$ should be small as well. Therefore, we intuitively select 
    \begin{equation}r_i=\sqrt{\frac{1}{C\times H\times W}\Vert {\bf y}-\mathcal{A}({\bm \mu}_i({\bf x}_i))\Vert_2^2},\label{eq:ridef}
    \end{equation}where $C, H, W$ are the channels, height and width of the image ${\bf x}_0$.

With the distribution $q_{0|i}({\bf x}_0|{\bf x}_i)$, we can estimate the score function $\tilde{\bf s}_i({\bf x}_i, i)$ in\eqref{eq:pgdef} using the Monte Carlo method. Let ${\bf x}_0^{(m)}\sim q_{0|i}({\bf x}_0|{\bf x}_i), m=1, \cdots, N_{\text{mc}}$ samples drawn i.i.d from distribution $q_{0|i}({\bf x}_0|{\bf x}_i)$. Following~\eqref{eq:vanillapg}, the $\tilde{\bf s}_i({\bf x}_i, i)$ can be computed by:
\begin{align}
    &\tilde{\bf{s}}_i({\bf x}_i, {\bf y})\nonumber\\
    \approx&\mathbb{E}_{q_{0|i}({\bf x}_0|{\bf x}_i)}\left[p_0({\bf y}|{\bf x}_0)\nabla_{{\bf x}_i}\log q_{0|i}({\bf x}_0|{\bf x}_i)\right]\nonumber\\
    =&\mathbb{E}_{q_{0|i}({\bf x}_0|{\bf x}_i)}\left[p_0({\bf y}|{\bf x}_0)\nabla_{{\bf x}_i}\left(-\frac{1}{2r_i^2}\Vert{\bf x}_0-{\bm \mu}_i({\bf x}_i)\Vert_2^2\right)\right]\nonumber\\
    =&-\frac{1}{2r_i^2N_{\text{mc}}}\sum_{m=1}^{N_{\text{mc}}}\left(p_0({\bf y}|{\bf x}_0^{(m)})\nabla_{{\bf x}_i}\Vert{\bf x}_0-{\bm \mu}_i({\bf x}_i)\Vert_2^2\right)\label{eq:mcvanilla}
\end{align}

\subsubsection{Selection of $p_0({\bf y}|{\bf x}_0)$}
    After sample ${\bf x}_0^{(m)}, m\in[N_{\text{mc}}]$ from distribution $q_{0|i}({\bf x}_0|{\bf x}_i)=\mathcal{N}({\bm \mu}_i({\bf x}_i), r_i^2{\bf I})$, we need to compute the conditional probability $p_0({\bf y}|{\bf x}_0^{(m)})$ for each data sample. Ideally, the conditional distribution $p_0({\bf y}|{\bf x}_0^{(m)})\propto\exp\left(-\frac{1}{Z}\ell_{\bf y}({\bf x}_0^{(m)})\right)$ should be a fixed distribution for all the generation steps $i$. However, in the initial steps of the diffusion generation process ($i$ is large), the prediction ${\bm \mu}_i({\bf x}_i)$ is far away from the target image, the reconstruction loss $\ell_{\bf y}({\bf x}_0)$ between the input image ${\bf y}$ and the posterior sampled image $\mathcal{A}({\bf x}_0^{(m)})$ can be very large. If $Z$ is too small, the conditional probability $p_{0}({\bf y}|{\bf x}_0)$ is always 0 for all the samples ${\bf x}_0^{(m)}$ in step $i$; while in later stage of the diffusion generation process ($i$ is small), the reconstruction loss $\ell_{\bf y}(\mathcal{A}({\bm \mu}_i({\bf x}_i))$ can be small, selecting a large $Z$ may lead to the fact that all the conditional probabilities of ${\bf x}_0^{(m)}\sim q_{0|i}({\bf x}_0|{\bf x}_i)$ is close to one. Selecting an invariant $Z$ for all the diffusion generation steps lead to the that all the conditional probability $p_0({\bf y}|{\bf x}_0^{(m)})$ are too close to 0 or 1, and the score function $\tilde{s}_i({\bf x}_i, {\bf y})$ is not accurate. To prevent these phenomena, we choose $Z_i$ adaptively and assign a different normalizing factor for the conditional probability $p_0({\bf y}|{\bf x}_0^{(m)})$ in each generation step $i$. For Gaussian noise, we select $p_0({\bf y}|{\bf x}_0)\propto\exp\left(-\frac{1}{Z_i}\Vert{\bf y}-\mathcal{A}({\bf x}_0)\Vert_2^2\right)$, where each $Z_i=r_i^2=\frac{1}{C\times H\times W}\Vert\mathbf{y}-\mathcal{A}({\bm \mu}_i({\bf x}_i))\Vert_2^2$, i.e., $p({\bf y}|{\bf x}_0)$ has the same variance as $q({\bf x}_0|{\bf x}_i)$ defined in \eqref{eq:ridef}; for Poisson noise, we select the conditional distribution $p_0({\bf y}|{\bf x}_0)\propto\exp(-\frac{1}{Z_i}\Vert{\bf y}-\mathcal{A}({\bf x}_0)\Vert_1)$ to be an exponential distribution, where $Z_i=\Vert {\bf y}-\mathcal{A}({\bm \mu}_i({\bf x}_i))\Vert_1$. 
\subsubsection{Reward Shaping}Similar to policy gradient in reinforcement learning, direct MC estimation of the policy gradient~\eqref{eq:vanillapg} suffers from high estimation variance. To reduce the estimation variance, we leverage the reward shaping technique \cite{rewardshaping}. The goal is to deduct the a bias term $b_i({\bf x}_i)=\mathbb{E}_{q_{0|i}({\bf x}_0|{\bf x}_i)}[p_0({\bf y}|{\bf x}_0)]$ from the cost $p_0({\bf y}|{\bf x}_0)$ of each sample ${\bf x}_0$. 
Notice that $b_i({\bf x}_i)$ should be a scalar independent of samples ${\bf x}_0$, therefore we deduct a different bias term $b^{(m)}$ from each  sample $p_0({\bf y}|{\bf x}_0^{(m)})$ using the leave-one-out cross-validation, i.e.,
\begin{equation}
    b_i^{(m)}:=\frac{1}{N_{\text{mc}}-1}\sum_{j=1, j\neq m}^{N_{\text{mc}}}p_0\left({\bf y}|{\bf x}_0^{(j)}\right). \label{eq:dpgbias}
\end{equation}
We can then improve the MC estimation from~\eqref{eq:mcvanilla} by:
\begin{align}
    \tilde{\bf{s}}_i({\bf x}_i, {\bf y})\!=\!-\!\frac{\sum_{m=1}^{N_{\text{mc}}}(p_0({\bf y}|{\bf x}_0^{(m)})\!-\!b^{(m)}_i)\nabla_{{\bf x}_i}\Vert{\bf x}_0^{(m)}\!-\!{\bm \mu}_i({\bf x}_i)\Vert_2^2}{2r_i^2N_{\text{mc}}}.\label{eq:dpgshaping}
\end{align}
\subsubsection{Score Function Re-scaling}
Notice that the score function computed from~\eqref{eq:dpgshaping} contains only direction information. The exact norm of the gradient $\nabla_{{\bf x}_i}\log p_i({\bf y}|{\bf x}_i)$ is unknown. We need to re-scale the computed score function $\tilde{\bf{s}}_i({\bf x}_i, {\bf y})$ with norm $B$, i.e., assume that $\sigma_i\nabla_{{\bf x}_i}\log p_i({\bf y}|{\bf x}_i)\approx B\cdot\frac{1}{\Vert \tilde{\bf s}_i({\bf x}_i, {\bf y})\Vert_2^2 }\tilde{\bf s}_i({\bf x}_i, {\bf y})$ and plug it into \eqref{eq:score} to compute the score function ${\bf s}_i({\bf x}_i, {\bf y})$, i.e., 
\begin{equation}
    {\bf s}_i({\bf x}_i, {\bf y})\approx{\bm \epsilon}_{\bf \theta}({\bf x}_i, i)+B\cdot\frac{\tilde{{\bf s}}_i({\bf x}_i, {\bf y})}{\Vert \tilde{{\bf s}}_i({\bf x}_i, {\bf y})\Vert_2^2}. \label{eq:normalizedfinalscore}
\end{equation}
\subsection{Algorithm Description}
Using the score function estimated by~\eqref{eq:normalizedfinalscore}, we can solve the image inversion problems with the conditional guided diffusion using the standard DDPM/DDIM sampling method. The image restoration restore the target image ${\bf x}_0$ is displayed in Algorithm~\ref{alg:cap}
\begin{algorithm}[h]
\caption{Diffusion Policy Gradient (DPG)}\label{alg:cap}
\begin{algorithmic}
\Require Number of generation steps $N$, input image ${\bf y}$, reconstruction loss function $\ell_{\bf y}(\mathcal{A}(\cdot))$
\State ${\bf x}_N\sim\mathcal{N}(0, {\bf I})$
\For{$i=N$ to $1$}
\State{${\bm \mu}_i({\bf x}_i)\leftarrow\frac{1}{\sqrt{\overline{\alpha}_i}}\left({\bf x}_t+\sqrt{1-\overline{\alpha}_i}{\bm \epsilon}_{{\bm \theta}}({\bf x}_i, i)\right)$}\Comment{\eqref{eq:tweedie}. }
\State{$r_i\leftarrow\sqrt{\frac{1}{C\times H\times W}\Vert{\bf y}-\mathcal{A}({\bm \mu}_i({\bf x}_i))\Vert_2^2}$}\Comment{\eqref{eq:ridef}}
\State{${\bf x}_0^{(m)}\gets\mathcal{N}({\bm \mu}_i({\bf x}_i), r_i^2{\bf I}), m=1, \cdots, N_{{\rm mc}}$,}\Comment{Sample from $q({\bf x}_0|{\bf x}_i)$}
\State{$b_i^{(m)}\gets\frac{1}{N_{{\rm mc}}-1}\sum_{j=1, j\neq m}^{N_{{\rm mc}}}p_0({\bf y}|{\bf x}_0^{(m)})$}\Comment{Reward Shaping \eqref{eq:dpgbias}}
\State{$\tilde{{\bf s}}_i({\bf x}_i, {\bf y})\!\gets\!-\frac{\sum_{m=1}^{N_{\rm{mc}}}(p_0({\bf y}|{\bf x}_0^{(m)})-b_i^{(m)})\nabla_{{\bf x}_i}\Vert {\bf x}_0^{(m)}\!-\!{\bm \mu}_i({\bf x}_i)\Vert_2^2}{2r_t^2N_{\rm mc}}$}\Comment{Diffusion Policy Gradient \eqref{eq:dpgshaping}}
\State{${\bf s}_i({\bf x}_i, {\bf y})\leftarrow{\bm \epsilon}_{{\bm \theta}}({\bf x}_i, i)+B\frac{\tilde{\bf s}_i({\bf x}_i, {\bf y})}{\Vert \tilde{\bf s}_i({\bf x}_i, {\bf y})\Vert_2^2}$}
\State{${\bf x}_{i-1}\leftarrow\frac{1}{\sqrt{\alpha_i}}{\bf x}_i+\frac{1-\alpha_i}{\sqrt{1-\overline{\alpha}_i}}{\bf s}_i({\bf x}_i, {\bf y})+\sqrt{\beta_i}\overline{{\bf z}}_i$.}\Comment{DDPM}
\State{Or ${\bf x}_{i-1}\gets\sqrt{\frac{\overline{\alpha}_{i-1}}{\overline{\alpha}_i}}{\bf x}_i+\sqrt{\overline{\alpha}_{i-1}}\left(\sqrt{\frac{1-\overline{\alpha}_{i-1}}{\overline{\alpha}_{i-1}}}-\sqrt{\frac{1-\overline{\alpha}_i}{\overline{\alpha}_i}}\right){\bf s}_i({\bf x}_i, {\bf y})$}\Comment{DDIM}
\EndFor
\State{Return image ${\bf x}_0$}
\end{algorithmic}
\end{algorithm}

\subsection{Connection with Diffusion Posterior Sampling (DPS)}
\begin{wrapfigure}{r}{0.25\textwidth}
  \begin{center}
  \vspace{-20pt}
  \includegraphics[width=.24\textwidth]{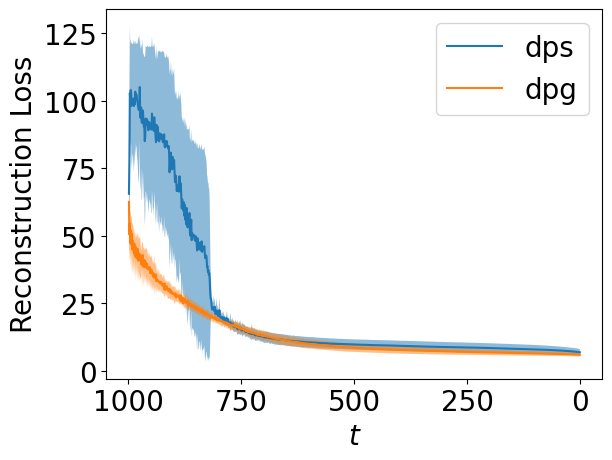}
  \vspace{-22pt}
    \caption{{\small Evolution of the reconstruction loss $\Vert{\bf y}-\mathcal{A}({\bm \mu}_i({\bf x}_i))\Vert_2$ of the DPS and DPG method. }}
    \vspace{-20pt}
    \label{fig:reconloss}
  \end{center}
\end{wrapfigure}
In this part, we discuss the relationship between the score function obtained by DPS \cite{chung2023diffusion} and our proposed method in Corollary~\ref{cor:equivalent-dps}.
\begin{Cor}\label{cor:equivalent-dps}
When $r_i\rightarrow 0$ and assume we have infinite number of MC samples, if $p_0({\bf y}|{\bf x}_0)$ is a Gaussian distribution, then the score function~\ref{eq:mcvanilla} is approximately
\begin{equation}
    \tilde{\bf s}_i({\bf x}_i, {\bf y})=\frac{1}{2\sigma_{\bf y}^2r_i}p_0({\bf y}|{\bf x}_0)\nabla_{{\bf x}_i}\ell_{\bf y}({\bf x}_i),
\end{equation}
whose direction is the same of the score function in DPS \cite{chung2023diffusion}.  
\end{Cor}
\begin{figure*}[h]
    \centering
    \begin{tabular}{c|cc@{}c@{}c@{}cc}
         &${\bm \mu}_{1000}({\bf x}_{1000})$& ${\bm \mu}_{900}({\bf x}_{900})$ &${\bm \mu}_{600}({\bf x}_{600})$ & ${\bm \mu}_{300}({\bf x}_{300})$ &${\bf x}_0$ (Final) &\\
         \includegraphics[width=.1\textwidth]{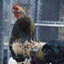}&\multirow{3}{*}{\includegraphics[width=.1\textwidth]{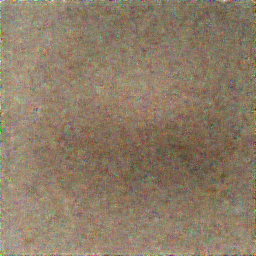}}& \includegraphics[width=.1\textwidth,frame]{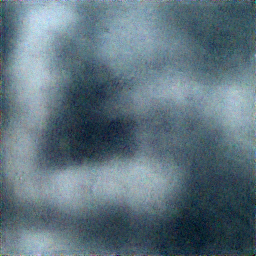}&\includegraphics[width=.1\textwidth, frame]{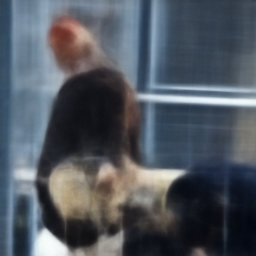}&\includegraphics[width=.1\textwidth, frame]{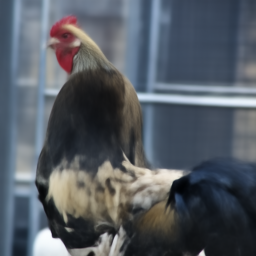}&\includegraphics[width=.1\textwidth, frame]{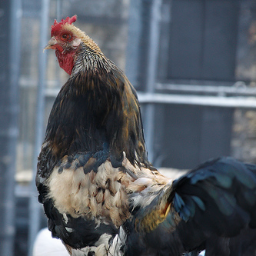}&\rotatebox{90}{\hspace{10pt}DPG}\\
         Input&&$\ell=51.0$&$\ell=15.5$&$\ell=9.64$&$\ell=6.56$\\
         \cline{3-6}\\
         \includegraphics[width=.1\textwidth]{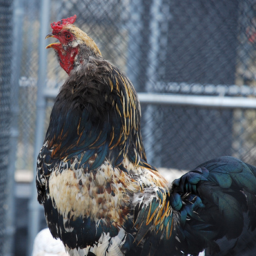}&&\includegraphics[width=.1\textwidth,frame]{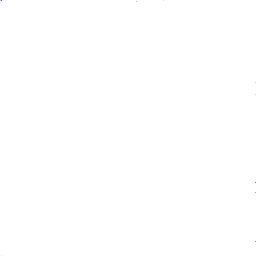}&\includegraphics[width=.1\textwidth, frame]{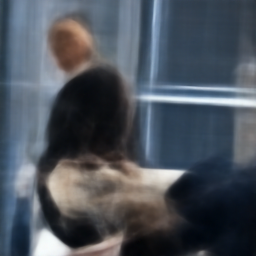}&\includegraphics[width=.1\textwidth, frame]{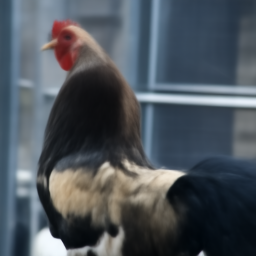}&\includegraphics[width=.1\textwidth, frame]{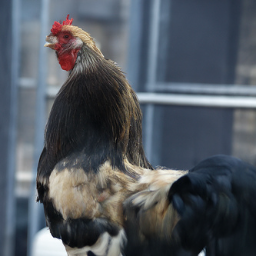}&\rotatebox{90}{\hspace{10pt} DPS}\\
         Ground Truth&&$\ell=144$&$\ell=15.8$&$\ell=10.9$&$\ell=7.99$
    \end{tabular}
    \caption{Image generation procedure and the reconstruction loss $\ell:=\Vert {\bf y}-\mathcal{A}({\bm \mu}_i({\bf x}_i))\Vert_2$ by using DPG and DPS methods in super-resolution. }
    \label{fig:recover-progress}
    \vspace{-20pt}
\end{figure*}

Proof of Corollary~\ref{cor:equivalent-dps} is provided in Appendix 5.2. Corollary~\ref{cor:equivalent-dps} shows that the score function $\tilde{\bf s}_i({\bf x}_i, {\bf y})\approx\nabla_{{\bf x}_i}\ell_{\bf y}({\bf x}_i)$ by DPS \cite{chung2023diffusion} is accurate when $r_i\rightarrow 0$, i.e., in later-stages of the diffusion generation process. 
However, in initial stages of the image generation (i.e., $i$ is large), the score function obtained by DPS is inaccurate. As is shown in Fig.~\ref{fig:reconloss}, the reconstruction loss $\ell_{\bf y}({\bf x}_i)$ by running the DPS algorithm in the initial image generation stages (i.e., $i\geq 750$) larger compared with our proposed DPG method. Fig.~\ref{fig:recover-progress} plots the intermediate recoverved figures during the diffusion process. Since DPG has a more accurate estimation of the guidance score function, the shape and the sketch of the image is recovered at an earlier stage compared with the DPS method (i.e., at step $i=900$, noisy image generated by DPG has the sketch of the chicken, while the image generated by DPS is blank.) More results for deblurring experiments can be found in the Appendix.

\section{Experiments}
\label{gen_inst}
\begin{figure*}[h]
\centering
\setlength\arrayrulewidth{2pt}
\renewcommand{\arraystretch}{0.0}
\vspace{-10pt}
\begin{tabular}{@{}c@{}c@{}c@{}c@{}c@{}c@{}|@{}c@{}c@{}@{}|@{}c}
\cline{7-8}
\specialrule{0em}{0.5pt}{0.5pt}
&{\tiny Input}&{\tiny DDRM}&{\tiny DDNM+}&{\tiny Reddif}&{\tiny DPS}&{\tiny {\bf DPG(DDPM)}}&{\tiny {\bf DPG(DDIM)}}&{\tiny Ground Truth}\\
\rotatebox{90}{\hspace{-25pt}Inpainting}&\includegraphics[width=.1\textwidth]{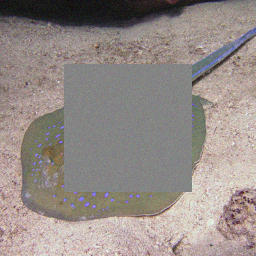}&\includegraphics[width=.1\textwidth]{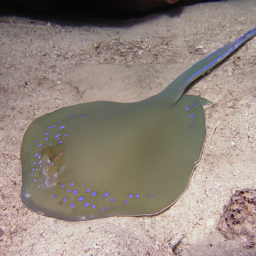}&\includegraphics[width=.1\textwidth]{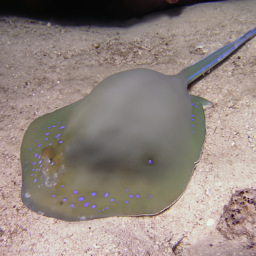}&\includegraphics[width=.1\textwidth]{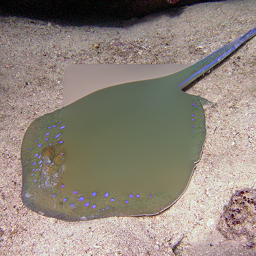}&\includegraphics[width=.1\textwidth]{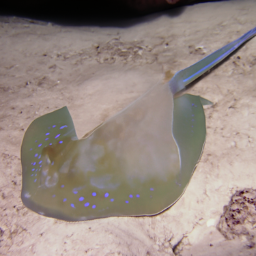}&\includegraphics[width=.1\textwidth]{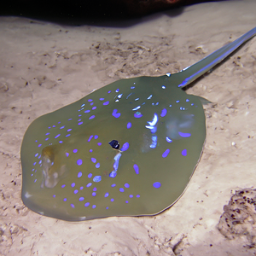}&\includegraphics[width=.1\textwidth]{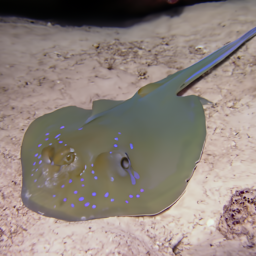}&\includegraphics[width=.1\textwidth]{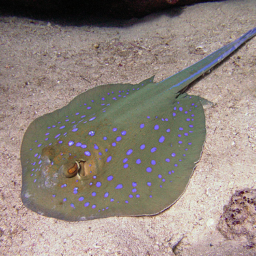}\\
&\includegraphics[width=.1\textwidth]{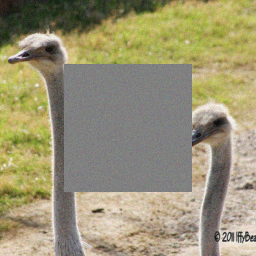}&\includegraphics[width=.1\textwidth]{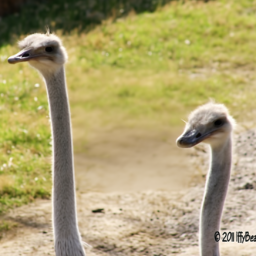}&\includegraphics[width=.1\textwidth]{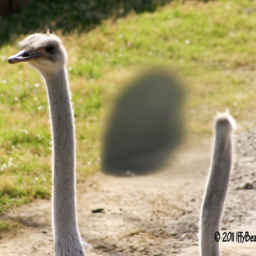}&\includegraphics[width=.1\textwidth]{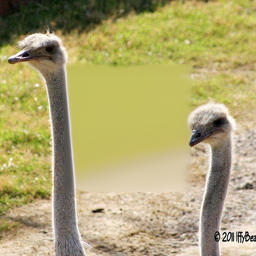}&\includegraphics[width=.1\textwidth]{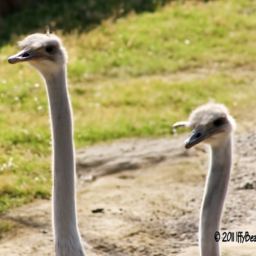}&\includegraphics[width=.1\textwidth]{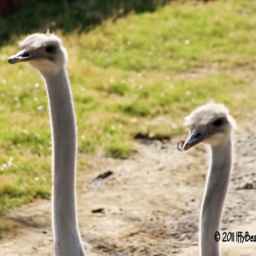}&\includegraphics[width=.1\textwidth]{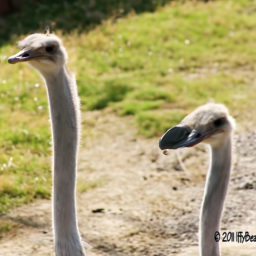}&\includegraphics[width=.1\textwidth]{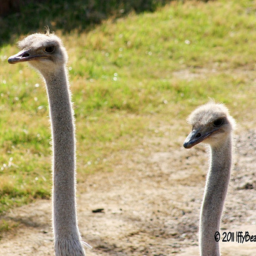}\\
\hline
\rotatebox{90}{\hspace{-35pt}Super-Resolution}&\includegraphics[width=.1\textwidth]{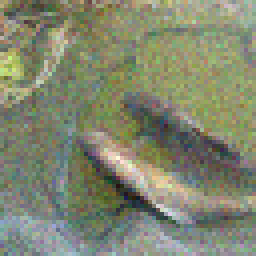}&\includegraphics[width=.1\textwidth]{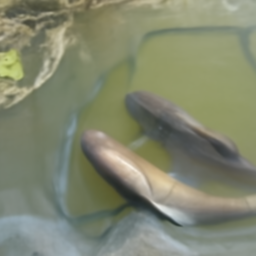}&\includegraphics[width=.1\textwidth]{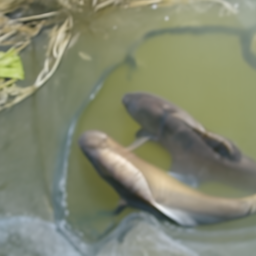}&\includegraphics[width=.1\textwidth]{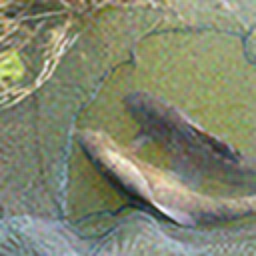}&\includegraphics[width=.1\textwidth]{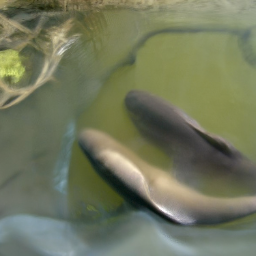}&\includegraphics[width=.1\textwidth]{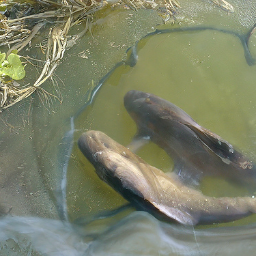}&\includegraphics[width=.1\textwidth]{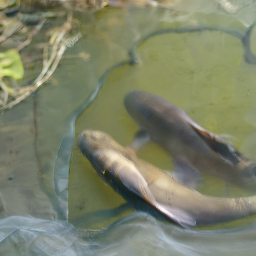}&\includegraphics[width=.1\textwidth]{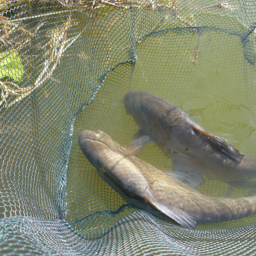}\\ 
&\includegraphics[width=.1\textwidth]{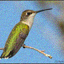}&\includegraphics[width=.1\textwidth]{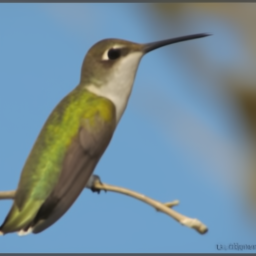}&\includegraphics[width=.1\textwidth]{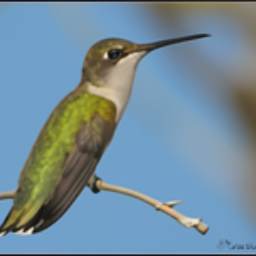}&\includegraphics[width=.1\textwidth]{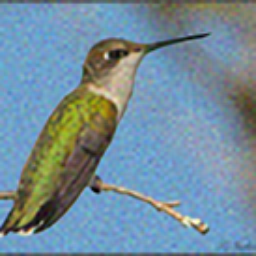}&\includegraphics[width=.1\textwidth]{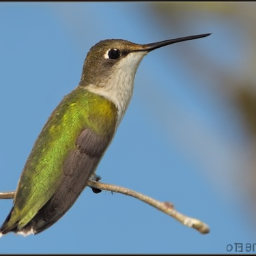}&\includegraphics[width=.1\textwidth]{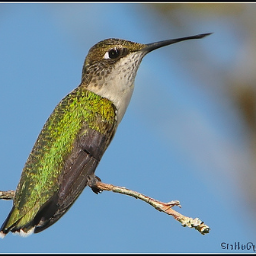}&\includegraphics[width=.1\textwidth]{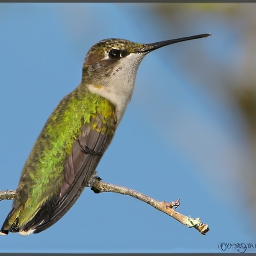}&\includegraphics[width=.1\textwidth]{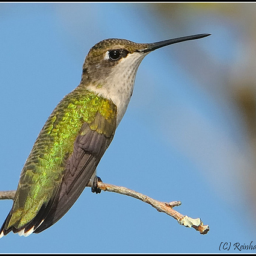}\\
\hline
\rotatebox{90}{\hspace{-40pt}Gaussian deblurring}&\includegraphics[width=.1\textwidth]{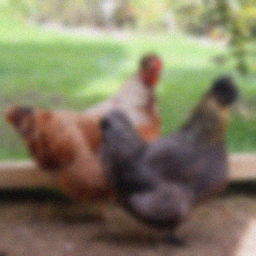}&\includegraphics[width=.1\textwidth]{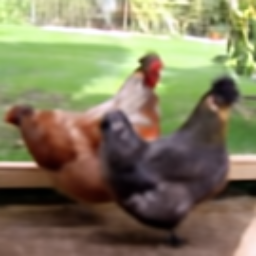}&\includegraphics[width=.1\textwidth]{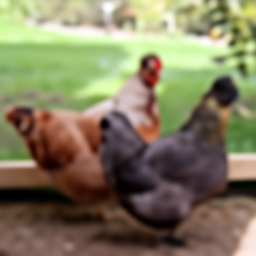}&\includegraphics[width=.1\textwidth]{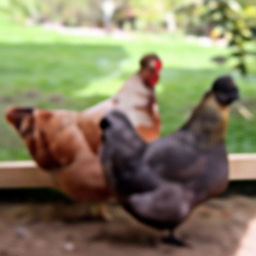}&\includegraphics[width=.1\textwidth]{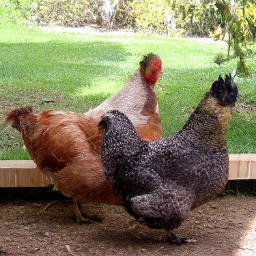}&\includegraphics[width=.1\textwidth]{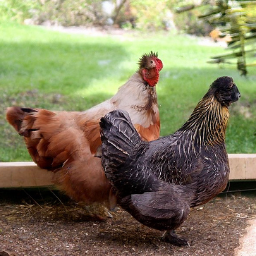}&\includegraphics[width=.1\textwidth]{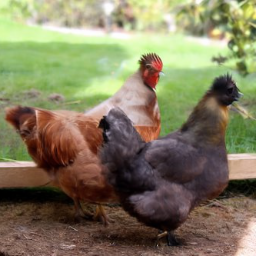}&\includegraphics[width=.1\textwidth]{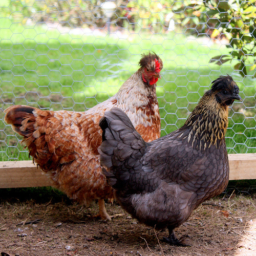}\\
&\includegraphics[width=.1\textwidth]{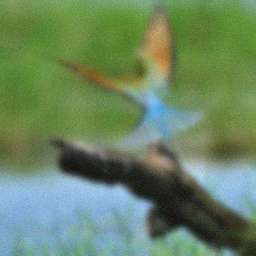}&\includegraphics[width=.1\textwidth]{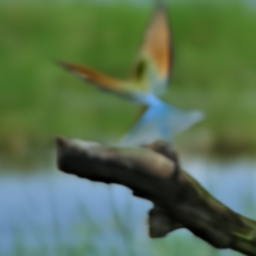}&\includegraphics[width=.1\textwidth]{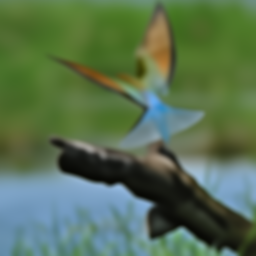}&\includegraphics[width=.1\textwidth]{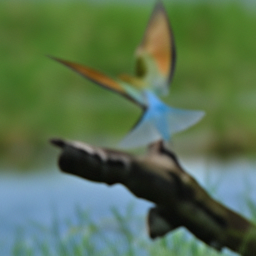}&\includegraphics[width=.1\textwidth]{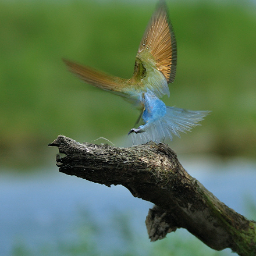}&\includegraphics[width=.1\textwidth]{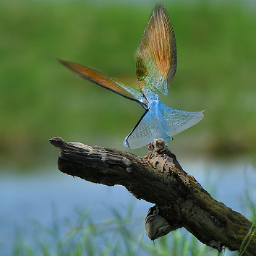}&\includegraphics[width=.1\textwidth]{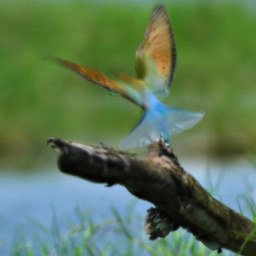}&\includegraphics[width=.1\textwidth]{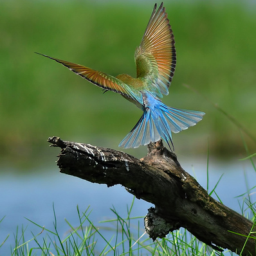}\\
\cline{7-8}
\end{tabular}
\vspace{-10pt}
\caption{Results on solving linear noisy inverse problems (inpainting, super-resolution and Gaussian deblurring) on ImageNet Dataset. The input image is distorted by random Gaussian noise $\sigma_{\bf y}=0.05$. }
\label{fig:linear-inverse}
\vspace{-20pt}
\end{figure*}

In this section, we show experimental results obtained by our DPG method and compare its performance with other state-of-the art methods. 
\subsection{Quantitative Results on Noisy Linear Inverse Problems}
\begin{table*}[h]
{\tiny \caption{Quantitative Results on Inpainting, Super-Resolution and Gaussian deblurringring on FFHQ, Imagenet and LSUN-Bedroom ($256\times256$) Datasets}
\label{results-ffhq}
\begin{center}
\begin{tabular}{c|c@{\hspace{0.1cm}}c@{\hspace{0.1cm}}c@{\hspace{0.1cm}}|c@{\hspace{0.1cm}}c@{\hspace{0.1cm}}c@{\hspace{0.1cm}}|c@{\hspace{0.1cm}}c@{\hspace{0.1cm}}c@{\hspace{0.1cm}}|c@{\hspace{0.1cm}}c@{\hspace{0.1cm}}c@{\hspace{0.1cm}}}
&\multicolumn{3}{c}{\bf Inpainting (Random)}&\multicolumn{3}{c}{\bf Super-Resolution (4$\times$)}&\multicolumn{3}{c}{\bf deblurring (Gauss)}&\multicolumn{3}{c}{\bf deblurring (Motion)}\\
{\bf Method}&FID$\downarrow$&LPIPS$\downarrow$&PSNR$\uparrow$&FID$\downarrow$&LPIPS$\downarrow$&PSNR$\uparrow$&FID$\downarrow$&LPIPS$\downarrow$&PSNR$\uparrow$&FID$\downarrow$&LPIPS$\downarrow$&PSNR$\uparrow$\\
\hline
\hline
\multicolumn{13}{c}{{\bf FFHQ 1k Validation Set}}\\
DPG (DDPM) &{\bf 22.44}&0.181&22.17
&\bf{22.49}&{\bf 0.214}&\uline{26.61}
&\bf{22.29}&{\bf 0.216}&\uline{26.02}
&{\bf 24.44}&{\bf 0.223}&{\bf 26.38}\\
DPS (DDPM) \cite{chung2023diffusion}&33.12&0.216&21.83
&\uline{39.35}&{\bf 0.214}&25.67
&\uline{44.05}&\uline{0.257}&24.93
&39.02&0.242&24.92\\
DDRM \cite{kawar2022denoising}&27.47&{\bf 0.172}&{\bf 23.44}&62.15&0.294&25.36&74.92&0.332&23.36&N/A&N/A&N/A\\
DDNM+ \cite{wang2023ddnm} &\uline{27.34}&\uline{0.173}&\uline{23.29}&46.13&0.260&{\bf 27.41}&63.19&0.301&{\bf 27.70}&N/A&N/A&N/A\\
\hline
\hline
\multicolumn{13}{c}{{\bf ImageNet 1k Validation Set}}\\
DPG (DDPM) &{\bf 41.86}&0.258&17.41
&\bf{31.02}&{\bf 0.293}&22.91
&{\bf 34.43}&{\bf 0.314}&22.10
&{\bf 36.15}&{\bf 0.343}&{\bf 21.67}\\
DPS (DDPM) \cite{chung2023diffusion}&\uline{45.95}&0.267&17.69
&\uline{43.60}&\uline{0.340}&23.10
&\uline{54.76}&0.386&20.04
&56.08&0.386&20.55\\
DDRM \cite{kawar2022denoising}&50.94&{\bf 0.246}&19.13
&51.77&0.355&{\bf 24.17}
&72.49&\uline{0.345}&22.62
&N/A&N/A&N/A\\
DDNM+ \cite{wang2023ddnm}&50.50&{\bf 0.246}&\uline{19.16} 
&51.08&0.362&\uline{24.00}
&71.74&0.410&{\bf 24.90}
&N/A&N/A&N/A\\
RED-Diff \cite{mardani2023variational} &192.96 &0.292 &{\bf 20.03}& 74.39 & 0.434 & 23.39 & 62.79 & 0.380 &\uline{23.65}& N/A & N/A & N/A\\
\hline
\hline
\multicolumn{13}{c}{{\bf LSUN-Bedroom Validation Set}}\\
DPG (DDPM) &{\bf 34.32}&0.218&18.90
&{\bf 31.44}&{\bf 0.262}&23.48
&{\bf 38.72}&{\bf 0.277}&22.39
&{\bf 34.44}&{\bf 0.284}&{\bf 22.82}\\
DPG (DDIM) &\uline{34.39}&0.209&\uline{19.74}
&\uline{33.86}&\uline{0.269}&23.99
&46.08&0.322&22.25
& \uline{45.08}&\uline{0.355}&\uline{21.97}\\
DPS \cite{chung2023diffusion}&35.91&0.218&18.58
&37.42&0.284&23.67
&\uline{48.10}&0.320&22.25
&50.09&0.358&21.73\\
DDRM \cite{kawar2022denoising}&37.61&\uline{0.205}&19.59
&50.96&0.310&24.10
&59.04&0.353&22.64
&N/A&N/A&N/A\\
DDNM+ \cite{wang2023ddnm}&37.03&{\bf 0.204}&19.55
&50.15&0.296&\uline{24.48}
&74.40&0.336&\uline{24.38}
&N/A&N/A&N/A\\
RED-Diff \cite{mardani2023variational} &44.35 & 0.240&{\bf 20.49}& 75.76 & 0.380  & {\bf 24.67} & 64.70 &\uline{0.314} &{\bf 25.27} & N/A &N/A & N/A
\end{tabular}
\vspace{-20pt}
\end{center}}
\label{tab:ffhq-eval}
\end{table*}
\begingroup
\setlength\arrayrulewidth{2pt}
\renewcommand{\arraystretch}{0.0}
\begin{figure}
\begin{tabular}{@{}c@{}c@{}|@{}c@{}c@{}|@{}c}
\cline{3-4}
\specialrule{0em}{0.5pt}{0.5pt}
    {\tiny  Input}&{\tiny  DPS}&{\tiny DPG(DDPM) }&{\tiny DPG(DDIM)}&{\tiny Ground Truth}\\
\includegraphics[width=.099\textwidth]{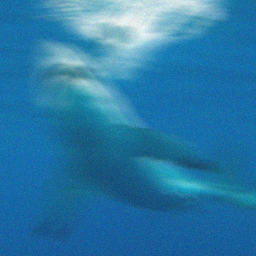}& \includegraphics[width=.099\textwidth]{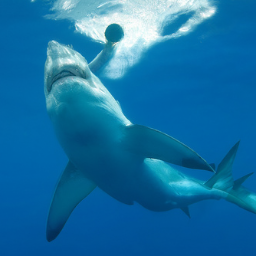}&\includegraphics[width=.099\textwidth]{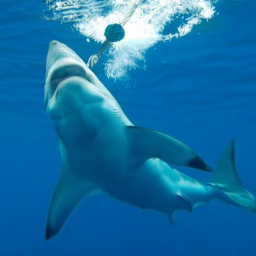}&\includegraphics[width=.099\textwidth]{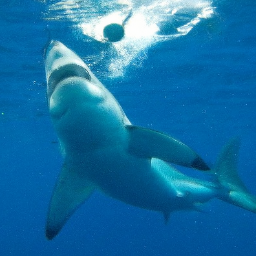}&\includegraphics[width=.099\textwidth]{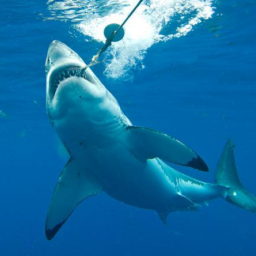}\\
\includegraphics[width=.099\textwidth]{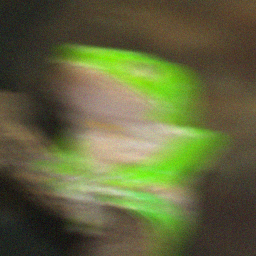}&\includegraphics[width=.099\textwidth]{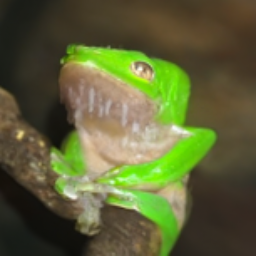}&\includegraphics[width=.099\textwidth]{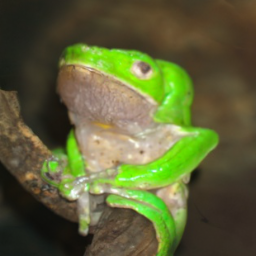}&\includegraphics[width=.099\textwidth]{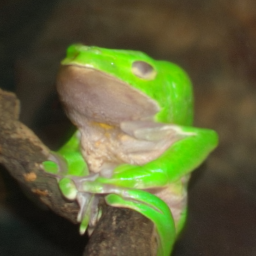}&\includegraphics[width=.099\textwidth]{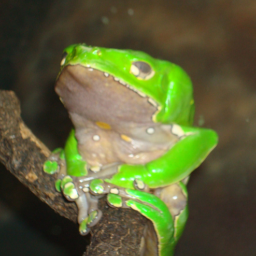}\\
\cline{3-4}
    \end{tabular}
    \vspace{-10pt}
    \caption{Image Restoration Results for Motion Deblurring on ImageNet256$\times$256.}
    \label{fig:linear-motion}
    \vspace{-20pt}
\end{figure}
\endgroup

\begingroup
\renewcommand{\arraystretch}{0.0}
\begin{figure}
\setlength\arrayrulewidth{2pt}
\begin{tabular}{c@{}c@{}c@{}|@{}c@{}|@{}c}
\cline{4-4}
\specialrule{0em}{1pt}{1pt}
    &{\small  Input}&{\small  DPS}&{\small \bf DPG}&{\small Ground Truth}\\
        \rotatebox{90}{\makecell[c]{\small Super-\\Resolution}}& \includegraphics[width=.1\textwidth]{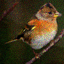}& \includegraphics[width=.1\textwidth]{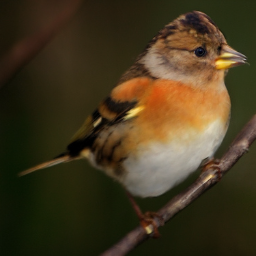}&\includegraphics[width=.1\textwidth]{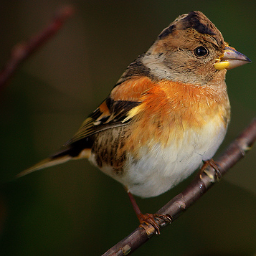}&\includegraphics[width=.1\textwidth]{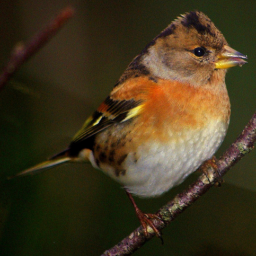}\\
        \rotatebox{90}{\makecell[c]{\small Gaussian-\\deblurring}}&\includegraphics[width=.1\textwidth]{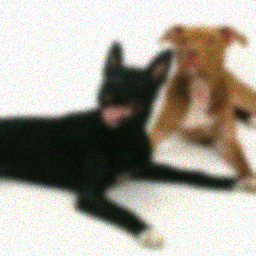}&\includegraphics[width=.1\textwidth]{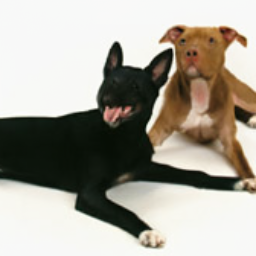}&\includegraphics[width=.1\textwidth]{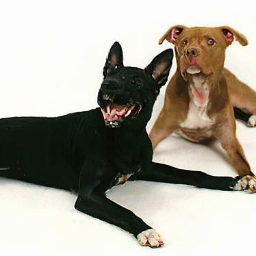}&\includegraphics[width=.1\textwidth]{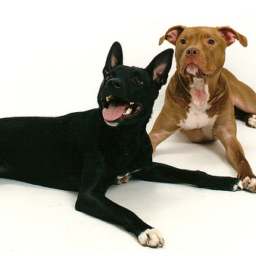}\\
         \rotatebox{90}{\makecell[c]{\small Motion-\\deblurring}}& \includegraphics[width=.1\textwidth]{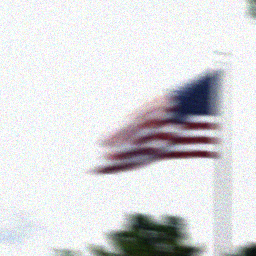}&\includegraphics[width=.1\textwidth]{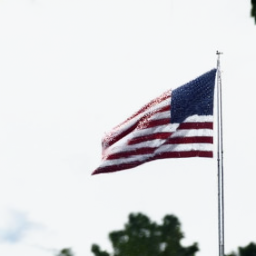} &\includegraphics[width=.1\textwidth]{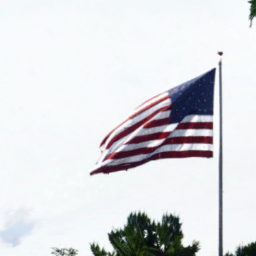} &\includegraphics[width=.1\textwidth]{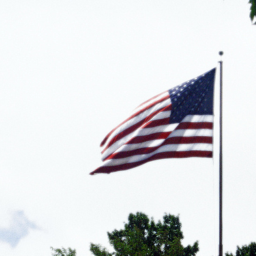}\\         \cline{4-4}
    \end{tabular}
    \vspace{-10pt}
    \caption{Image Restoration Results on ImageNet with Poisson Noise $\lambda=1.0$.}
    \vspace{-20pt}
    \label{fig:linear-inverse-poisson}
\end{figure}
\endgroup
{\bf Experiment Setup} Similar to \cite{chung2023diffusion,song2023pseudoinverseguided}, we test the performance of our proposed algorithm on three datasets: the FFHQ 256$\times$256 dataset~\cite{karras2019style}, the ImageNet dataset~\cite{imagenet_cvpr09} and the LSUN-Bedroom dataset~\cite{yu15lsun}. We consider four types of image inverse tasks: (1) Inpainting, where a size of 128$\times$128 mask is added randomly on the input image; (2) 4$\times$super-resolution with average pooling; (3) Gaussian deblurring with kernel size $61\times 61$ and standard deviation of $3.0$; (4) Motion deblurring with kernel size of 61 and intensity value 0.5 generated by\footnote{\url{https://github.com/LeviBorodenko/motionblur}}. We consider that the input image is noisy, i.e., Gaussian noise with variance $\sigma_{\bf y}=0.05$ or Poisson noise with rate $\lambda=1.0$ is added on the input image. For FFHQ experiments, we use the pretrained model from \cite{chung2023diffusion} (trained on 4.9k images on FFHQ) and test the performance of 1k validation set; For Imagenet experiments, we use the unconditional Imagenet 256$\times$256 generation model from \cite{dhariwal2021diffusion} and the 1k images are selected from \footnote{\url{https://github.com/XingangPan/deep-generative-prior/blob/master/scripts/imagenet_val_1k.txt}}; For LSUN experiments, we use the pretrained LSUN-Bedroom model from \cite{dhariwal2021diffusion} and test the performance on LSUN-Bedroom validation set containing 300 images. All experiments are run on an NVIDIA A100 GPU. 

We compare the performance with the following methods: Denoising Diffusion Null Space models (DDNM+) \cite{wang2023ddnm} for noisy problems, Diffusion Posterior Sampling (DPS) \cite{chung2022improving}, Denoising Diffusion Restoration Models (DDRM) \cite{kawar2022denoising} and the Reddiff \cite{mardani2023variational}. The key parameters for different methods are displayed in Appendix 6. For LSUN dataset, we also report the results obtained by the DDIM sampler with 200 steps generation. 

{\bf Evaluation Metrics} We measure both the image restoration quality and consistency compared with the ground-truth image. For image restoration quality, we compute the Fréchet inception distance between the restored images and the ground truth images; For image restoration consitency, we computes the LPIPS score \cite{zhang2018perceptual} (VGG Net) and the Peak Signal to Noise Ratio (PSNR) between the restored image and the ground truth image. Quantative evaluation results are displayed in Table~\ref{tab:ffhq-eval}. Selected image restoration samples when the observation noise are Gaussian and Poisson are displayed in Fig.~\ref{fig:linear-inverse} and Fig.~\ref{fig:linear-inverse-poisson}. 

{\bf Analysis} The FID and LPIPS score of our proposed DPG method is smaller than DPS method, indicating that DPG has a better image restoration quality than DPS method. This is because the estimation of the score function by DPG is more accurate than DPS, especially in the initial stages of the diffusion generation process. Therefore, the shape and structure of the image can be recovered in an earlilier stage of the diffusion process, this gives room to recover high frequency details in later stage of the image generation. 
Notice that DDNM+ and DDRM uses a plug-in estimation, i.e., the known pixels in ${\bf y}$ are directly used in the generation process. Therefore, the PSNR of DDNM+ and DDRM are better than the proposed DPG method, but the recovered high frequency detailed features are less than our proposed DPG method. DPG has a smaller LPIPS score and FID score in most tasks. The Reddif method is not robust to input noise, the quality of the restored image degrades significantly. Moreover, since DPG has a better estimation of the guidance score, we can combine DPG with the DDIM solver. Results on LSUN dataset shows there is little quality loss compared with the 1000 steps DDPM generation. 

\subsection{Quantitative Experiments on Non-Linear Image Inverse Problems}
Our method does not require the operator $\mathcal{A}$ to be linear, and thus can be applied to non-linear image inversion problems. We consider the input image is distorted by the non-linear blur kernel \cite{tran2021explore}, and compare the image restoration results of our DPG method with the DPS \cite{chung2023diffusion}. Selected image restoration results are displayed in Fig.~\ref{fig:nonlinear} and quantitative evaluations are displayed in Table~\ref{tab:nonlinear-eval}. According to Fig.~\ref{fig:nonlinear}, DPG can restore more details of the original image, and this results in a smaller LPIPS score. 

\begin{table}
{\small \caption{Quantitative Results on Non-linear deblurring on Imagenet ($256\times256$) Datasets}
\label{results-ffhq}
\begin{center}
\begin{tabular}{c|ccc}
{\bf Method}&FID$\downarrow$&LPIPS$\downarrow$&PSNR$\uparrow$\\
\hline
DPG (Ours) &{\bf 88.15}&{\bf 0.464}&{\bf 20.73}\\
DPS & 120.79 & 0.484 & 17.79
\end{tabular}
\end{center}}
\label{tab:nonlinear-eval}
\end{table}

\begingroup
\renewcommand{\arraystretch}{0.0}
\begin{figure}[h]
\centering
\setlength\arrayrulewidth{2pt}
\begin{tabular}{@{}c@{}c@{}|@{}c@{}|@{}c@{}}
\cline{3-3}
\specialrule{0em}{1pt}{1pt}
{\tiny Input}&{\tiny DPS}&{\tiny \bf DPG}&{\tiny Ground Truth}\\
\includegraphics[width=.099\textwidth]{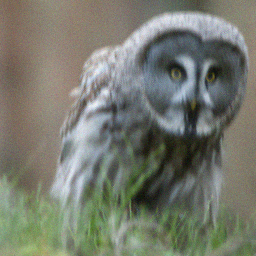}&\includegraphics[width=.099\textwidth]{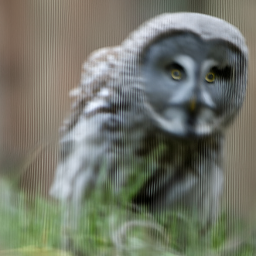}&\includegraphics[width=.099\textwidth]{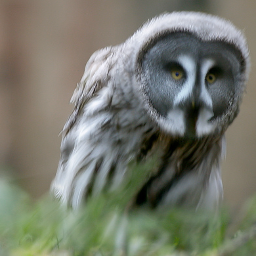}&\includegraphics[width=.099\textwidth]{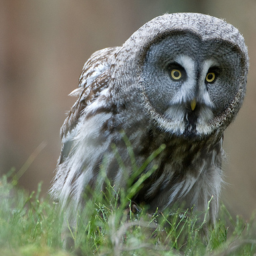}\\
\includegraphics[width=.099\textwidth]{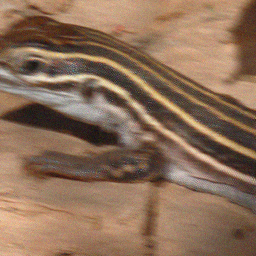}&\includegraphics[width=.099\textwidth]{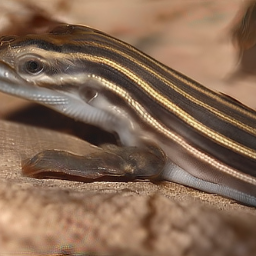}&\includegraphics[width=.099\textwidth]{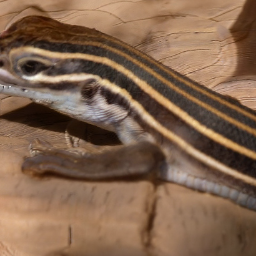}&\includegraphics[width=.099\textwidth]{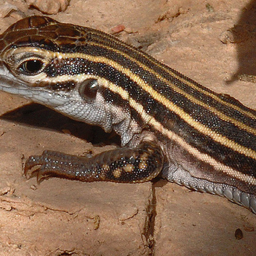}\\
        \cline{3-3}
        \end{tabular}
        \vspace{-10pt}
    \caption{Image Restoration Results for Non-linear Deblurring on ImageNet. }
    \label{fig:nonlinear}
\end{figure}
\vspace{-20pt}
\endgroup

\section{Conclusions and Limitations}

In this paper, we proposed a new method to estimate the guidance score function to solve image inverse problems using pre-trained diffusion generative models. Our method is robust when the input image is perturbed by random noise, and can be used for solving non-linear inverse problems. Experiments demonstrate that the proposed method can improve image restoration quality in both human eye evaluation and quantitative metrics for a wide range of tasks such as inpainting, super-resolution and deblurring. 

While the score function computed by DPG using \eqref{eq:normalizedfinalscore} eliminates the necessity for a differentiable loss function $\ell_{{\bf y}}({\bf x}_0)$, we observe, in the specific case of JPEG restoration, that the quality of the generated image  is less stable and far worse compared to the results achieved by $\Pi$GDM~\cite{song2023pseudoinverseguided}. 

Acknowledging the superiority of our method over alternatives like DPS and DDNM, it is essential to note that the generation speed of our approach DPG is slower when the operator becomes complex (i.e., the non-linear deblurring kernel). Moreover, current deblurring experiments are carried out when the blur kernel is known. In the future, we would like to explore solving deblurring problems with unknown kernels and apply the method on latent diffusion models \cite{rombach2021highresolution}.

\bibliography{iclr2021_conference}

\onecolumn
\aistatstitle{Solving General Noisy Inverse Problem via Posterior Sampling: \\A Policy Gradient Viewpoint Supplementary Materials}
\setcounter{section}{4}
\section{Proofs}
\subsection{Proof of Theorem 1}\label{pf:leibniz}
To prove Theorem 1, we need to verify the following three condition holds for density function $p_t({\bf x}_t|{\bf y})$:
For fixed ${\bf x}_0$ and any time $\tau<T$, if the conditional density function $p_{0|t}({\bf x}_0|{\bf x}_t={\bf x})$ satisfy the following conditions: (1) function $p_{0|t}({\bf x}_0|{\bf x}_t={{\bf x}})$ is a Lebesgue-integrable of $t$ for each ${\bf x}$; (2) the gradient $\nabla_{{\bf x}_t} p_{0|t}({\bf x}_0|{\bf x}_t), \forall {\bf x}$ exists for almost all $t\in[0, T]$; (3) there is an integral function $g(t)$ so that $\Vert \nabla_{{\bf x}_t} p_{0|t}({\bf x}_0|{\bf x}_t)\Vert\leq g(t)$
We will first verify each condition in Theorem 1 respectively and then provide detailed derivations of \eqref{eq:vanillapg}.

(1) To show function $p_{0|t}({\bf x}_0|{\bf x}_t)$ is integrable of $t$, we show that $p_{0|t}({\bf x}_0|{\bf x}_t)$ is bounded. Recall that $p_0({\bf x}_0)$ is the probability density function of the high-quality images and let $\mu_0(\text{d}{\bf x}_0)$ be the probability measure. Then density function $p_{0|t}({\bf x}_0|{\bf x}_t)$ can be computed by:
\begin{equation}
    p_{0|t}({\bf x}_0|{\bf x}_t)=\frac{\mu_0(\text{d}{\bf x}_0)p_{t|0}({\bf x}_t|{\bf x}_0)}{\text{d}{\bf x}_0\int p_{t|0}({\bf x}_t|{\bf x}_0)\mu_0(\text{d}{\bf x}_0)}. 
\end{equation}
According to \cite[Eq.~(29)]{song2021scorebased}, distribution $p({\bf x}_t|{\bf x}_0)=\mathcal{N}({\bf x}_t|\sqrt{\overline{\alpha}(t)}{\bf x}_0, (1-\overline{\alpha}(t)){\bf I})$ is Gaussian, hence $p_{t|0}({\bf x}_t|{\bf x}_0)$ is bounded. Therefore, $p_{0|t}({\bf x}_0|{\bf x}_t)<1, \forall t$ is bounded on $[0, T]$ and is hence Lebesgue-integrable of $t$. 

(2) The gradient of conditional density function $\nabla_{{\bf x}_t}p_{0|t}({\bf x}_0|{\bf x}_t)$ can be decomposed by:
\begin{align}
    \nabla_{{\bf x}_t}p_{0|t}({\bf x}_0|{\bf x}_t)=\nabla_{{\bf x}_t}\left(\frac{p_{t, 0}({\bf x}_t, {\bf x}_0)}{p_t({\bf x}_t)}\right)=\frac{1}{p_t({\bf x}_t)^2}\left(\nabla_{{\bf x}_t}p({\bf x}_t|{\bf x}_0)\cdot p_t({\bf x}_t)-\nabla_{{\bf x}_t}p_t({\bf x}_t)\cdot p({\bf x}_t|{\bf x}_0)\right). \label{eq:posteriordecompose}
\end{align}
According to \cite[Eq.~(29)]{song2021scorebased}, distribution $p_{t|0}({\bf x}_t|{\bf x}_0)=\mathcal{N}({\bf x}_t|\sqrt{\overline{\alpha}(t)}{\bf x}_0, (1-\overline{\alpha}(t)){\bf I})$ is Gaussian, therefore, $p_t({\bf x}_t)$ is non-zero $\forall {\bf x}_t\in\mathbb{R}, t<T$ and the gradient $\nabla_{{\bf x}_t}p({\bf x}_t|{\bf x}_0)$ exists for all ${\bf x}_t, \forall t$. It then remains to prove that gradient $\nabla_{{\bf x}_t}p_t({\bf x}_t)$ exists for almost all $t\in[0, T]$. Since $p_t({\bf x}_t)=\mathbb{E}_{p_0}\left[p_{t|0}({\bf x}_t|{\bf x}_0)\right]$ and density function $p({\bf x}_t|{\bf x}_0)$ is a Gaussian, as function $p({\bf x}_t|{\bf x}_0)$ is continuous on ${\bf x}_t$, function $p_t({\bf x}_t)$ is continuous for all $t\in[0, T)$ and hence $\nabla_{{\bf x}_t}p_t({\bf x}_t)$ exists for almost all $t\in[0, T]$. 

(3) The probability density function $p({\bf x}_0|{\bf x}_t)$ can be computed by:
\begin{equation}
    p_{0|t}({\bf x}_0|{\bf x}_t)=\frac{p_0({\bf x}_0)p_{t|0}({\bf x}_t|{\bf x}_0)}{\mathbb{E}_{p_0}\left[p_{t|0}({\bf x}_t|{\bf x}_0)\right]}. \label{eq:p0t}
\end{equation}
Notice that $p_{t|0}({\bf x}_t|{\bf x}_0)$ is a Gaussian distribution, function $p_{0|t}({\bf x}_0|{\bf x}_t)$ is continuous for ${\bf x}_t$. To verify condition (3), it remains to upper bound function $p_{0|t}({\bf x}_t)$. Notice that the high quality image $p_0({\bf x}_0)$ consists of $N_{\text{train}}$ samples. Therefore, for each ${\bf x}_0$, we have:
\begin{equation}
    p_0({\bf x}_0)p_{t|0}({\bf x}_t|{\bf x}_0)\leq N_{\text{train}}\int p_0({\bf x}_0)p_{t|0}({\bf x}_t|{\bf x}_t)\text{d}{\bf x}_0. \label{eq:p0tdenorm}
\end{equation}

Plugging \eqref{eq:p0tdenorm} into \eqref{eq:p0t}, we can upper bound $p_{0|t}({\bf x}_0|{\bf x}_t)$ as follows:
\begin{equation}
    p_{0|t}({\bf x}_0|{\bf x}_t)\leq N_{\text{train}}. 
\end{equation}

Since function $p_{0|t}({\bf x}_0|{\bf x}_t)$ is continuous and bounded, $\Vert \nabla_{{\bf x}_t}p_{0|t}({\bf x}_0|{\bf x}_t)\Vert$ is integrable. Therefore, we can apply the Leibniz rule and compute the score function $\tilde{\bf s}_t({\bf x}_t, {\bf y})$ in \eqref{eq:graddecompose} as follows:
\begin{align}
    \tilde{\bf{s}}_t({\bf x}_t, {\bf y})=&\nabla_{{\bf x}_t}\left(\int p_{0|t}({\bf x}_0|{\bf x}_t) p_0({\bf y}|{\bf x}_0)\text{d}{\bf x}_0\right)\nonumber\\
    \overset{(b)}{=}&\int \nabla_{{\bf x}_t}p_{0|t}({\bf x}_0|{\bf x}_t)p_0({\bf y}|{\bf x}_0)\text{d}{\bf x}_0\nonumber\\
    =&\int p_{0|t}({\bf x}_0|{\bf x}_t)\left(\frac{1}{p_{0|t}({\bf x}_0|{\bf x}_t)}\nabla_{{\bf x}_t}p_{0|t}({\bf x}_0|{\bf x}_t)\right)p_0({\bf y}|{\bf x}_0)\text{d}{\bf x}_0\nonumber\\
    \overset{(c)}{=}&\int p_{0|t}({\bf x}_0|{\bf x}_t)\nabla_{{\bf x}_t}\log p_{0|t}({\bf x}_0|{\bf x}_t)p_0({\bf y}|{\bf x}_0)\text{d}{\bf x}_0\nonumber\\
    =&\mathbb{E}_{p_{0|t}({\bf x}_0|{\bf x}_t)}\left[p_0({\bf y}|{\bf x}_0)\nabla_{{\bf x}_t}\log p_{0|t}({\bf x}_0|{\bf x}_t)\right],
\end{align}
where equation $(b)$  is obtained by exchanging the integration and gradient operator; equation $(c)$ is obtained because $\nabla_{{\bf x}_t}\log p_{0|t}({\bf x}_0|{\bf x}_t)=\frac{1}{p_{0|t}({\bf x}_0|{\bf x}_t)}p_{0|t}({\bf x}_0|{\bf x}_t)$. 

\subsection{Proof of Corollary 1}\label{pf:equivalent-dps}
The score function from 
\begin{align}
    &\tilde{\bf s}_t({\bf x}_t, {\bf y})\nonumber\\
    =&\mathbb{E}_{q({\bf x}_0|{\bf x}_t)}\left[p_0({\bf y}|{\bf x}_0)\nabla_{{\bf x}_t}\left(-\frac{1}{2r_t^2}\Vert{\bf x}_0-\hat{{\bf x}}_0({\bf x}_t)\Vert_2^2\right)\right]\nonumber\\
    =&\mathbb{E}_{q({\bf x}_0|{\bf x}_t)}\left[-\frac{1}{r_t^2}p_0({\bf y}|{\bf x}_0)\left((\hat{\bf x}_0({\bf x}_t)-{\bf x}_0)^T\frac{\partial \hat{\bf x}_0({\bf x}_t)}{\partial {\bf x}_t}\right)^T\right]\nonumber\\
        \overset{(a)}{=}&\mathbb{E}_{{\bm \xi}\sim\mathcal{N}(0, {\bf I})}\left[-\frac{1}{r_t^2}p_0({\bf y}|\hat{\bf x}_0({\bf x}_t)+r_t{\bm \xi})\left({\bm \xi}^T\frac{\partial \hat{\bf x}_0({\bf x}_t)}{\partial {\bf x}_t}\right)^T\right]\nonumber\\
    \overset{(b)}{\approx}&\mathbb{E}_{{\bm \xi}\sim\mathcal{N}(0, {\bf I})}\left[-\frac{1}{r_t^2}\left(p_0({\bf y}|\hat{\bf x}_0)+r_t\nabla_{{\bf x}_0}^Tp_0({\bf y}|\hat{\bf x}_0){\bm \xi}\right)\left({\bm \xi}^T\frac{\partial \hat{\bf x}_0({\bf x}_t)}{\partial {\bf x}_t}\right)^T\right]\nonumber\\
    \overset{(c)}{=}&\cancel{\mathbb{E}_{{\bm \xi}\sim\mathcal{N}(0, {\bf I})}\left[-\frac{1}{r_t^2}p_0({\bf y}|\hat{\bf x}_0)\left({\bm \xi}^T\frac{\partial \hat{\bf x}_0({\bf x}_t)}{\partial {\bf x}_t}\right)^T\right]}\nonumber\\
    &+\mathbb{E}_{{\bm \xi}\sim\mathcal{N}(0, {\bf I})}\left[-\frac{1}{r_t}\nabla_{{\bf x}_0}^Tp_0({\bf y}|\hat{\bf x}_0){\bm \xi}\cdot\left({\bm \xi}^T\frac{\partial \hat{\bf x}_0({\bf x}_t)}{\partial {\bf x}_t}\right)^T\right]\nonumber\\
    \overset{(d)}{=}&\frac{1}{2\sigma_{\bf y}^2r_t}p_0({\bf y}|{\bf x}_0)\mathbb{E}_{{\bm \xi}\sim\mathcal{N}(0, {\bf I})}\left[(\nabla_{\bf {x}_0}\ell({\bf y}, \mathcal{A}({\bf x}_0))^T{\bm \xi})\cdot\left({\bm \xi}^T\frac{\partial \hat{\bf x}_0({\bf x}_t)}{\partial {\bf x}_t}\right)^T\right]\nonumber\\
    =&\frac{1}{2\sigma_{\bf y}^2r_t}p_0({\bf y}|{\bf x}_0)\mathbb{E}_{{\bm \xi}\sim\mathcal{N}(0, {\bf I})}\left[({\bm\xi}^T\nabla_{\bf {x}_0}\ell({\bf y}, \mathcal{A}({\bf x}_0))\cdot\left((\frac{\partial \hat{\bf x}_0({\bf x}_t)}{\partial {\bf x}_t})^T{\bm \xi}\right)\right]\nonumber\\
    =&\frac{1}{2\sigma_{\bf y}^2r_t}p_0({\bf y}|{\bf x}_0)\mathbb{E}_{{\bm \xi}\sim\mathcal{N}(0, {\bf I})}\left[\text{Tr}\left({\bm \xi \bm \xi}^T\nabla_{\bf {x}_0}\ell({\bf y}, \mathcal{A}({\bf x}_0))\cdot(\frac{\partial \hat{\bf x}_0({\bf x}_t)}{\partial {\bf x}_t})^T\right)\right]\nonumber\\
    =&\frac{1}{2\sigma_{\bf y}^2r_t}p_0({\bf y}|{\bf x}_0)\nabla_{{\bf x}_t}\ell({\bf y}, \mathcal{A}({\bf x}_0))
\end{align}
where equality $(a)$ is obtained because $q({\bf x}_0|{\bf x}_0)$ is a Gaussian distribution; approximation $(b)$ is obtained via the first order Taylor expansion and is accurate when $r_t$ is small; equation $(c)$ is obtained because $\mathbb{E}[{\bm \xi}]=0$ and equation $(d)$ is obtained because
 $p_0({\bf y}|{\bf x}_0)$ is a Gaussian distribution with mean $\mathcal{A}({\bf x}_0)$, then denote $\ell({\bf y}, \mathcal{A}({\bf x}_0))=\Vert{\bf y}-\mathcal{A}({\bf x}_0)\Vert_2^2$ be the reconstruction loss,  the gradient $\nabla_{{\bf x}_0}p_0({\bf y}|{\bf x}_0)=-\frac{1}{2\sigma_{\bf y}^2}p_0({\bf y}|{\bf x}_0)\nabla \ell({\bf y}, \mathcal{A}({\bf x}_0))$. 
\section{Key parameters for experiments}\label{appendix:params}
\subsection{DPG}
For FFHQ dataset:
\begin{itemize}
    \item Gaussian Noise $\sigma=0.05$:
    \begin{itemize}
        \item Inpainting: $N_{\text{mc}}=5000$, $B=180$
        \item Super-Resolution: $N_{\text{mc}}=800$, $B=160$
        \item Gaussian Deblurring: $N_{\text{mc}}=800$, $B=200$
        \item Motion Deblurring: $N_{\text{mc}}=500$, $B=200$
    \end{itemize}
    \item Poisson Noise $\lambda=1.0$:
    \begin{itemize}
        \item Inpainting: $N_{\text{mc}}=5000$, $C=180$
        \item Super-Resolution: $N_{\text{mc}}=500$, $C=150$
        \item Gaussian Deblurring: $N_{\text{mc}}=500$, $C=150$
        \item Motion Deblurring: $N_{\text{mc}}=500$, $C=150$
    \end{itemize}
    \item Nonlinear Inversion Task with Gaussian noise $\sigma=0.05$:
    \begin{itemize}
        \item Phase Retrieval: $N_{\text{mc}}=1000$, $C=200$
        \item Non-Linear Deblurring: $N_{\text{mc}}=500$, $C=150$
    \end{itemize}
\end{itemize}

For ImageNet\&LSUN-Bedroom dataset:

Because the structure of the diffusion model used for ImageNet and LSUN is the same. We use the same parameters for experiments. 

\begin{itemize}
    \item Gaussian Noise $\sigma=0.05$:
\begin{itemize}
    \item Inpainting: $N=5000, B=250$
    \item Super-Resolution: $N=500$, $B=160$
    \item Gaussian Deblurring: $N=500$, $B=200$
    \item Motion Deblurring: $N=500$, $B=200$
\end{itemize}
\item Poisson Noise $\lambda=1.0$:
\begin{itemize}
    \item Super-Resolution: $N=500$, $B=200$
    \item Gaussian Deblurring: $N=500$, $B=200$
    \item Motion Deblurring: $N=500$, $B=200$
\end{itemize}
    \item Nonlinear Inversion Task with Gaussian noise $\sigma=0.05$:
    \begin{itemize}
        \item Phase Retrieval: $N=1000$, $B=200$
        \item Non-Linear Deblurring: $N=500$, $B=150$
    \end{itemize}
\end{itemize}
All the images are restored with DDPM solver using 1000 steps. 
\subsection{Key parameters for other methods}
For DPS \cite{chung2023diffusion}, we use the default setting \cite[Appendix D.1]{chung2022improving} with a DDPM solver using 1000 steps. \\
For DDNM \cite{wang2023ddnm} experiments, we use the DDNM+ algorithm (with traversal) in super-resolution and inpainting tasks. For deblurring experiments, we found that the result with DDNM+ is worse than DDNM (similar phenomena has been observed in \cite{meng2022diffusion}. Therefore, we the quantitative and visualization results for Gaussian deblurring are produced using DDNM algorithm. All the results of DDNM are obtained with an DDIM type solver using 100 steps. \\
For DDRM \cite{kawar2022denoising}, we use 100 steps for inpainting tasks, and 20 steps for super-resolution and deblurring tasks. \\
For Reddiff \cite{mardani2023variational}, we use the Adam optimizer with 1000 steps as suggested by the authors. \\
All the experiments are performed using the same pretrained diffusion generative models. 

\subsection{Runtime Comparisons}
All the experiments are implemented using NVIDIA A100 GPU. We display the running time for each image in Table~\ref{tab:running}. Notice that by using the DDPM solver, DPG takes more time than DPS due to MC sampling. However, the guidance score estimated by DPG is more accurate. Therefore, we have use the DDIM solver to retrieve images with less running time. 
\begin{table}[h]
\centering
\begin{tabular}{c|c}
    Method & Running Time \\
    \hline
    DPS & 188s\\
    DPG (DDPM) & 208s\\
    DPG (DDIM) & 53.4s\\
    DDRM & 2.1s\\
    DDNM+ & 18.5s
\end{tabular}
\caption{Running time comparison for different method}
\label{tab:running}
\end{table}
\section{More Experiement Results}
\subsection{Results on the Image Restoration Process}\label{sec:exp-progress}
\begin{figure}[h]
    \centering
    \includegraphics[width=.32\textwidth]{figures/progress/progress-sr.png}
    \includegraphics[width=.32\textwidth]{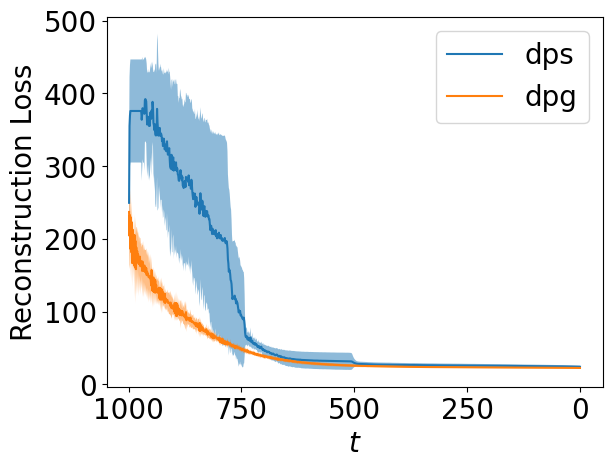}
    \includegraphics[width=.32\textwidth]{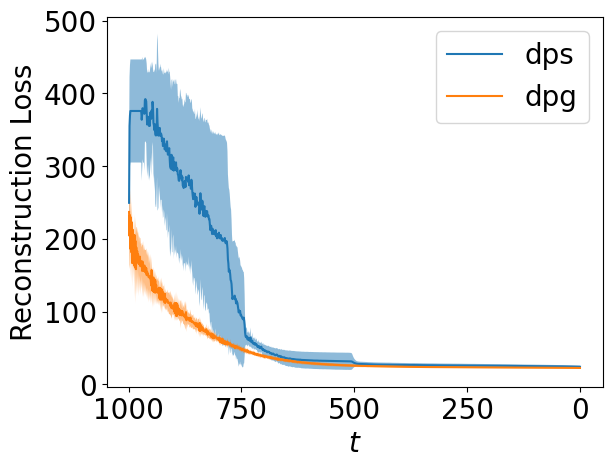}
    \caption{Evolution of the reconstruction error in the image restoration process of super-resolution (left), Gaussian Deblurring (middle) and Motion Deblurring (Right). }
    \label{fig:reconloss-more}
\end{figure}

\begin{figure}[h]
    \centering
    \begin{longtable}{c|cc|cc|cc}\renewcommand{\arraystretch}{0.0}
         & \multicolumn{2}{c}{Super-Resolution} & \multicolumn{2}{c}{Gaussian Deblurring}& \multicolumn{2}{c}{Motion Deblurring}\\
         \cline{2-7}
         & DPS& DPG & DPS& DPG& DPS & DPG\\
         \rotatebox{90}{$\hat{\bf x}_0({\bf x}_{900})$}&\includegraphics[width=.11\textwidth,frame]{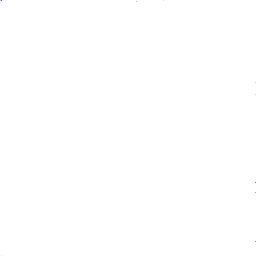}&\includegraphics[width=.11\textwidth]{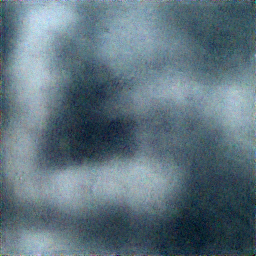}&\includegraphics[width=.11\textwidth]{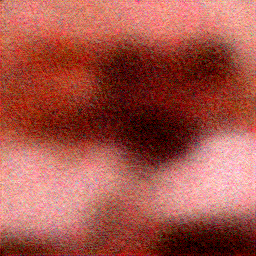}&\includegraphics[width=.11\textwidth]{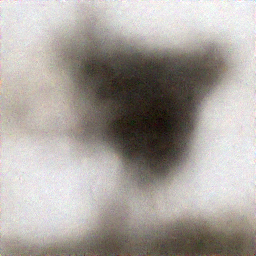}&\includegraphics[width=.11\textwidth]{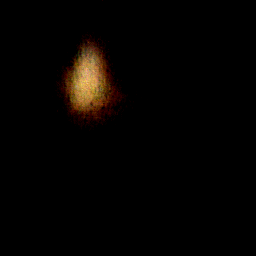}&\includegraphics[width=.11\textwidth]{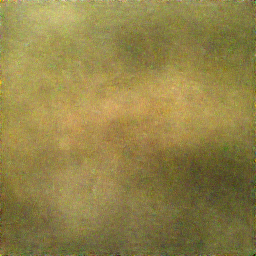}\\
        \rotatebox{90}{$\hat{\bf x}_0({\bf x}_{800})$}&\includegraphics[width=.11\textwidth]{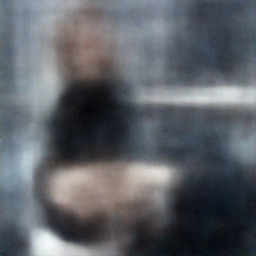}&\includegraphics[width=.11\textwidth]{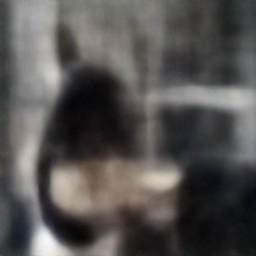}&\includegraphics[width=.11\textwidth]{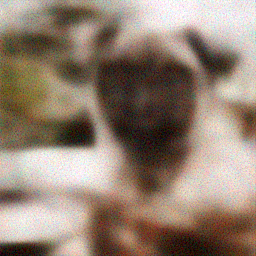}&\includegraphics[width=.11\textwidth]{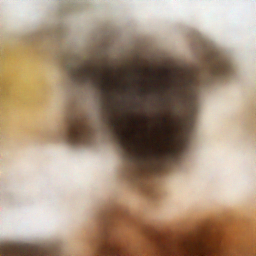}&\includegraphics[width=.11\textwidth]{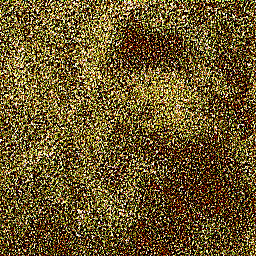}&\includegraphics[width=.11\textwidth]{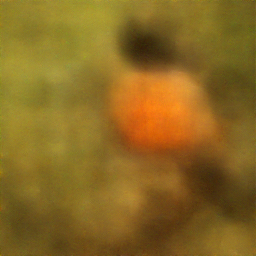}\\       
        \rotatebox{90}{$\hat{\bf x}_0({\bf x}_{700})$}&\includegraphics[width=.11\textwidth]{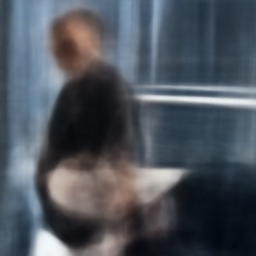}&\includegraphics[width=.11\textwidth]{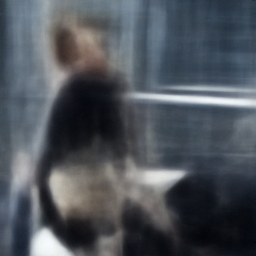}&\includegraphics[width=.11\textwidth]{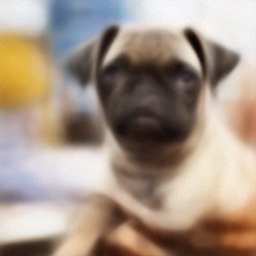}&\includegraphics[width=.11\textwidth]{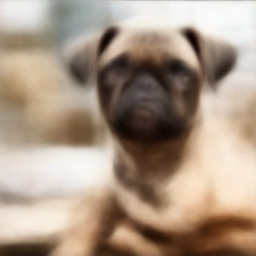}&\includegraphics[width=.11\textwidth]{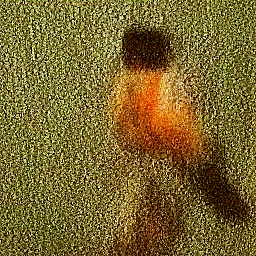}&\includegraphics[width=.11\textwidth]{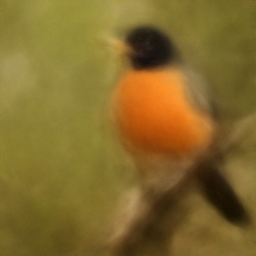}\\    
        \rotatebox{90}{$\hat{\bf x}_0({\bf x}_{600})$}&\includegraphics[width=.11\textwidth]{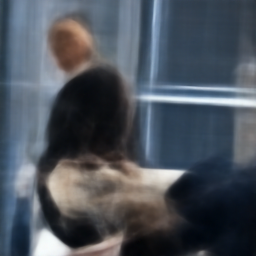}&\includegraphics[width=.11\textwidth]{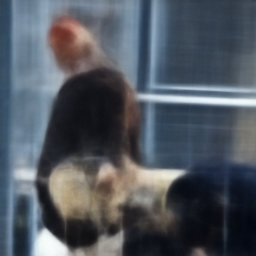}&\includegraphics[width=.11\textwidth]{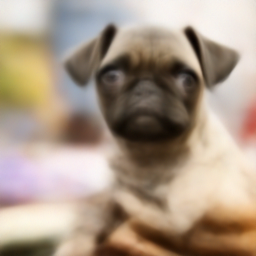}&\includegraphics[width=.11\textwidth]{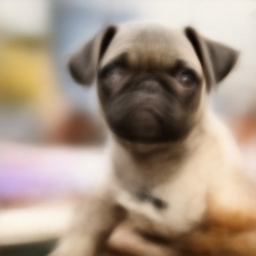}&\includegraphics[width=.11\textwidth]{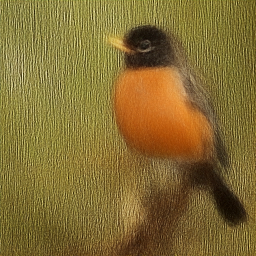}&\includegraphics[width=.11\textwidth]{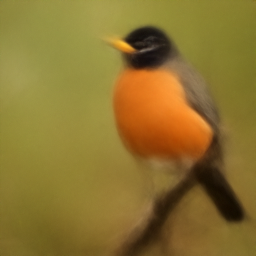}\\
        \rotatebox{90}{$\hat{\bf x}_0({\bf x}_{500})$}&\includegraphics[width=.11\textwidth]{figures/progress/sr/dps/x_0600.png}&\includegraphics[width=.11\textwidth]{figures/progress/sr/dpg/x_0600.png}&\includegraphics[width=.11\textwidth]{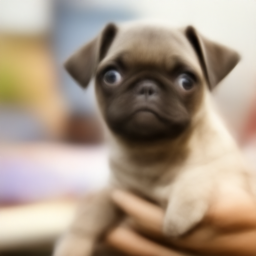}&\includegraphics[width=.11\textwidth]{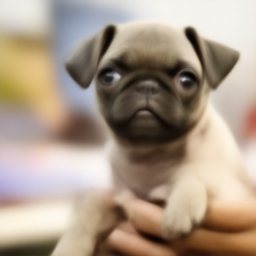}&\includegraphics[width=.11\textwidth]{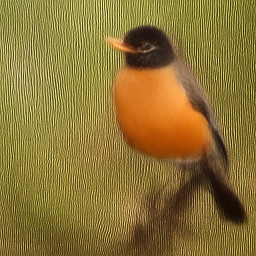}&\includegraphics[width=.11\textwidth]{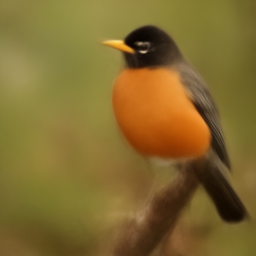}\\ 
        \rotatebox{90}{$\hat{\bf x}_0({\bf x}_{400})$}&\includegraphics[width=.11\textwidth]{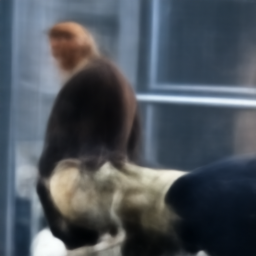}&\includegraphics[width=.11\textwidth]{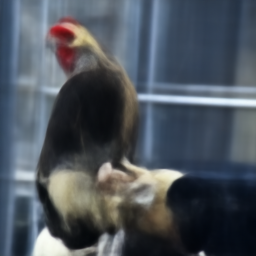}&\includegraphics[width=.11\textwidth]{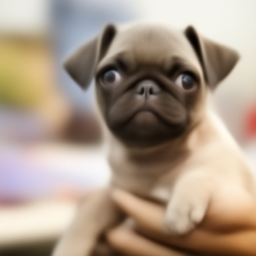}&\includegraphics[width=.11\textwidth]{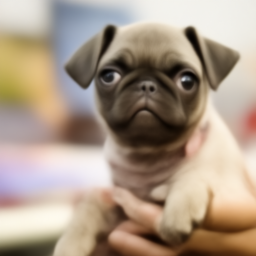}&\includegraphics[width=.11\textwidth]{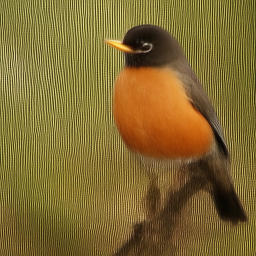}&\includegraphics[width=.11\textwidth]{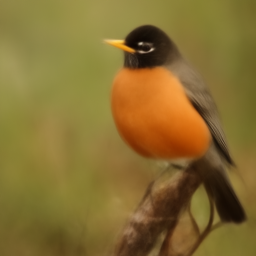}\\
        \rotatebox{90}{$\hat{\bf x}_0({\bf x}_{300})$}&\includegraphics[width=.11\textwidth]{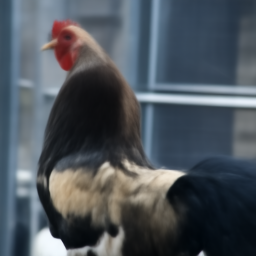}&\includegraphics[width=.11\textwidth]{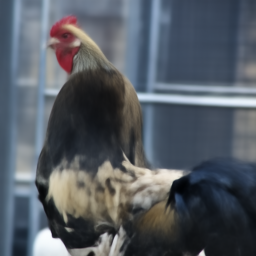}&\includegraphics[width=.11\textwidth]{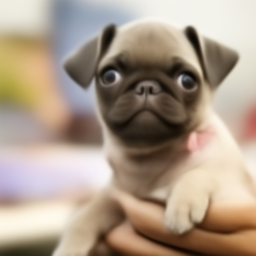}&\includegraphics[width=.11\textwidth]{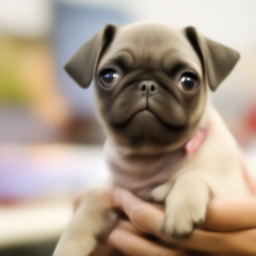}&\includegraphics[width=.11\textwidth]{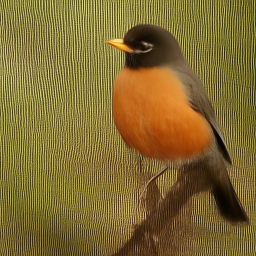}&\includegraphics[width=.11\textwidth]{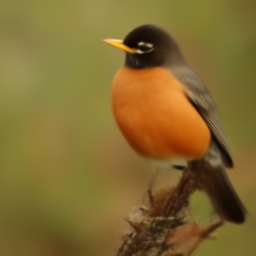}\\
        \rotatebox{90}{$\hat{\bf x}_0({\bf x}_{200})$}&\includegraphics[width=.11\textwidth]{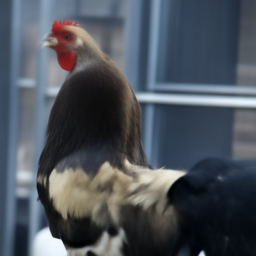}&\includegraphics[width=.11\textwidth]{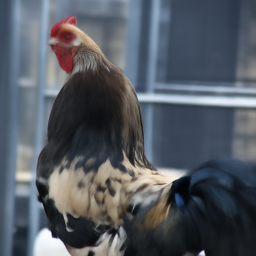}&\includegraphics[width=.11\textwidth]{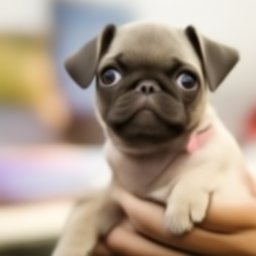}&\includegraphics[width=.11\textwidth]{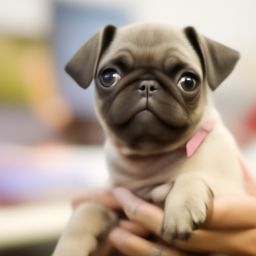}&\includegraphics[width=.11\textwidth]{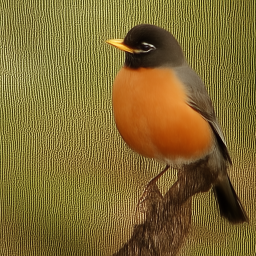}&\includegraphics[width=.11\textwidth]{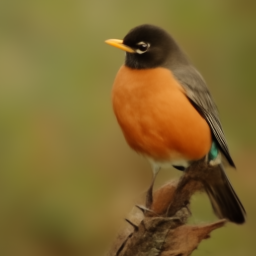}\\ 
        \rotatebox{90}{$\hat{\bf x}_0({\bf x}_{100})$}&\includegraphics[width=.11\textwidth]{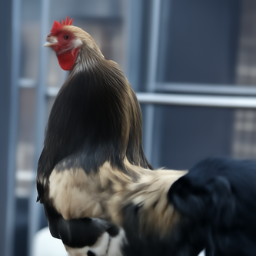}&\includegraphics[width=.11\textwidth]{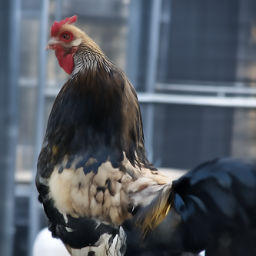}&\includegraphics[width=.11\textwidth]{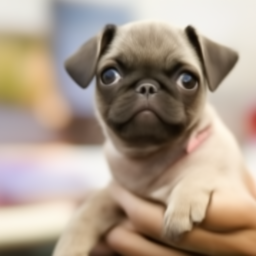}&\includegraphics[width=.11\textwidth]{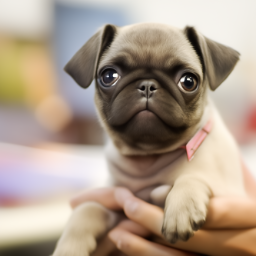}&\includegraphics[width=.11\textwidth]{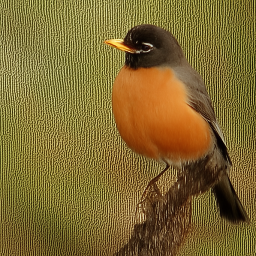}&\includegraphics[width=.11\textwidth]{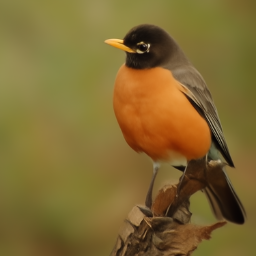}\\
        \rotatebox{90}{$\hat{\bf x}_0({\bf x}_{000})$}&\includegraphics[width=.11\textwidth]{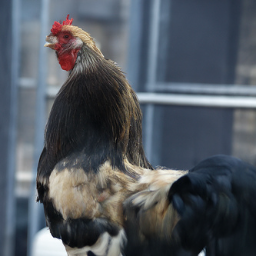}&\includegraphics[width=.11\textwidth]{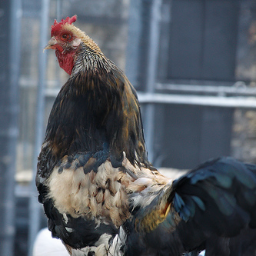}&\includegraphics[width=.11\textwidth]{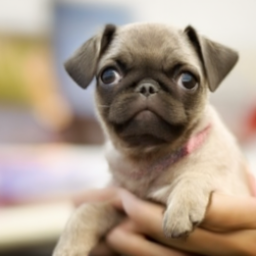}&\includegraphics[width=.11\textwidth]{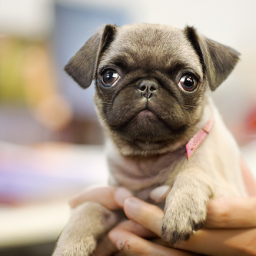}&\includegraphics[width=.11\textwidth]{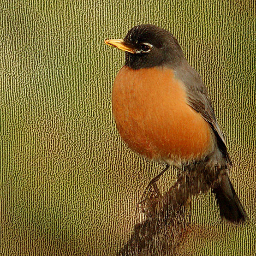}&\includegraphics[width=.11\textwidth]{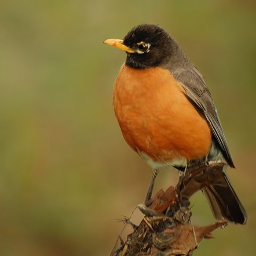}\\    
        \hline
        &\includegraphics[width=.11\textwidth]{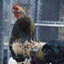}&\includegraphics[width=.11\textwidth]{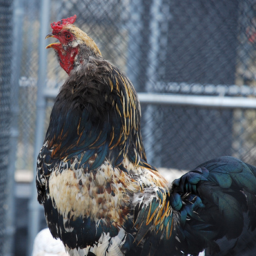}&\includegraphics[width=.11\textwidth]{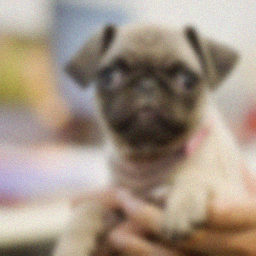}&\includegraphics[width=.11\textwidth]{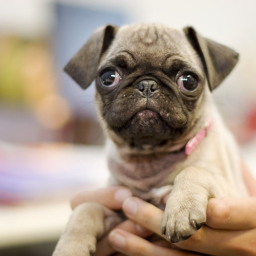}&\includegraphics[width=.11\textwidth]{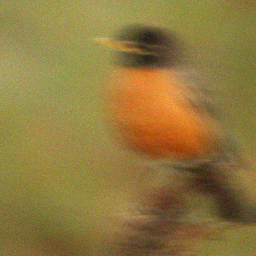}&\includegraphics[width=.11\textwidth]{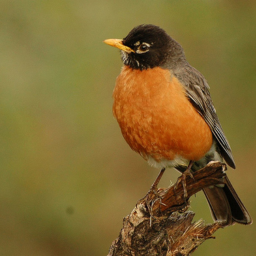}\\
        & Input & Label & Input & Label & Input & Label
    \end{longtable}
    \caption{The image restoration process for super-resolution, gaussian deblurring and motion deblurring task. }
    \label{fig:restore-process}
\end{figure}

We plot the resonctruction loss evolution of DPG (red) and DPS (blue) method in Fig.~\ref{fig:reconloss-more}. Shaded area depicts the confidence interval. The mean and confidence interval are obtained by taking the average of 10 runs. 
According to Fig.~\ref{fig:reconloss-more}, DPG always have a smaller reconstruction error in the earlier stages, and evolution of the reconstruction error is more stable. This leads to the observation that DPG can restore image sketches at an earlier stage, which provides room and opportunity to improve and generation detailed figures in later stage of the diffusion generation process. The image generation results for super-resolution and deblurring tasks are illustrated in Fig.~\ref{fig:restore-process}. 

\subsection{More Results on Image Restoration}
Image restoration results for super-resolution with different downsampling scale are displayed in Fig.~\ref{fig:appendix-sr-imagenet-mult} and~\ref{fig:appendix-sr-lsun-mult}. More image restoration results on super-resolution for ImageNet and LSUN dataset are displayed in Fig.~\ref{fig:appendix-sr-imagenet} and \ref{fig:appendix-sr-lsun}. Results for Gaussian Deblurring are displayed in Fig.~\ref{fig:appendix-deblur-g-imagenet} and \ref{fig:appendix-deblur-g-lsun}. More results for Motion Deblurring are displayed in Fig.~\ref{fig:appendix-deblur-motion}. 

\begingroup
\renewcommand{\arraystretch}{0.0}
\begin{figure}[h]
\setlength\arrayrulewidth{2pt}
\centering

\hspace{-21pt}\begin{tabular}{@{}c@{}c@{}c@{}c@{\hspace{0.1cm}}c@{}c@{}c@{\hspace{0.1cm}}c@{}c@{}c}
\specialrule{0em}{1pt}{1pt}
    {\small  Ground Truth}&{\small Input (2$\times$)}&{\small DPS}&{\small DPG}&{\small  Input (4$\times$)}&{\small DPS}&{\small DPG}&{\small Input (8$\times$)}&{\small DPS}&{\small DPG}\\
    \includegraphics[width=.1\textwidth]{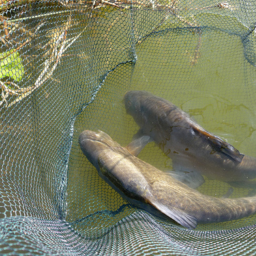}& \includegraphics[width=.1\textwidth]{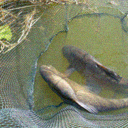}&\includegraphics[width=.1\textwidth]{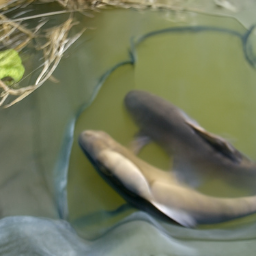}&\includegraphics[width=.1\textwidth]{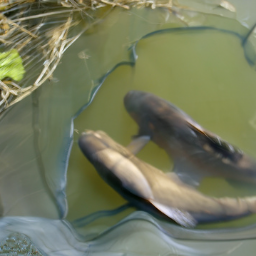}&\includegraphics[width=.1\textwidth]{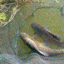}&\includegraphics[width=.1\textwidth]{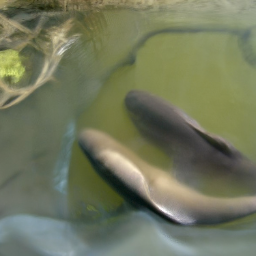}&\includegraphics[width=.1\textwidth]{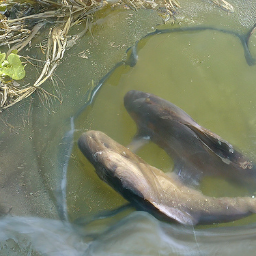}&\includegraphics[width=.1\textwidth]{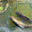}&\includegraphics[width=.1\textwidth]{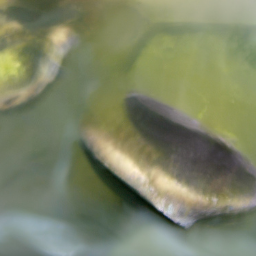}&\includegraphics[width=.1\textwidth]{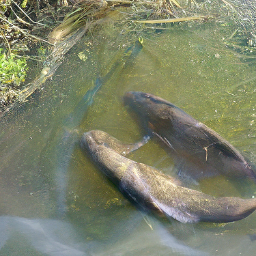}\\
    \includegraphics[width=.1\textwidth]{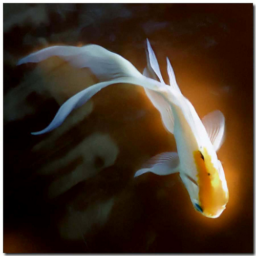}& \includegraphics[width=.1\textwidth]{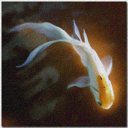}&\includegraphics[width=.1\textwidth]{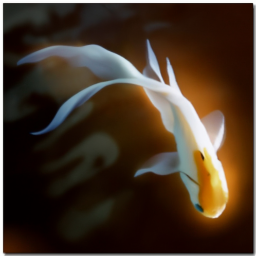}&\includegraphics[width=.1\textwidth]{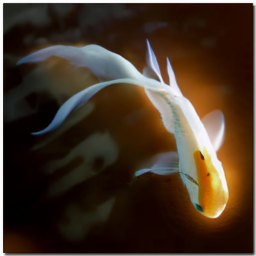}&\includegraphics[width=.1\textwidth]{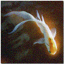}&\includegraphics[width=.1\textwidth]{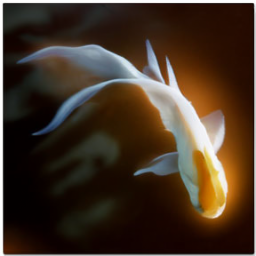}&\includegraphics[width=.1\textwidth]{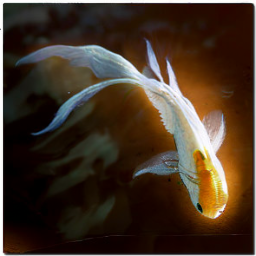}&\includegraphics[width=.1\textwidth]{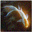}&\includegraphics[width=.1\textwidth]{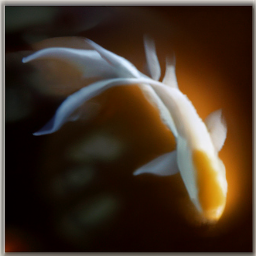}&\includegraphics[width=.1\textwidth]{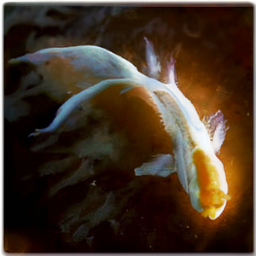}\\
    \includegraphics[width=.1\textwidth]{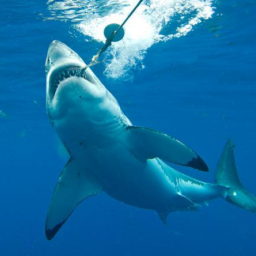}& \includegraphics[width=.1\textwidth]{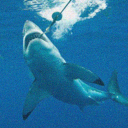}&\includegraphics[width=.1\textwidth]{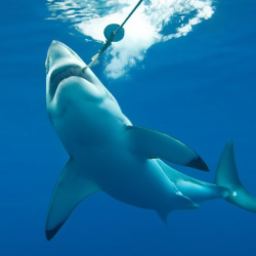}&\includegraphics[width=.1\textwidth]{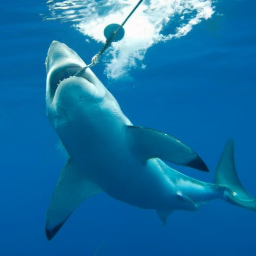}&\includegraphics[width=.1\textwidth]{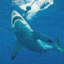}&\includegraphics[width=.1\textwidth]{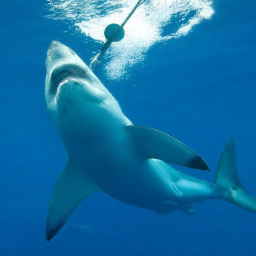}&\includegraphics[width=.1\textwidth]{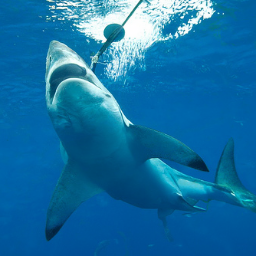}&\includegraphics[width=.1\textwidth]{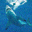}&\includegraphics[width=.1\textwidth]{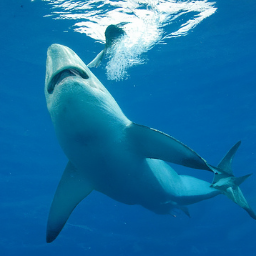}&\includegraphics[width=.1\textwidth]{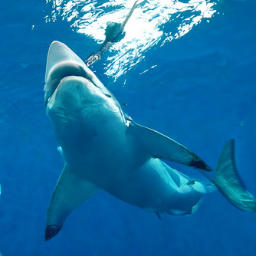}\\
    \includegraphics[width=.1\textwidth]{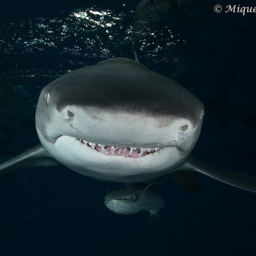}& \includegraphics[width=.1\textwidth]{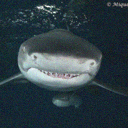}&\includegraphics[width=.1\textwidth]{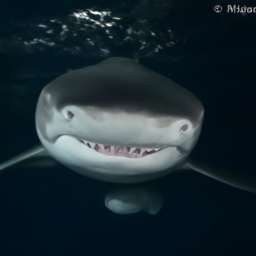}&\includegraphics[width=.1\textwidth]{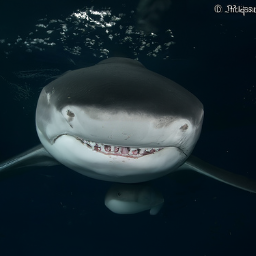}&\includegraphics[width=.1\textwidth]{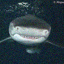}&\includegraphics[width=.1\textwidth]{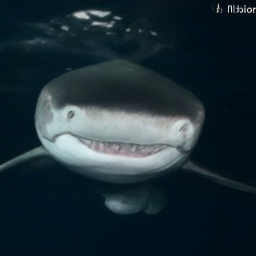}&\includegraphics[width=.1\textwidth]{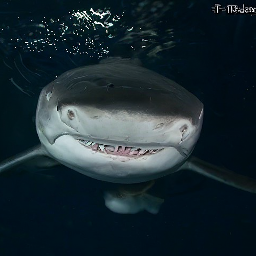}&\includegraphics[width=.1\textwidth]{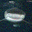}&\includegraphics[width=.1\textwidth]{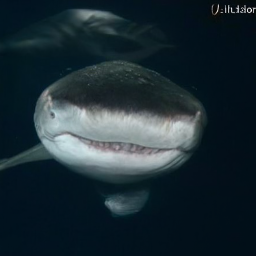}&\includegraphics[width=.1\textwidth]{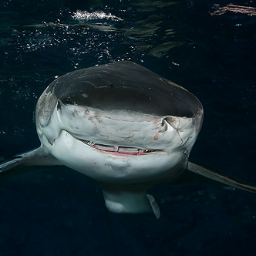}\\
    \includegraphics[width=.1\textwidth]{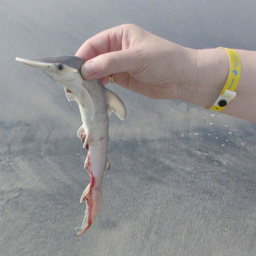}& \includegraphics[width=.1\textwidth]{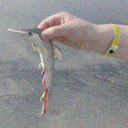}&\includegraphics[width=.1\textwidth]{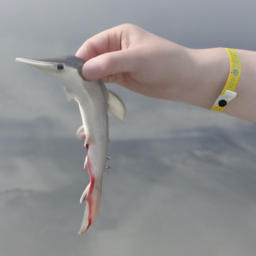}&\includegraphics[width=.1\textwidth]{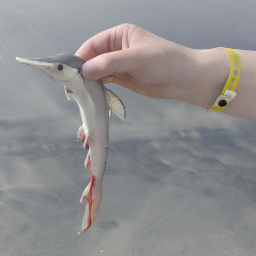}&\includegraphics[width=.1\textwidth]{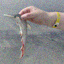}&\includegraphics[width=.1\textwidth]{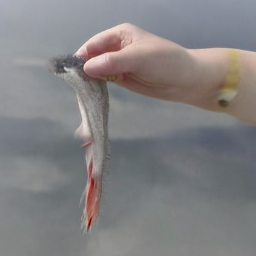}&\includegraphics[width=.1\textwidth]{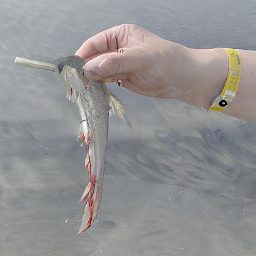}&\includegraphics[width=.1\textwidth]{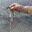}&\includegraphics[width=.1\textwidth]{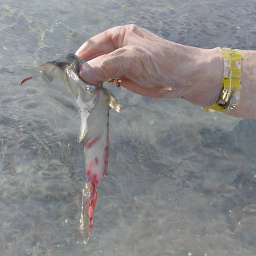}&\includegraphics[width=.1\textwidth]{figures/appendix/sr-imagenet/sr8-dpg-4.png}
    \end{tabular}
    \caption{Image Restoration Results on Super-Resolution with different factors on ImageNet dataset.}
    \label{fig:appendix-sr-imagenet-mult}
\end{figure}
\endgroup

\begingroup
\renewcommand{\arraystretch}{0.0}
\begin{figure}[h]
\setlength\arrayrulewidth{2pt}
\centering
\hspace{-21pt}\begin{tabular}{@{}c@{}c@{}c@{}c@{\hspace{0.1cm}}c@{}c@{}c@{\hspace{0.1cm}}c@{}c@{}c}
\specialrule{0em}{1pt}{1pt}
    {\small  Ground Truth}&{\small Input (2$\times$)}&{\small DPS}&{\small DPG}&{\small  Input (4$\times$)}&{\small DPS}&{\small DPG}&{\small Input (8$\times$)}&{\small DPS}&{\small DPG}\\
    \includegraphics[width=.1\textwidth]{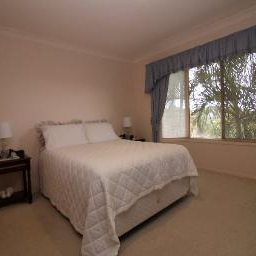}& \includegraphics[width=.1\textwidth]{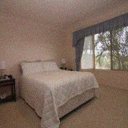}&\includegraphics[width=.1\textwidth]{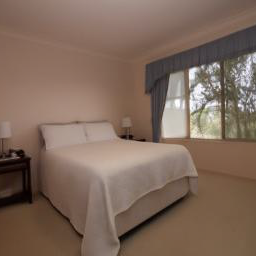}&\includegraphics[width=.1\textwidth]{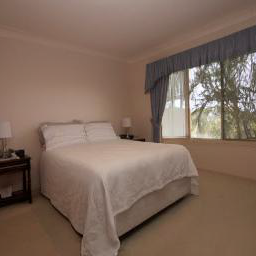}&\includegraphics[width=.1\textwidth]{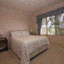}&\includegraphics[width=.1\textwidth]{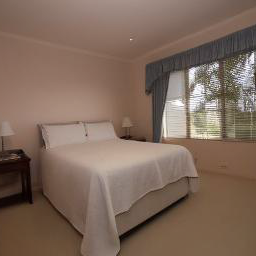}&\includegraphics[width=.1\textwidth]{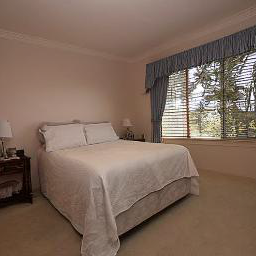}&\includegraphics[width=.1\textwidth]{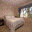}&\includegraphics[width=.1\textwidth]{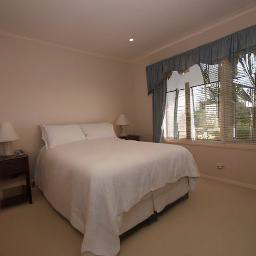}&\includegraphics[width=.1\textwidth]{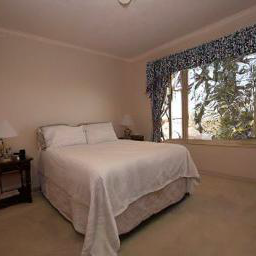}\\
    \includegraphics[width=.1\textwidth]{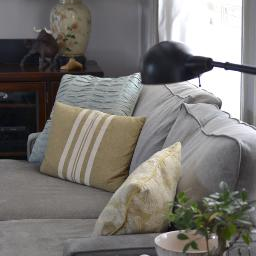}& \includegraphics[width=.1\textwidth]{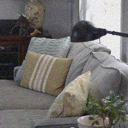}&\includegraphics[width=.1\textwidth]{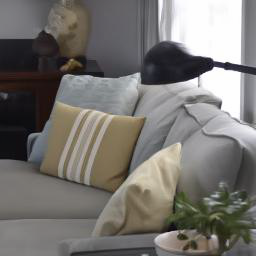}&\includegraphics[width=.1\textwidth]{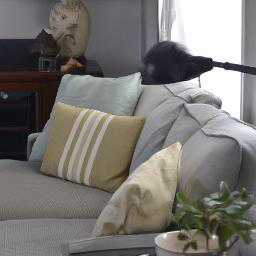}&\includegraphics[width=.1\textwidth]{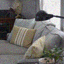}&\includegraphics[width=.1\textwidth]{figures/appendix/sr-imagenet/sr4-dps-1.png}&\includegraphics[width=.1\textwidth]{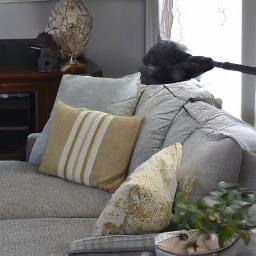}&\includegraphics[width=.1\textwidth]{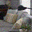}&\includegraphics[width=.1\textwidth]{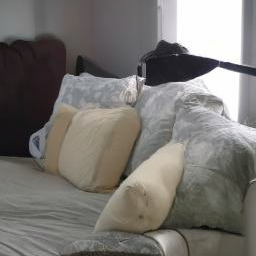}&\includegraphics[width=.1\textwidth]{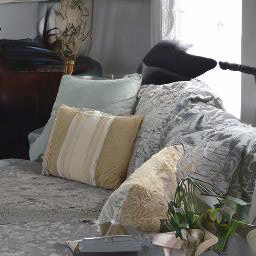}\\
    \includegraphics[width=.1\textwidth]{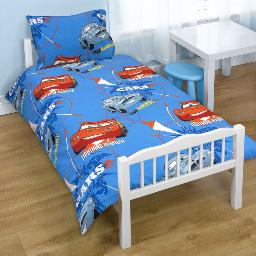}& \includegraphics[width=.1\textwidth]{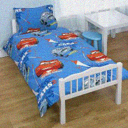}&\includegraphics[width=.1\textwidth]{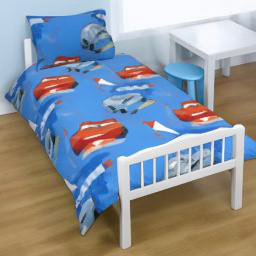}&\includegraphics[width=.1\textwidth]{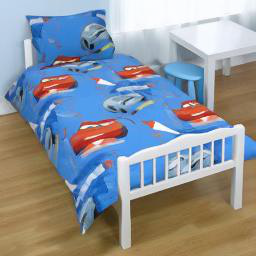}&\includegraphics[width=.1\textwidth]{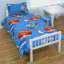}&\includegraphics[width=.1\textwidth]{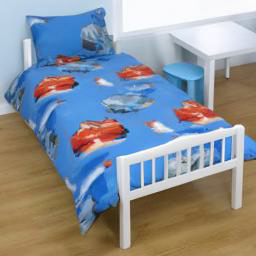}&\includegraphics[width=.1\textwidth]{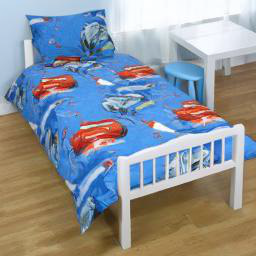}&\includegraphics[width=.1\textwidth]{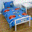}&\includegraphics[width=.1\textwidth]{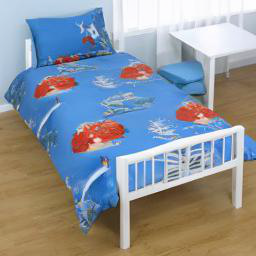}&\includegraphics[width=.1\textwidth]{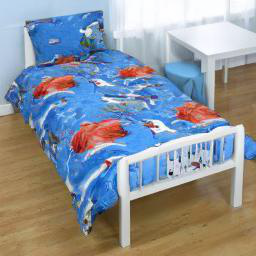}\\
    \includegraphics[width=.1\textwidth]{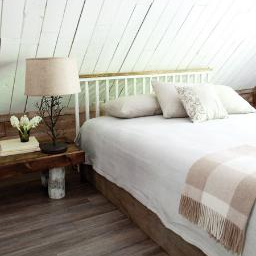}& \includegraphics[width=.1\textwidth]{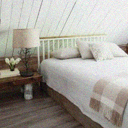}&\includegraphics[width=.1\textwidth]{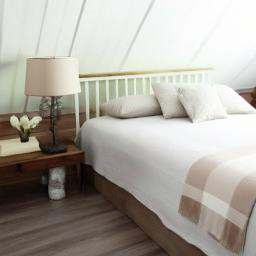}&\includegraphics[width=.1\textwidth]{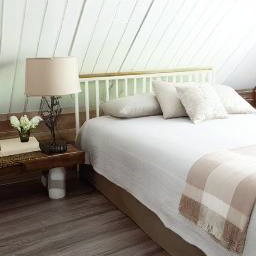}&\includegraphics[width=.1\textwidth]{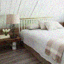}&\includegraphics[width=.1\textwidth]{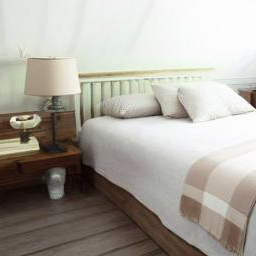}&\includegraphics[width=.1\textwidth]{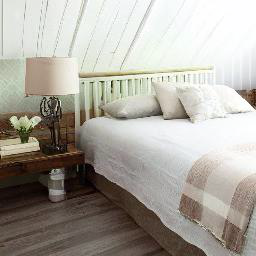}&\includegraphics[width=.1\textwidth]{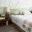}&\includegraphics[width=.1\textwidth]{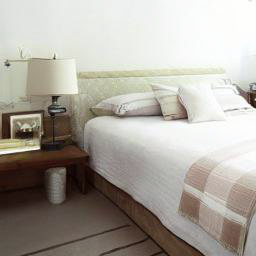}&\includegraphics[width=.1\textwidth]{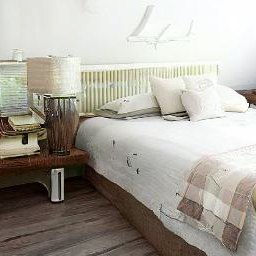}\\
    \includegraphics[width=.1\textwidth]{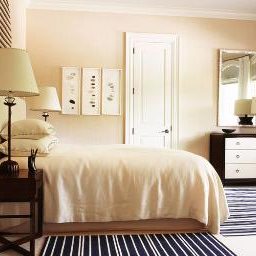}& \includegraphics[width=.1\textwidth]{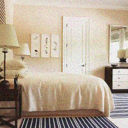}&\includegraphics[width=.1\textwidth]{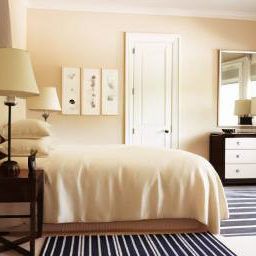}&\includegraphics[width=.1\textwidth]{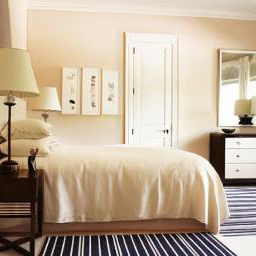}&\includegraphics[width=.1\textwidth]{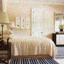}&\includegraphics[width=.1\textwidth]{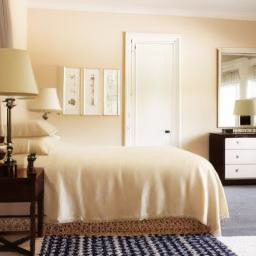}&\includegraphics[width=.1\textwidth]{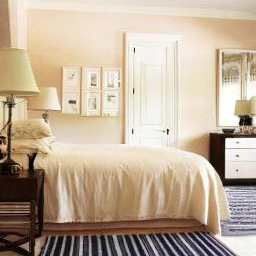}&\includegraphics[width=.1\textwidth]{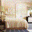}&\includegraphics[width=.1\textwidth]{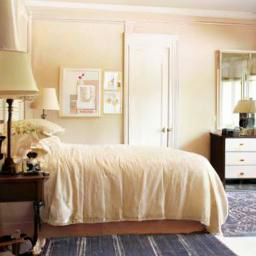}&\includegraphics[width=.1\textwidth]{figures/appendix/sr-lsun/sr8-dpg-4.png}
    \end{tabular}
    \caption{Image Restoration Results on Super-Resolution with different factors on LSUN dataset.}
    \label{fig:appendix-sr-lsun-mult}
\end{figure}
\endgroup

\begingroup
\renewcommand{\arraystretch}{0.0}
\begin{figure}[h]
\setlength\arrayrulewidth{2pt}
\centering
\begin{tabular}{c@{}c@{}c@{}c@{}c@{}|@{}c@{}|@{}c}
\cline{6-6}
\specialrule{0em}{1pt}{1pt}
    {\small  Input}&{\small DDRM}&{\small DDNM+}&{\small Reddiff}&{\small  DPS}&{\small \bf DPG}&{\small Ground Truth}\\
    \includegraphics[width=.12\textwidth]{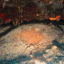}& \includegraphics[width=.12\textwidth]{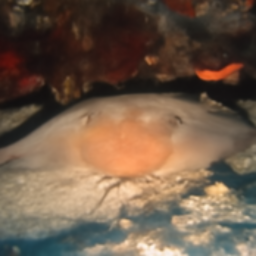}&\includegraphics[width=.12\textwidth]{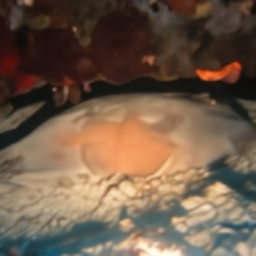}&\includegraphics[width=.12\textwidth]{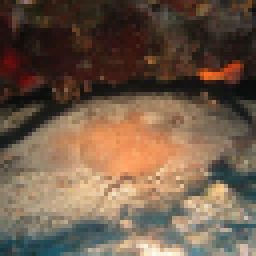}&\includegraphics[width=.12\textwidth]{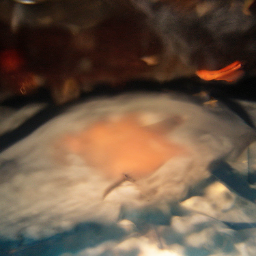}&\includegraphics[width=.12\textwidth]{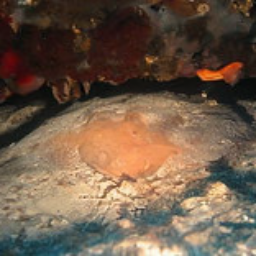}&\includegraphics[width=.12\textwidth]{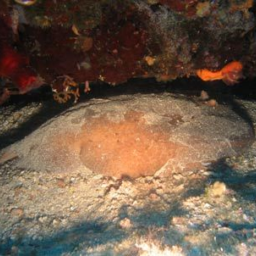}\\
    \includegraphics[width=.12\textwidth]{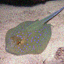}& \includegraphics[width=.12\textwidth]{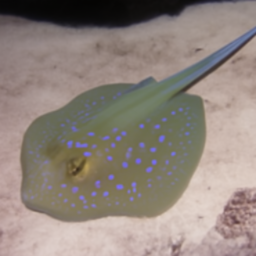}&\includegraphics[width=.12\textwidth]{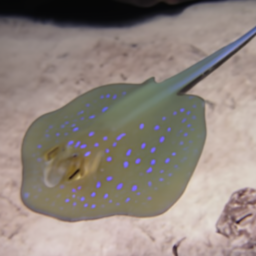}&\includegraphics[width=.12\textwidth]{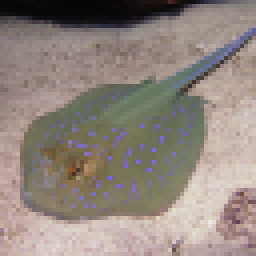}&\includegraphics[width=.12\textwidth]{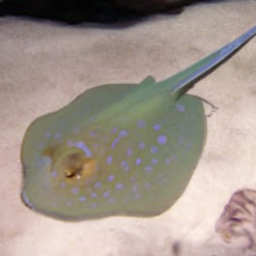}&\includegraphics[width=.12\textwidth]{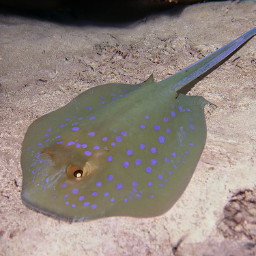}&\includegraphics[width=.12\textwidth]{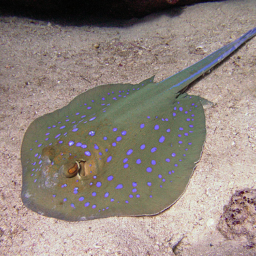}\\
        \includegraphics[width=.12\textwidth]{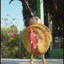}& \includegraphics[width=.12\textwidth]{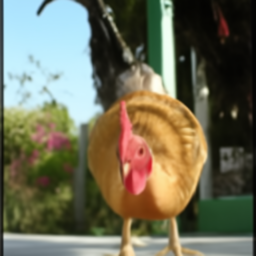}&\includegraphics[width=.12\textwidth]{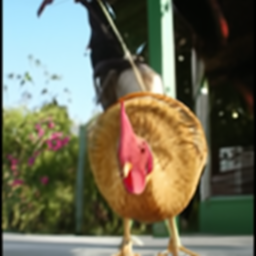}&\includegraphics[width=.12\textwidth]{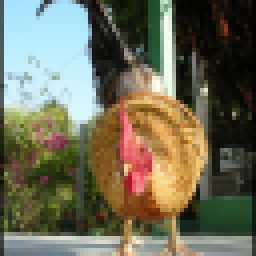}&\includegraphics[width=.12\textwidth]{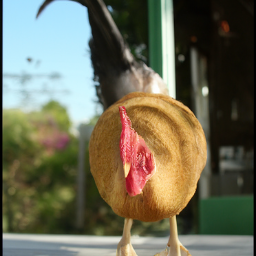}&\includegraphics[width=.12\textwidth]{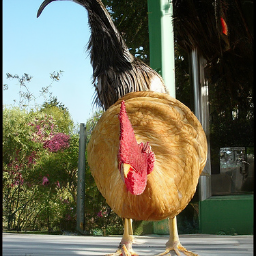}&\includegraphics[width=.12\textwidth]{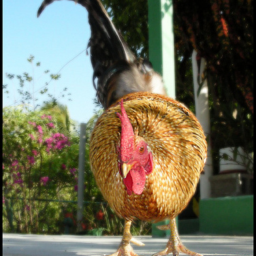}\\
            \includegraphics[width=.12\textwidth]{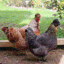}& \includegraphics[width=.12\textwidth]{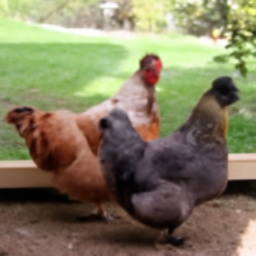}&\includegraphics[width=.12\textwidth]{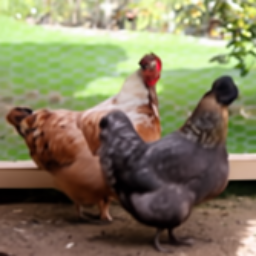}&\includegraphics[width=.12\textwidth]{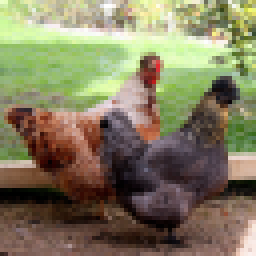}&\includegraphics[width=.12\textwidth]{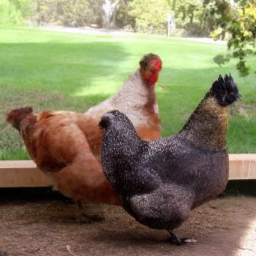}&\includegraphics[width=.12\textwidth]{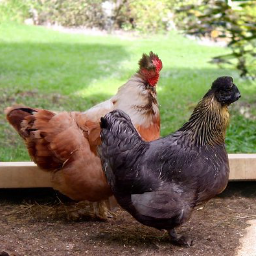}&\includegraphics[width=.12\textwidth]{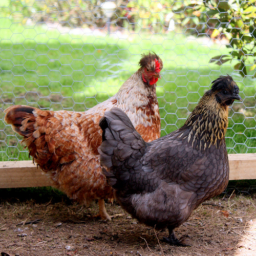}\\
                \includegraphics[width=.12\textwidth]{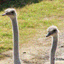}& \includegraphics[width=.12\textwidth]{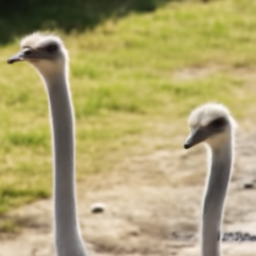}&\includegraphics[width=.12\textwidth]{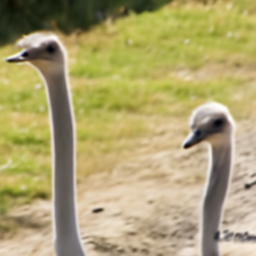}&\includegraphics[width=.12\textwidth]{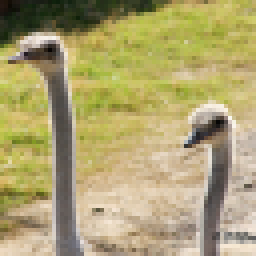}&\includegraphics[width=.12\textwidth]{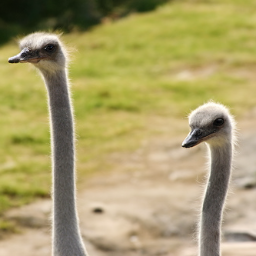}&\includegraphics[width=.12\textwidth]{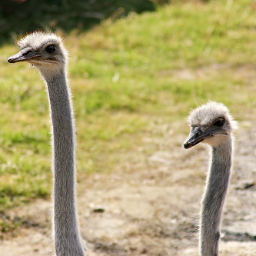}&\includegraphics[width=.12\textwidth]{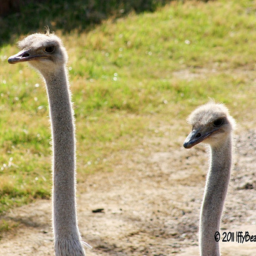}\\
                \cline{6-6}
    \end{tabular}
    \caption{Image Restoration Results on 4$\times$ Super-Resolution on ImageNet Dataset. }
    \label{fig:appendix-sr-imagenet}
\end{figure}
\endgroup

\begingroup
\renewcommand{\arraystretch}{0.0}
\begin{figure}[h]
\setlength\arrayrulewidth{2pt}
\centering
\begin{tabular}{c@{}c@{}c@{}c@{}|@{}c@{}|@{}c}
\cline{5-5}
\specialrule{0em}{1pt}{1pt}
    {\small  Input}&{\small DDRM}&{\small DDNM+}&{\small  DPS}&{\small \bf DPG}&{\small Ground Truth}\\
    \includegraphics[width=.12\textwidth]{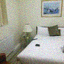}& \includegraphics[width=.12\textwidth]{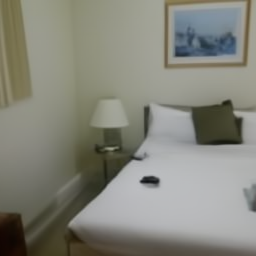}&\includegraphics[width=.12\textwidth]{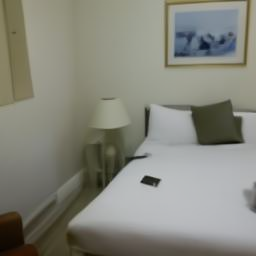}&\includegraphics[width=.12\textwidth]{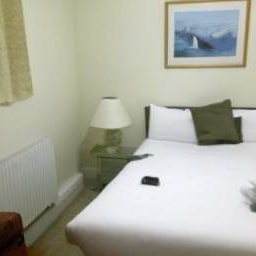}&\includegraphics[width=.12\textwidth]{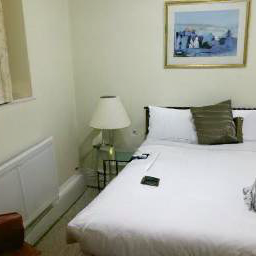}&\includegraphics[width=.12\textwidth]{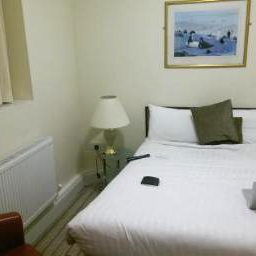}\\
    \includegraphics[width=.12\textwidth]{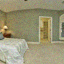}& \includegraphics[width=.12\textwidth]{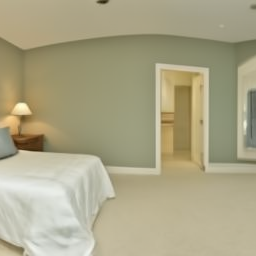}&\includegraphics[width=.12\textwidth]{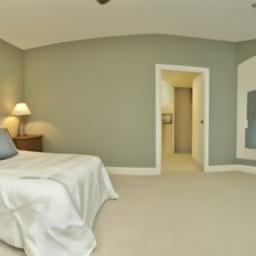}&\includegraphics[width=.12\textwidth]{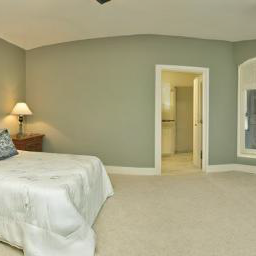}&\includegraphics[width=.12\textwidth]{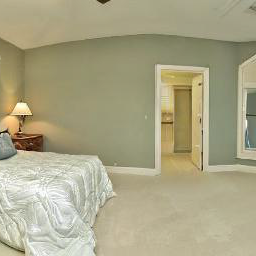}&\includegraphics[width=.12\textwidth]{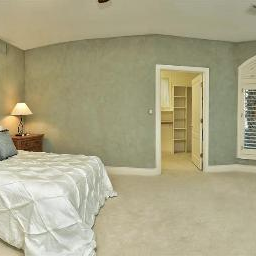}\\
        \includegraphics[width=.12\textwidth]{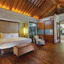}& \includegraphics[width=.12\textwidth]{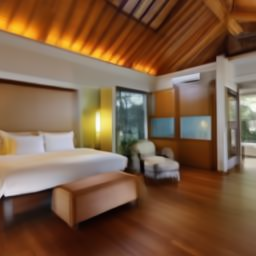}&\includegraphics[width=.12\textwidth]{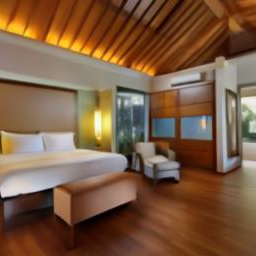}&\includegraphics[width=.12\textwidth]{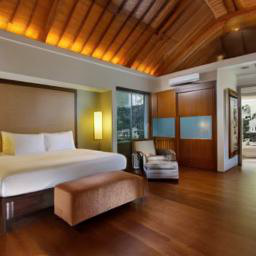}&\includegraphics[width=.12\textwidth]{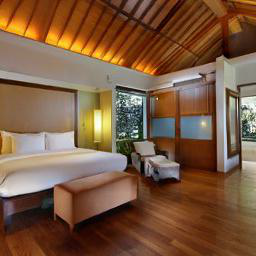}&\includegraphics[width=.12\textwidth]{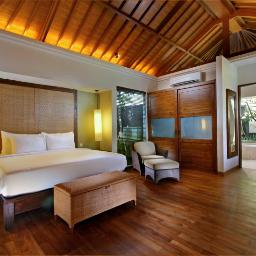}\\
            \includegraphics[width=.12\textwidth]{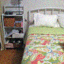}& \includegraphics[width=.12\textwidth]{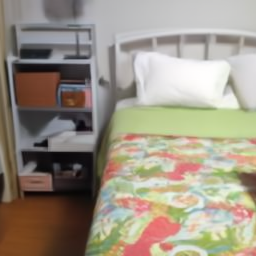}&\includegraphics[width=.12\textwidth]{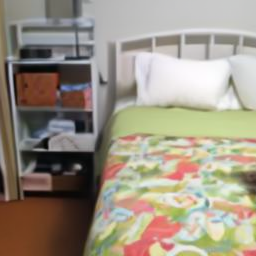}&\includegraphics[width=.12\textwidth]{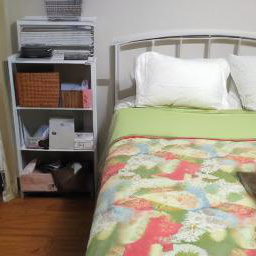}&\includegraphics[width=.12\textwidth]{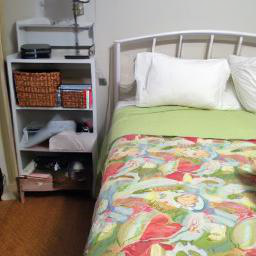}&\includegraphics[width=.12\textwidth]{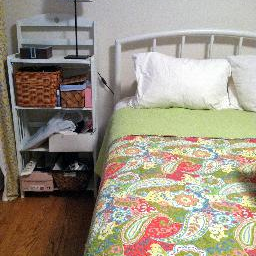}\\
                \includegraphics[width=.12\textwidth]{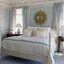}& \includegraphics[width=.12\textwidth]{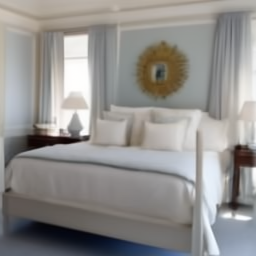}&\includegraphics[width=.12\textwidth]{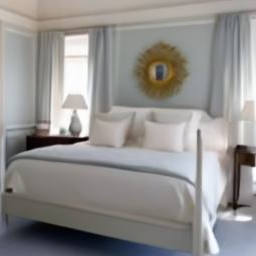}&\includegraphics[width=.12\textwidth]{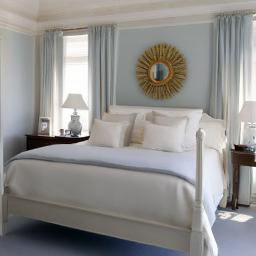}&\includegraphics[width=.12\textwidth]{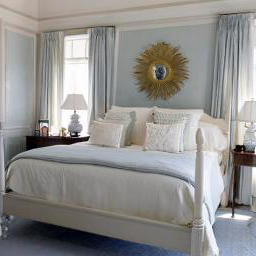}&\includegraphics[width=.12\textwidth]{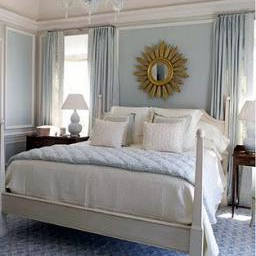}\\\cline{5-5}
    \end{tabular}
    \caption{Image Restoration Results on 4$\times$ Super-Resolution on LSUN Dataset. }
    \label{fig:appendix-sr-lsun}
\end{figure}
\endgroup
\begingroup
\renewcommand{\arraystretch}{0.0}
\begin{figure}[h]
\setlength\arrayrulewidth{2pt}
\centering
\begin{tabular}{c@{}c@{}c@{}c@{}c@{}|@{}c@{}|@{}c}
\cline{6-6}
\specialrule{0em}{1pt}{1pt}
    {\small  Input}&{\small DDRM}&{\small DDNM+}&{\small Reddiff}&{\small  DPS}&{\small \bf DPG}&{\small Ground Truth}\\
    \includegraphics[width=.12\textwidth]{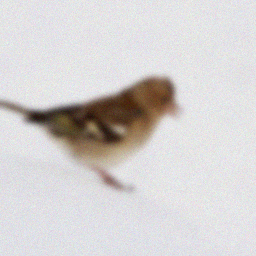}& \includegraphics[width=.12\textwidth]{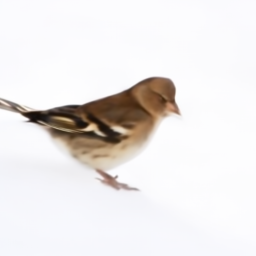}&\includegraphics[width=.12\textwidth]{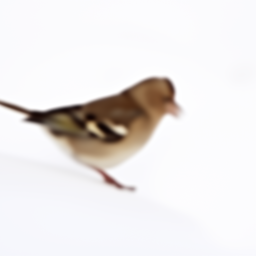}&\includegraphics[width=.12\textwidth]{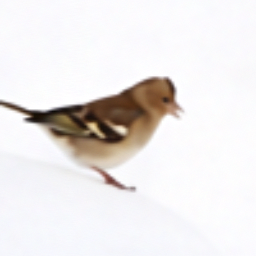}&\includegraphics[width=.12\textwidth]{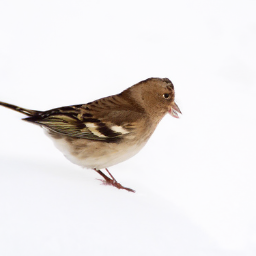}&\includegraphics[width=.12\textwidth]{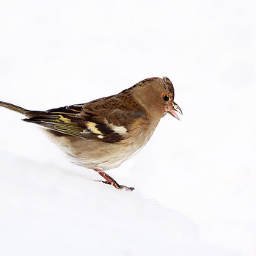}&\includegraphics[width=.12\textwidth]{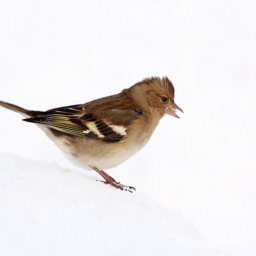}\\
    \includegraphics[width=.12\textwidth]{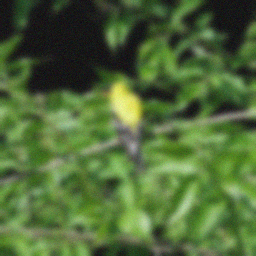}& \includegraphics[width=.12\textwidth]{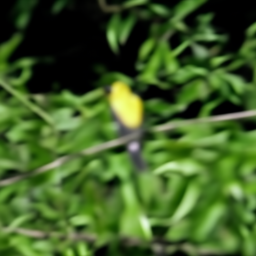}&\includegraphics[width=.12\textwidth]{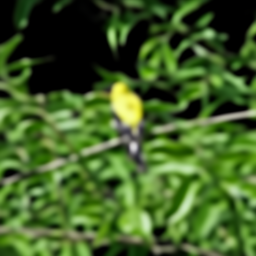}&\includegraphics[width=.12\textwidth]{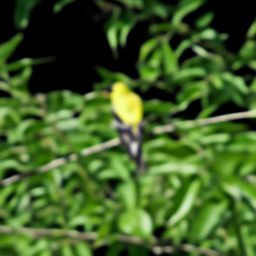}&\includegraphics[width=.12\textwidth]{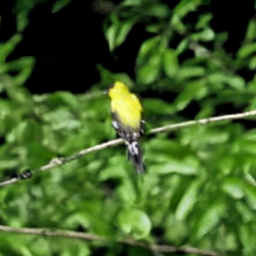}&\includegraphics[width=.12\textwidth]{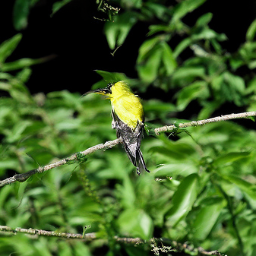}&\includegraphics[width=.12\textwidth]{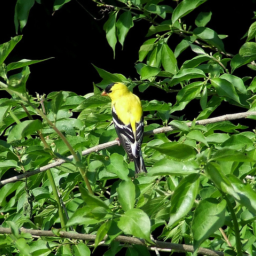}\\
        \includegraphics[width=.12\textwidth]{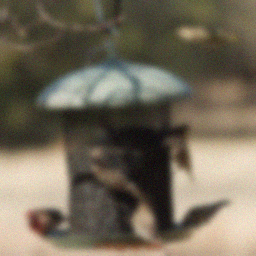}& \includegraphics[width=.12\textwidth]{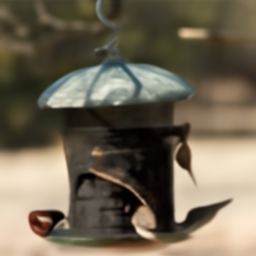}&\includegraphics[width=.12\textwidth]{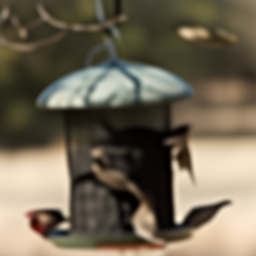}&\includegraphics[width=.12\textwidth]{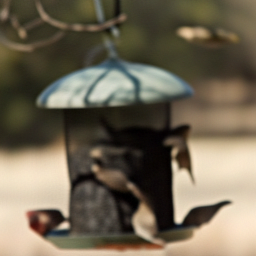}&\includegraphics[width=.12\textwidth]{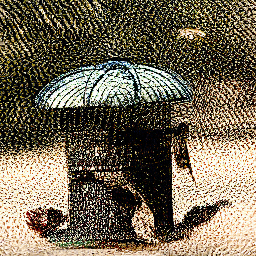}&\includegraphics[width=.12\textwidth]{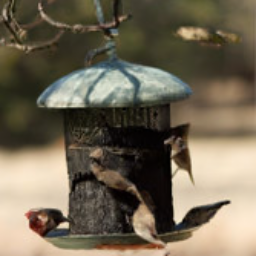}&\includegraphics[width=.12\textwidth]{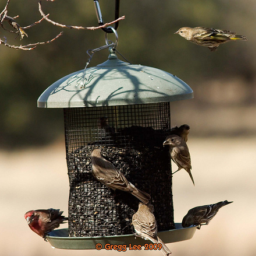}\\
            \includegraphics[width=.12\textwidth]{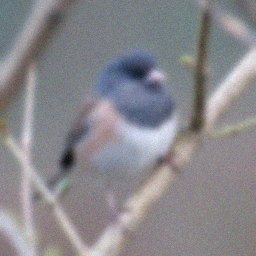}& \includegraphics[width=.12\textwidth]{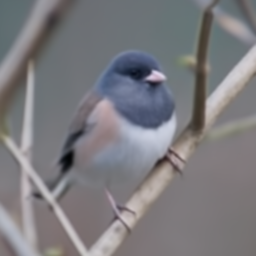}&\includegraphics[width=.12\textwidth]{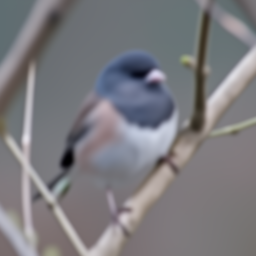}&\includegraphics[width=.12\textwidth]{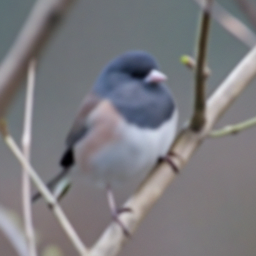}&\includegraphics[width=.12\textwidth]{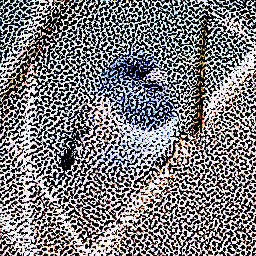}&\includegraphics[width=.12\textwidth]{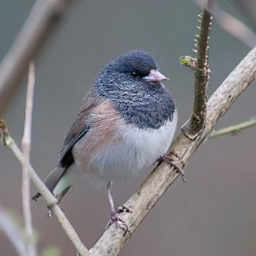}&\includegraphics[width=.12\textwidth]{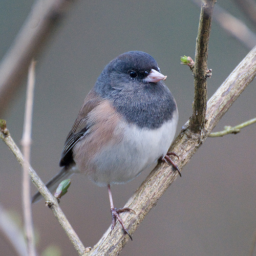}\\
                \includegraphics[width=.12\textwidth]{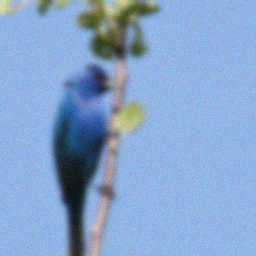}& \includegraphics[width=.12\textwidth]{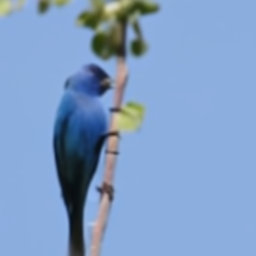}&\includegraphics[width=.12\textwidth]{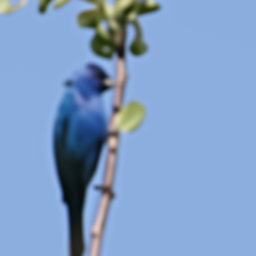}&\includegraphics[width=.12\textwidth]{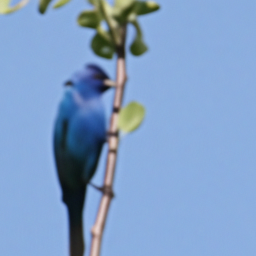}&\includegraphics[width=.12\textwidth]{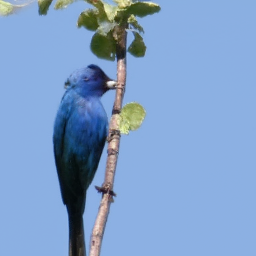}&\includegraphics[width=.12\textwidth]{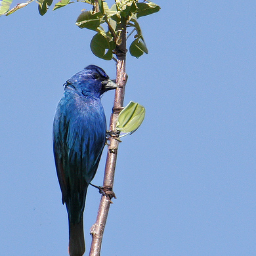}&\includegraphics[width=.12\textwidth]{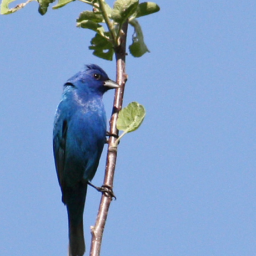}\\
                \cline{6-6}
    \end{tabular}
    \caption{Image Restoration Results on Gaussian Deblurring on ImageNet Dataset. }
    \label{fig:appendix-deblur-g-imagenet}
\end{figure}
\endgroup

\begingroup
\renewcommand{\arraystretch}{0.0}
\begin{figure}[h]
\setlength\arrayrulewidth{2pt}
\centering
\begin{tabular}{c@{}c@{}c@{}c@{}|@{}c@{}|@{}c}
\cline{5-5}
\specialrule{0em}{1pt}{1pt}
    {\small  Input}&{\small DDRM}&{\small DDNM+}&{\small  DPS}&{\small \bf DPG}&{\small Ground Truth}\\
    \includegraphics[width=.12\textwidth]{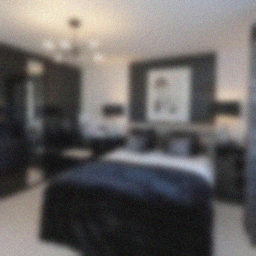}& \includegraphics[width=.12\textwidth]{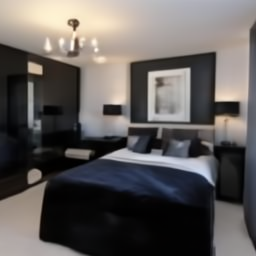}&\includegraphics[width=.12\textwidth]{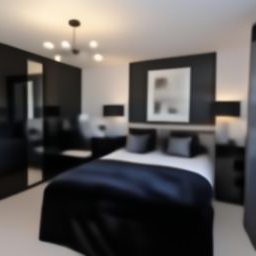}&\includegraphics[width=.12\textwidth]{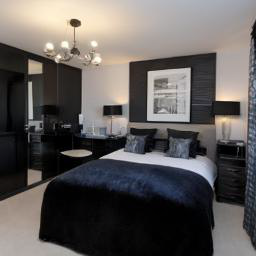}&\includegraphics[width=.12\textwidth]{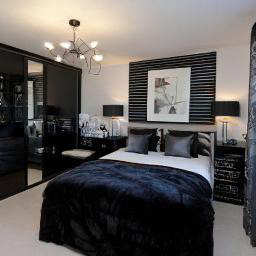}&\includegraphics[width=.12\textwidth]{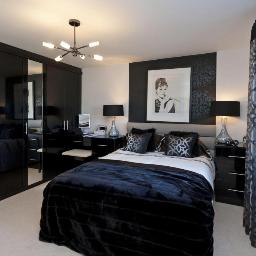}\\
    \includegraphics[width=.12\textwidth]{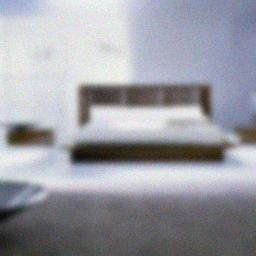}& \includegraphics[width=.12\textwidth]{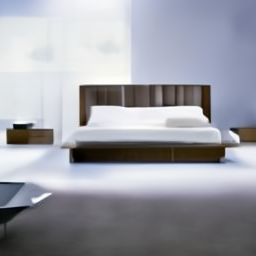}&\includegraphics[width=.12\textwidth]{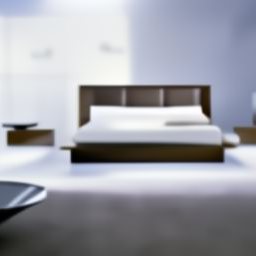}&\includegraphics[width=.12\textwidth]{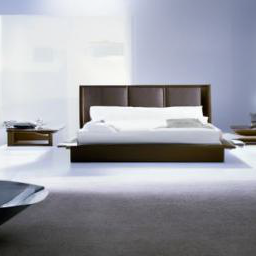}&\includegraphics[width=.12\textwidth]{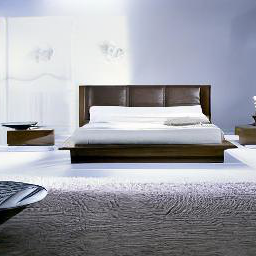}&\includegraphics[width=.12\textwidth]{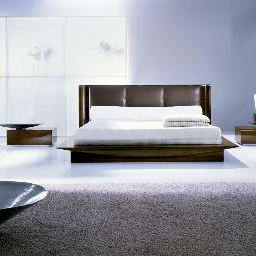}\\
        \includegraphics[width=.12\textwidth]{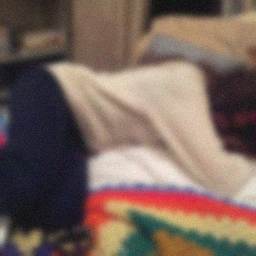}& \includegraphics[width=.12\textwidth]{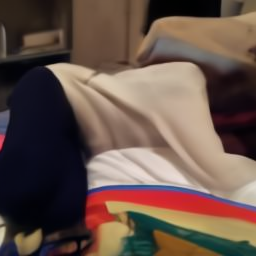}&\includegraphics[width=.12\textwidth]{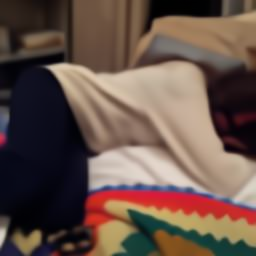}&\includegraphics[width=.12\textwidth]{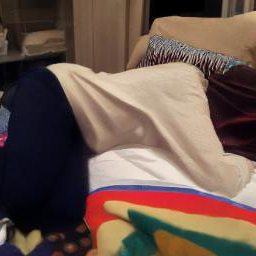}&\includegraphics[width=.12\textwidth]{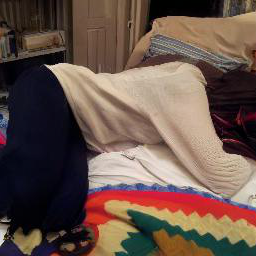}&\includegraphics[width=.12\textwidth]{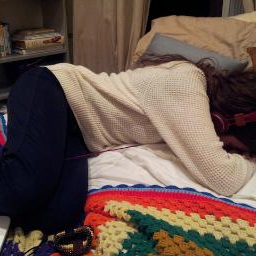}\\
            \includegraphics[width=.12\textwidth]{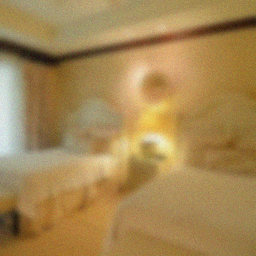}& \includegraphics[width=.12\textwidth]{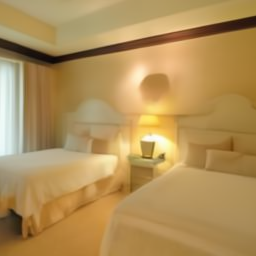}&\includegraphics[width=.12\textwidth]{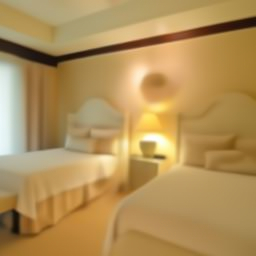}&\includegraphics[width=.12\textwidth]{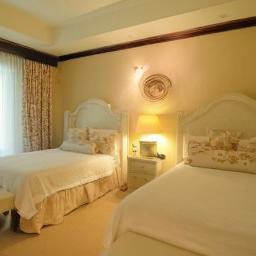}&\includegraphics[width=.12\textwidth]{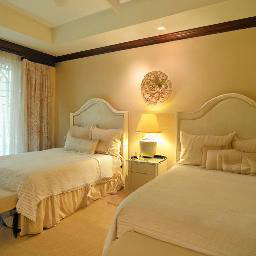}&\includegraphics[width=.12\textwidth]{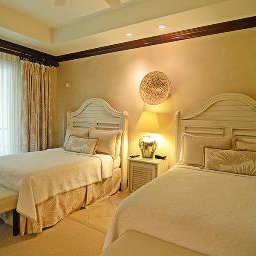}\\
                \includegraphics[width=.12\textwidth]{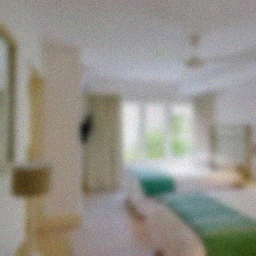}& \includegraphics[width=.12\textwidth]{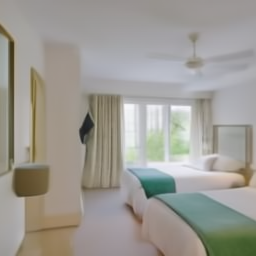}&\includegraphics[width=.12\textwidth]{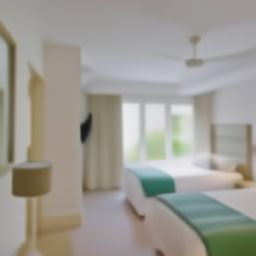}&\includegraphics[width=.12\textwidth]{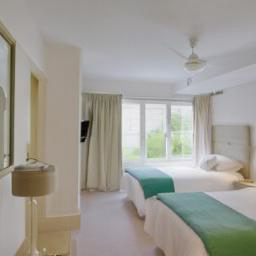}&\includegraphics[width=.12\textwidth]{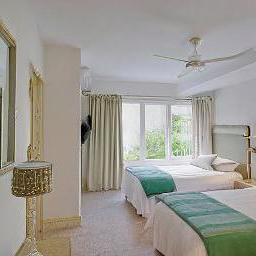}&\includegraphics[width=.12\textwidth]{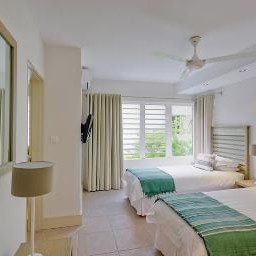}\\\cline{5-5}
    \end{tabular}
    \caption{Image Restoration Results for Gaussian Deblurring on LSUN Dataset. }
    \label{fig:appendix-deblur-g-lsun}
\end{figure}
\endgroup

\begingroup
\renewcommand{\arraystretch}{0.0}
\begin{figure}[h]
\setlength\arrayrulewidth{2pt}
\centering
\begin{tabular}{c@{}c@{}|@{}c@{}|@{}cc@{}c@{}|@{}c@{}|@{}c}
\cline{3-3}
\cline{7-7}
\specialrule{0em}{1pt}{1pt}
    {\small  Input}&{\small  DPS}&{\small \bf DPG}&{\small Ground Truth}&{\small  Input}&{\small  DPS}&{\small \bf DPG}&{\small Ground Truth}\\
    \includegraphics[width=.12\textwidth]{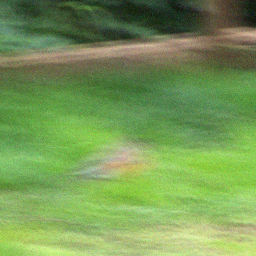}& \includegraphics[width=.12\textwidth]{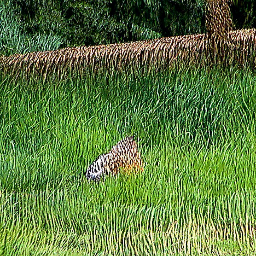}&\includegraphics[width=.12\textwidth]{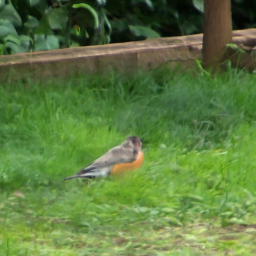}&\includegraphics[width=.12\textwidth]{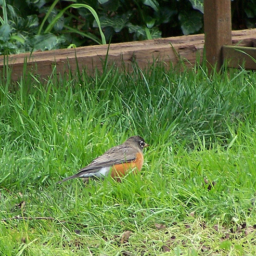}&
    \includegraphics[width=.12\textwidth]{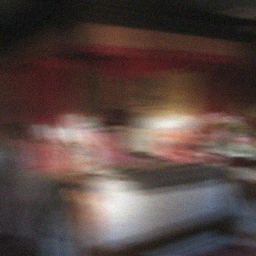}& \includegraphics[width=.12\textwidth]{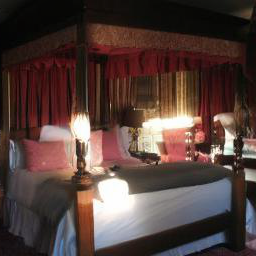}&\includegraphics[width=.12\textwidth]{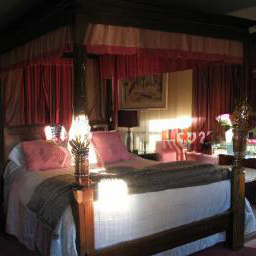}&\includegraphics[width=.12\textwidth]{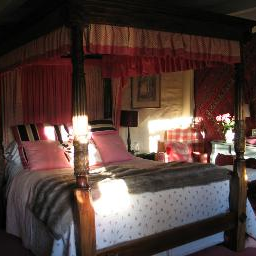}\\
    \includegraphics[width=.12\textwidth]{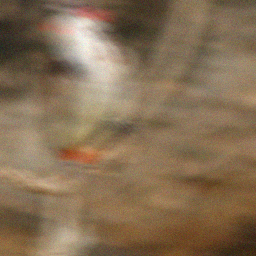}& \includegraphics[width=.12\textwidth]{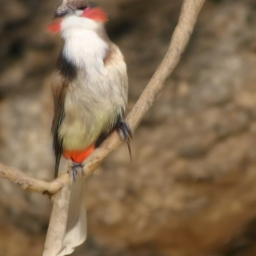}&\includegraphics[width=.12\textwidth]{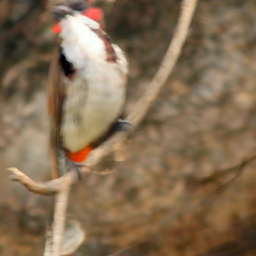}&\includegraphics[width=.12\textwidth]{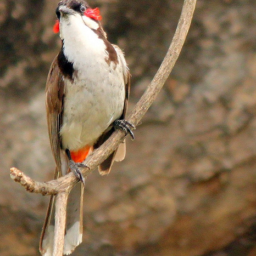}&
    \includegraphics[width=.12\textwidth]{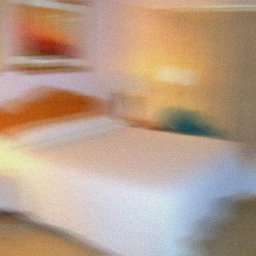}& \includegraphics[width=.12\textwidth]{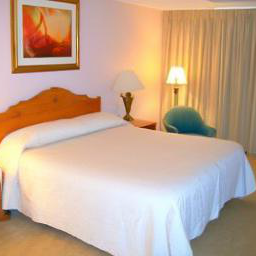}&\includegraphics[width=.12\textwidth]{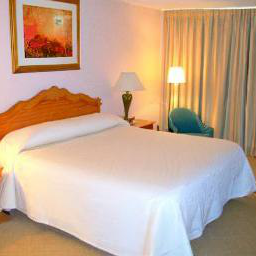}&\includegraphics[width=.12\textwidth]{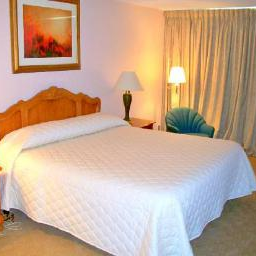}\\
        \includegraphics[width=.12\textwidth]{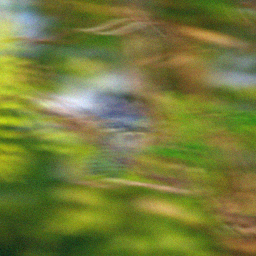}& \includegraphics[width=.12\textwidth]{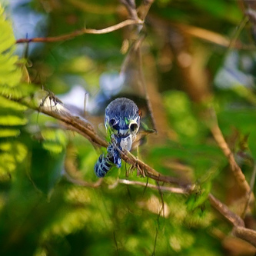}&\includegraphics[width=.12\textwidth]{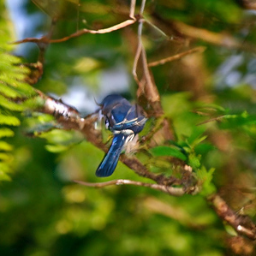}&\includegraphics[width=.12\textwidth]{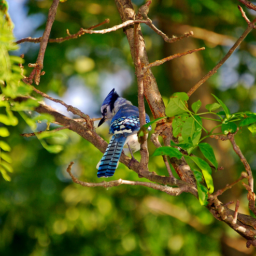}&
    \includegraphics[width=.12\textwidth]{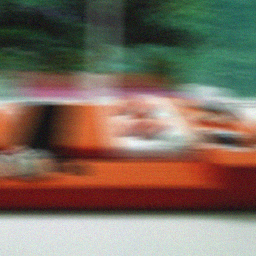}& \includegraphics[width=.12\textwidth]{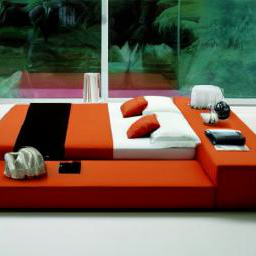}&\includegraphics[width=.12\textwidth]{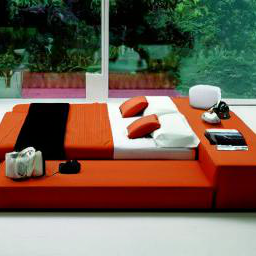}&\includegraphics[width=.12\textwidth]{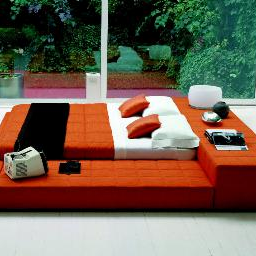}\\
        \includegraphics[width=.12\textwidth]{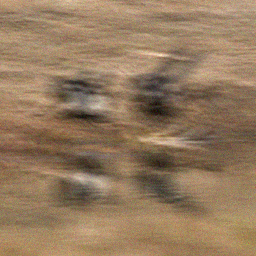}& \includegraphics[width=.12\textwidth]{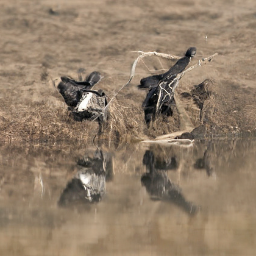}&\includegraphics[width=.12\textwidth]{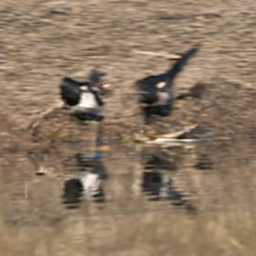}&\includegraphics[width=.12\textwidth]{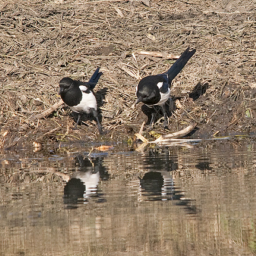}&
    \includegraphics[width=.12\textwidth]{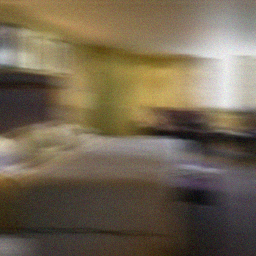}& \includegraphics[width=.12\textwidth]{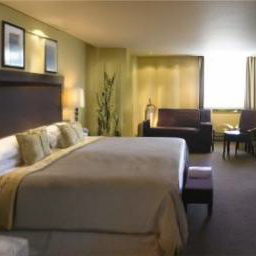}&\includegraphics[width=.12\textwidth]{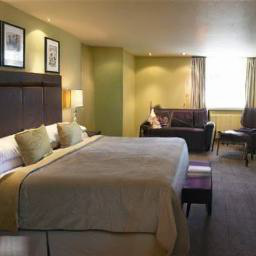}&\includegraphics[width=.12\textwidth]{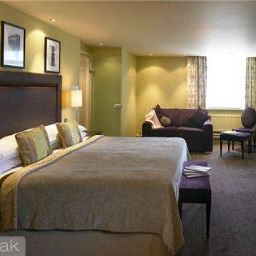}\\
        \includegraphics[width=.12\textwidth]{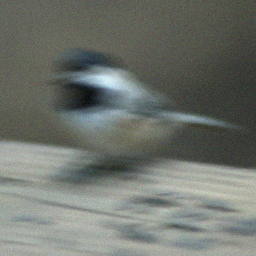}& \includegraphics[width=.12\textwidth]{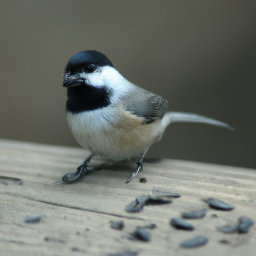}&\includegraphics[width=.12\textwidth]{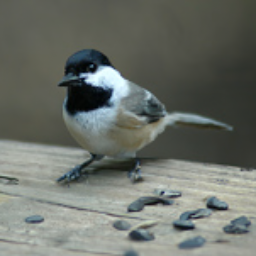}&\includegraphics[width=.12\textwidth]{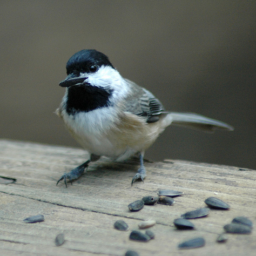}&
    \includegraphics[width=.12\textwidth]{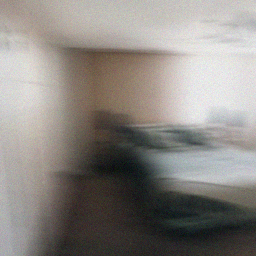}& \includegraphics[width=.12\textwidth]{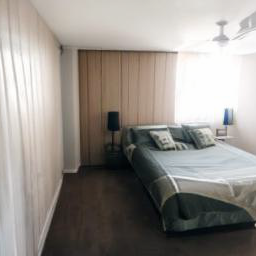}&\includegraphics[width=.12\textwidth]{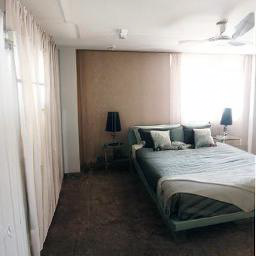}&\includegraphics[width=.12\textwidth]{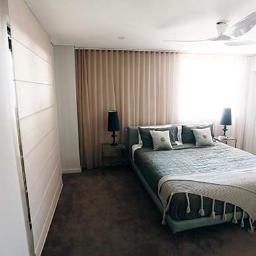}\\
    \cline{3-3}
    \cline{7-7}
    \end{tabular}
    \caption{Image Restoration Results for Motion Deblurring on ImageNet and LSUN dataset. }
    \label{fig:appendix-deblur-motion}
\end{figure}
\endgroup

\begingroup
\renewcommand{\arraystretch}{0.0}
\begin{figure}[h]
\setlength\arrayrulewidth{2pt}
\centering
\begin{tabular}{c@{}c@{}|@{}c@{}|@{}cc@{}c@{}|@{}c@{}|@{}c}
\cline{3-3}
\cline{7-7}
\specialrule{0em}{1pt}{1pt}
    {\small  Input}&{\small  DPS}&{\small \bf DPG}&{\small Ground Truth}&{\small  Input}&{\small  DPS}&{\small \bf DPG}&{\small Ground Truth}\\
    \includegraphics[width=.12\textwidth]{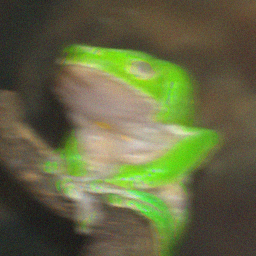}& \includegraphics[width=.12\textwidth]{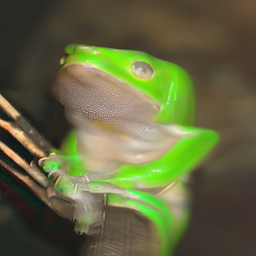}&\includegraphics[width=.12\textwidth]{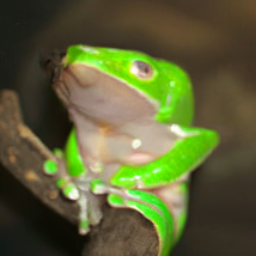}&\includegraphics[width=.12\textwidth]{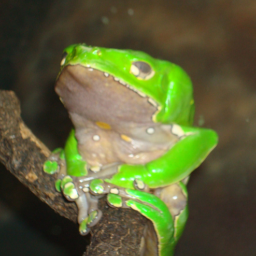}&\includegraphics[width=.12\textwidth]{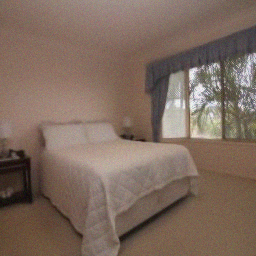}& \includegraphics[width=.12\textwidth]{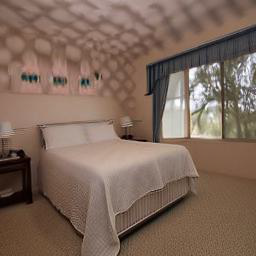}&\includegraphics[width=.12\textwidth]{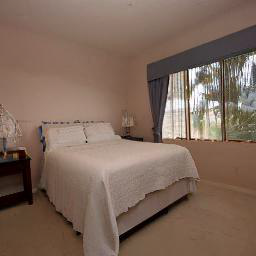}&\includegraphics[width=.12\textwidth]{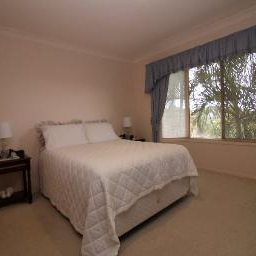}\\
    \includegraphics[width=.12\textwidth]{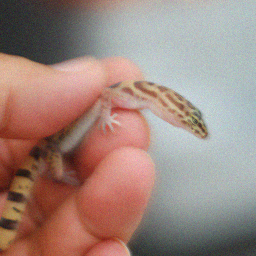}& \includegraphics[width=.12\textwidth]{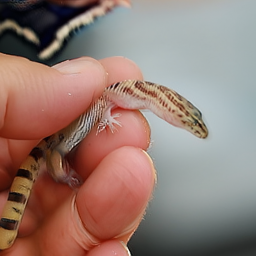}&\includegraphics[width=.12\textwidth]{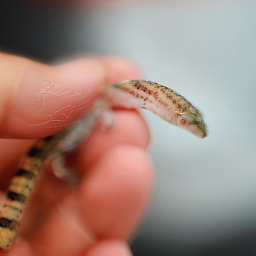}&\includegraphics[width=.12\textwidth]{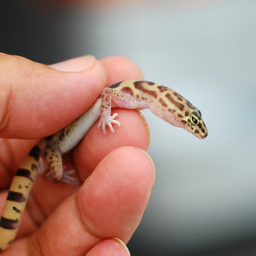}&\includegraphics[width=.12\textwidth]{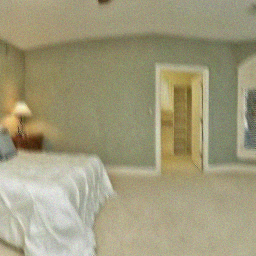}& \includegraphics[width=.12\textwidth]{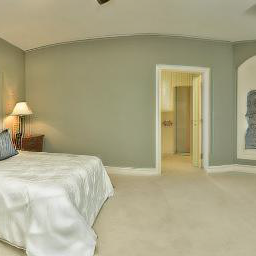}&\includegraphics[width=.12\textwidth]{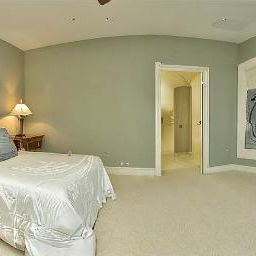}&\includegraphics[width=.12\textwidth]{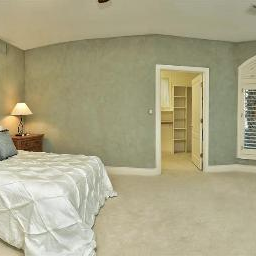}\\
        \includegraphics[width=.12\textwidth]{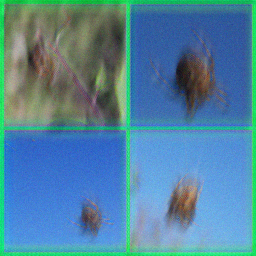}& \includegraphics[width=.12\textwidth]{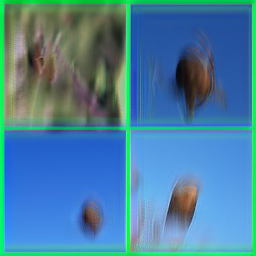}&\includegraphics[width=.12\textwidth]{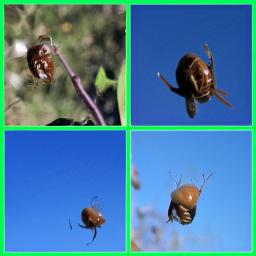}&\includegraphics[width=.12\textwidth]{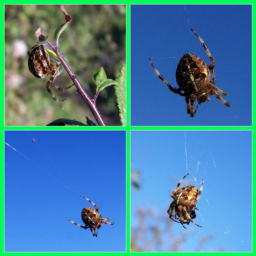}&        \includegraphics[width=.12\textwidth]{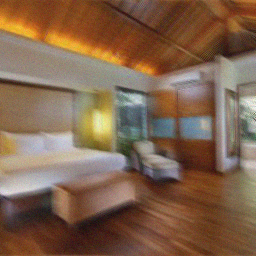}& \includegraphics[width=.12\textwidth]{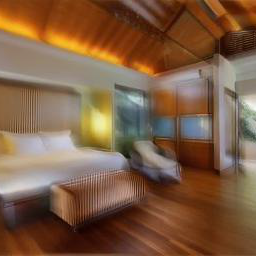}&\includegraphics[width=.12\textwidth]{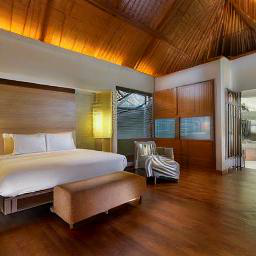}&\includegraphics[width=.12\textwidth]{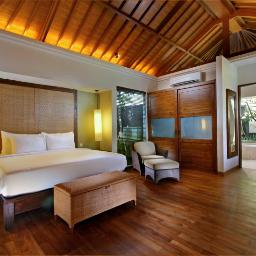}\\
            \includegraphics[width=.12\textwidth]{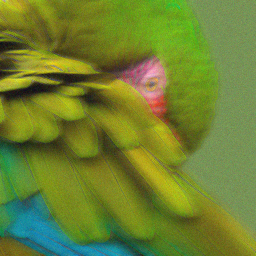}& \includegraphics[width=.12\textwidth]{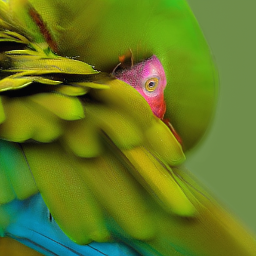}&\includegraphics[width=.12\textwidth]{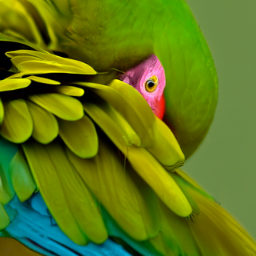}&\includegraphics[width=.12\textwidth]{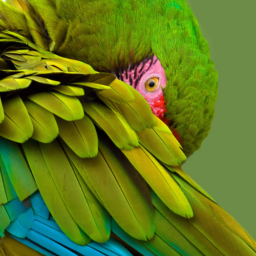}&            \includegraphics[width=.12\textwidth]{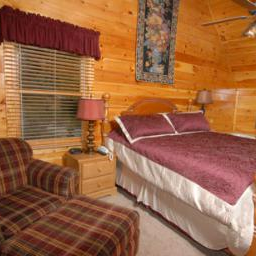}& \includegraphics[width=.12\textwidth]{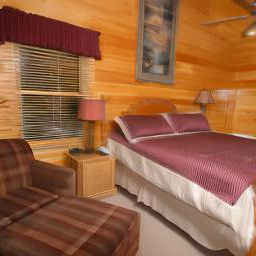}&\includegraphics[width=.12\textwidth]{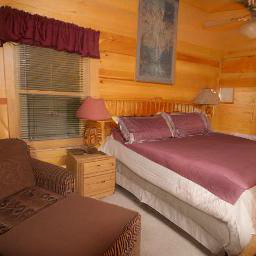}&\includegraphics[width=.12\textwidth]{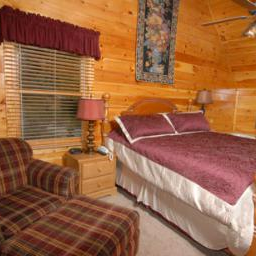}\\
                \includegraphics[width=.12\textwidth]{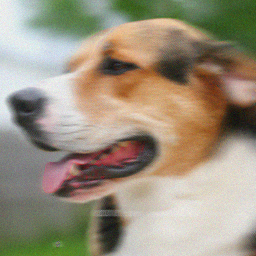}& \includegraphics[width=.12\textwidth]{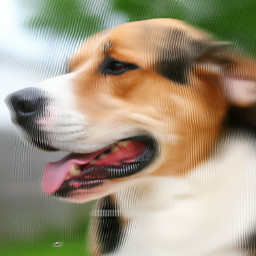}&\includegraphics[width=.12\textwidth]{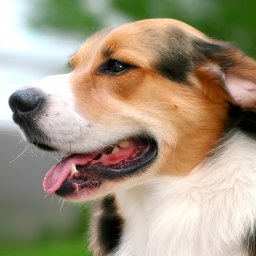}&\includegraphics[width=.12\textwidth]{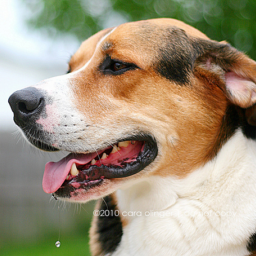}&                \includegraphics[width=.12\textwidth]{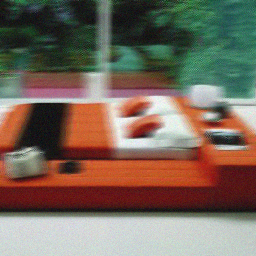}& \includegraphics[width=.12\textwidth]{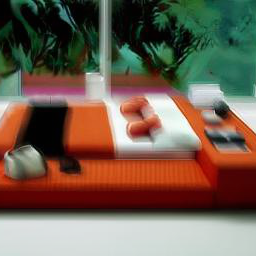}&\includegraphics[width=.12\textwidth]{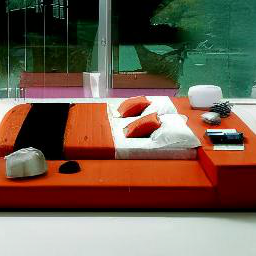}&\includegraphics[width=.12\textwidth]{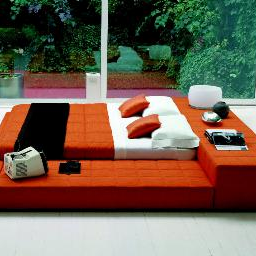}\\
\cline{3-3}
\cline{7-7}
\end{tabular}
    \caption{Image Restoration Results for non-linear deblurring on ImageNet and LSUN dataset. }
    \label{fig:appendix-nonlinear}
\end{figure}
\endgroup



\end{document}